\documentclass[fleqn,10pt]{wlscirep}
\usepackage{amsmath,amsthm,amssymb}
\usepackage[utf8]{inputenc}
\usepackage[T1]{fontenc}
\usepackage{bm}
\usepackage{amsmath,amsthm,amssymb}
\usepackage{physics}
\usepackage[capitalise]{cleveref}
\usepackage{siunitx}
\usepackage{subfigure}
\usepackage{multicol}
\usepackage[normalem]{ulem}
\usepackage{tcolorbox}
\tcbuselibrary{skins, breakable}
\usepackage[font=small]{caption}
\newtcolorbox{naturebox}[1]{
    enhanced,
    breakable,
    colback=gray!5,      
    colframe=black,      
    coltitle=black,
    colbacktitle=gray!5,
    left=6pt,
    right=6pt,
    top=6pt,
    bottom=6pt,
    before skip=12pt,
    after skip=12pt,
    fonttitle=\bfseries\sffamily,
    title=#1,
    attach title to upper,
    boxrule=0.5pt        
}

\title{New frontiers in quantum science and technology using van der Waals Josephson junctions}
\author[1]{Joydip Sarkar}
\author[1]{Ayshi Mukherjee}
\author[1]{Amit Basu}
\author[2]{Ritajit Kundu}
\author[2]{Arijit Kundu}
\author[1,*]{Mandar M. Deshmukh}
\affil[1]{Department of Condensed Matter Physics and Materials Science, Tata Institute of Fundamental Research, Homi Bhabha Road, Mumbai 400005, India.}
\affil[2]{Department of Physics, Indian Institute of Technology Kanpur, Kanpur 208016, India.}

\affil[*]{E-mail: deshmukh@tifr.res.in}

\begin{abstract}
Over the last decade, the development of Josephson devices based on van der Waals (vdW) materials has advanced rapidly, representing a paradigm shift driven by the advent of 2D materials. The diverse vdW materials library, combined with advanced fabrication techniques, enables the integration of materials with vastly disparate properties for scientific exploration. The vdW Josephson junctions (JJs) offer a unique route to explore novel functionalities and associated physics that remain inaccessible in conventional JJs, which have reached an industrial level in terms of fabrication.
Beyond material diversity, vdW crystalline materials offer fundamental new control over device symmetries, enabling the realization of Hamiltonians unique to 2D systems. Furthermore, the long relaxation times of myriad excitations in 2D heterostructures open possibilities for creating exquisite quantum sensors, with the 2D material itself acting as an efficient bus for transmitting excitations to the active sensing element. This creative explosion in vdW-based superconducting electronics is rapidly growing, and our review highlights the resulting devices and physics. The confluence of vdW JJs with twistronics and topology has the potential to redefine superconducting quantum technology, enabling applications from quantum computation to ultra-sensitive hybrid sensors. While opportunities abound with vdW JJs, the challenge of scalability must be surmounted for translation into real-world devices. This review synthesizes current developments and offers a roadmap for researchers navigating this burgeoning field.

\end{abstract}

\begin{document}

\flushbottom
\maketitle

\thispagestyle{empty}

\section*{Introduction}

The discovery of new materials has always been crucial for human prehistory and technological progress, and the ongoing quantum era is no exception.  The quantum revolution intends to leverage quantum mechanics for new technologies, with potential applications in computing, secure communication, and sensing~\cite{deutsch_harnessing_2020}. Significant advances have been made in the fields of quantum computing and sensing~\cite{devoret_superconducting_2013,degen_quantum_2017}. As a result, it has now become feasible to simulate complex many-body problems on a quantum processor~\cite{google_ai_quantum_and_collaborators_hartree-fock_2020,browaeys_many-body_2020} and build highly sensitive quantum sensors~\cite{finkler_scanning_2012,vasyukov_scanning_2013,rovny_nanoscale_2024}. Although these platforms are state-of-the-art, they each have their own advantages and limitations, leading researchers to look for systems with improved performance in terms of coherence time~\cite{place_new_2021}, scalability, and device functionality, motivating the search for improved architectures~\cite{siddiqi_engineering_2021}.
This search has naturally extended into the field of materials science, where the properties of the underlying building blocks can set the limits of device performance~\cite{noauthor_qubits_2021}. In particular, there is a focus on the use of van der Waals (vdW) materials for realizing compact devices with novel control knobs~\cite{liu_two-dimensional_2020,montblanch_layered_2023,wang2024two}.

Superconducting devices are central to these efforts, with the Josephson junction (JJ) lying at the heart of many superconducting devices. In circuit quantum electrodynamics (cQED) architectures, JJ is the key component forming the foundation of much of today’s quantum technology~\cite{kim_josephson_2025}, ranging from a wide variety of superconducting qubits~\cite{devoret_superconducting_2013,krantz_quantum_2019}, parametric amplifiers~\cite{vijay_invited_2009,aumentado_superconducting_2020}, on-chip circulators~\cite{sliwa_reconfigurable_2015,navarathna_passive_2023}, superconducting quantum interference device (SQUID) magnetometers~\cite{drung_integrated_1996,hatridge_dispersive_2011,vasyukov_scanning_2013}, radiation sensors~\cite{lee_graphene-based_2020,walsh_josephson_2021}, and as a voltage standard in metrology~\cite{benz_application_2004}, among others. Being a building block, understanding the physics and limitations of a JJ is central to advancing the field. A JJ is a quantum element formed by coupling two superconductors through a weak link, enabling coherent transfer of Cooper pairs -- the charge carriers in the macroscopic quantum state of superconductors~\cite{Golubov_2004}, see \hypersetup{linkcolor=black}\hyperref[box1]{Theory Box}. The weak link allows for the superconducting phase to have an abrupt drop across a JJ. The dissipationless current flowing via such a weak link is a function of the phase drop, and it is described by the current-phase relation (CPR).
The weak link can take various structural forms and utilizes different architectures – S-I-S, where a thin insulating barrier separates the superconductors; S-N-S, where a normal metal bridges the superconductors; and S-S$^{\prime}$-S, where a narrow superconductor with a reduced critical current acts as a weak link. Electrically, the JJ is analogous to an inductor; however, it is distinct because its inductance is a function of the current flowing through it, rendering it a non-linear inductor. Non-linearity is the crucial reason the JJ is at the heart of current superconducting quantum circuits. Combined with shunt capacitors, JJs form artificial atoms, structured as an anharmonic LC oscillator. Such circuit-based architecture forms the basis for many superconducting qubits~\cite{blais_circuit_2021}. A pedagogical description of the JJ is provided in the \hypersetup{linkcolor=black}\hyperref[box1]{Theory Box}.

Since their inception about 70 years ago, JJs have been instrumental in probing quantum phenomena and in the realization of quantum circuits -- from observation of macroscopic quantum tunneling~\cite{clarke_quantum_1988} to present-day quantum processors~\cite{arute_quantum_2019}. The conventional JJs, mainly based on aluminum/niobium and their self-limiting amorphous oxide, have dominated due to their reproducibility and compatibility with lithographic techniques~\cite{van_damme_advanced_2024}. The rapid growth of superconducting quantum circuits is a testament to the power of JJs. However, this technological platform faces some challenges. In practice, the spatial non-uniformity of tunnel barriers leads to variation in the qubit properties within a single processor~\cite{zeng_direct_2015,zeng_atomic_2016}. Studies have shown that tunnel barriers are typically composed of amorphous oxides, which host lossy two-level defects that lead to decoherence in the qubits~\cite{martinis_decoherence_2005,shalibo_lifetime_2010}. In addition, the material choice is restricted to a small subset of superconductors with an isotropic order parameter. As a result, these JJs offer only a few knobs for tuning their properties -- for instance, the magnetic field is often the only available control knob. In recent years, vdW materials have emerged as a powerful alternative platform for realizing Josephson devices.

\section{Context to vdW Josephson Junctions} 
\label{sec:1}
In the last decade, there has been rapid progress in the variety of Josephson devices made using vdW materials, allowing new device architectures with exquisite tunability. vdW materials have an extensive library \cite{geim_van_2013,mounet_two-dimensional_2018} -- semimetals such as graphene that can be proximitized by contacting a superconductor, s-wave superconductors, d-wave superconductors, ferromagnets, and insulators to form tunnel barriers. Combining magnetic materials with superconductors to form abrupt atomic interfaces, it is possible to realize superconducting devices that are challenging to realize using conventional materials, where interfacial mixing is typically detrimental to performance. The ability to tune the density of charge carriers electrostatically in semimetals, like graphene, and semiconductors, is an avenue to tune JJs electrostatically rather than with a magnetic field.
This approach offers a significant advantage by mitigating crosstalk in intricate superconducting circuits, thereby enhancing their scalability, performance, and reliability.
In some of the device architectures, materials such as graphene not only host the active sensor but also function as a "bus" to carry excitations, such as photons, phonons, and magnons, because of a large relaxation length~\cite{walsh_graphene-based_2017,tombros_electronic_2007}. 
The use of \textit{d}-wave high-T$_c$ superconductors further expands the landscape of possible JJ architectures, as the relative twist angle between layers can strongly influence the superconducting order parameter and thereby, the symmetries of the junction due to anisotropy~\cite{can_high-temperature_2021}.  
Similarly, graphene JJs provide a platform to implement the Floquet Hamiltonian~\cite{rudner_band_2020,Park_2022}, because of the favorable energy scale and the ability to "pump" the system. These examples illustrate the unique opportunities offered by vdW materials that did not exist a few years ago.

Over the past ten years, creative techniques have been developed to assemble, fabricate, and probe vdW JJs, and our review will cover this rapidly growing area of superconducting electronics and associated physics. 
We first highlight the significant advantages that the vdW materials library provides to quantum device architectures in \hypersetup{linkcolor=black}\hyperref[sec:2]{section 2}. 
Following that, \hypersetup{linkcolor=black}\hyperref[sec:3]{section 3} discusses the graphene S-N-S JJs and their applications. From there, we transition into JJs with moiré superconductors in \hypersetup{linkcolor=black}\hyperref[sec:4]{section 4}. \hypersetup{linkcolor=black}\hyperref[sec:5]{Section 5} focuses on fully vdW S-I-S junctions, and \hypersetup{linkcolor=black}\hyperref[sec:6]{section 6} explores topological vdW JJs. Together with the rapidly emerging field of twistronics and topology are likely to enable the development of new quantum technologies, enabling applications in quantum computation, quantum hardware design, and sensing.
Finally, we address the challenges associated with the successful integration of vdW JJs in quantum devices and outline a prospective direction for the evolution of this field, highlighting intriguing avenues for future research in \hypersetup{linkcolor=black}\hyperref[sec:7]{section 7} and \hypersetup{linkcolor=black}\hyperref[sec:8]{8}, respectively.

\section{Unique Opportunities Provided by vdW JJs }
\label{sec:2}
\begin{figure*}[h!]
    \centering
	\includegraphics[width=17cm]{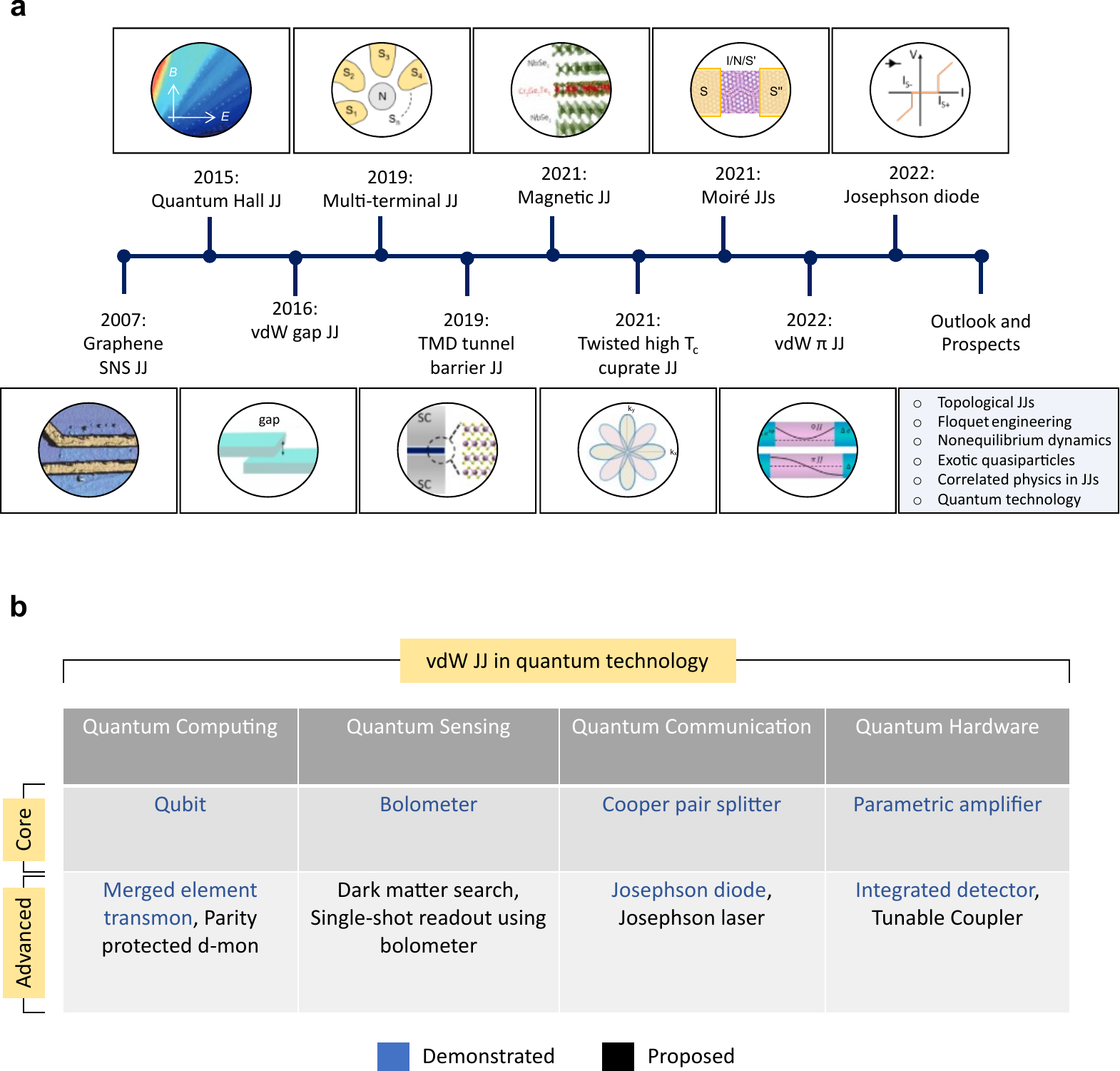}
    \caption{ \label{fig:roadmap} {\textbf{Roadmap of vdW Josephson junctions.} 
    \textbf{a,}~Shows the developmental timeline of the different vdW Josephson junctions. First experimental realization of graphene JJ~\cite{heersche_bipolar_2007}, quantum Hall physics explored in graphene JJ~\cite{amet_supercurrent_2016}, JJs utilizing the interfacial gap between two vertically stacked vdW flakes~\cite{yabuki_supercurrent_2016}, multi-terminal JJs for exploring topological matter~\cite{Draelos_2019,Riwar_2016,meyer_nontrivial_2017}, crystalline insulator based vdW tunnel JJs~\cite{lee_two-dimensional_2019}, JJs incorporating magnetic insulators~\cite{ai_van_2021}, twisted high $T_c$ cuprate JJs~\cite{lee_twisted_2021}, moiré JJs in twisted graphene~\cite{rodan-legrain_highly_2021}, vdW $\pi$-JJ~\cite{kang_van_2022}, Josephson diode effect~\cite{wu_field-free_2022,nadeem_superconducting_2023} among others. The light-blue text box highlights prospective research directions that motivate further explorations -- such as topological JJs~\cite{choi2020evidence}, Floquet engineering and non-equilibrium dynamics in JJs~\cite{Park_2022,Merboldt_2025}, Majorana quasiparticle in JJs~\cite{jiang_unconventional_2011,pino_unpaired_2015}, correlation-driven flat band physics in JJs~\cite{PhysRevResearch.7.023273 ,diez2025probing}, and pathways toward advancing state-of-the-art quantum technologies.
    \textbf{b,}~Shows the application opportunities of vdW JJs in quantum technology. Various quantum devices have been realized based on vdW JJs, opening up possibilities for future technology, starting from superconducting qubits~\cite{wang_coherent_2019}, bolometers~\cite{lee_graphene-based_2020,kokkoniemi_bolometer_2020,walsh_josephson_2021}, Cooper pair splitter~\cite{pandey_ballistic_2021}, quantum-noise-limited amplifiers~\cite{sarkar_quantum-noise-limited_2022,butseraen_gate-tunable_2022}, which further opens up possibilities for new device ideas, such as compact merge element transmon qubits~\cite{balgley_crystalline_2025}, parity protected superconducting qubits~\cite{patel_d_2024,brosco_superconducting_2024}, dark matter search using graphene JJ bolometers~\cite{ braine_extended_2020,paolucci_fully_2021}, single-shot readout of quantum states using bolometers~\cite{opremcak_measurement_2018,gunyho_single-shot_2024}, Josephson diode~\cite{wu_field-free_2022,diez-merida_symmetry-broken_2023,zhao_time-reversal_2023,ghosh_high-temperature_2024}, Josephson laser~\cite{welp_superconducting_2013,cassidy_demonstration_2017,simon_theory_2018}, integrated detectors incorporating parametric amplifiers~\cite{sarkar_kerr_2025}, tunable inductor coupler~\cite{schmidt_ballistic_2018,jha2025large}, among others.}}
    
\end{figure*}

Conventional JJs employing oxide tunnel barriers are commonly used as a nonlinear inductor, where the inductance value is tailored by tuning the oxide barrier properties, microwave current bias, or magnetic flux. In recent years, semiconductor-based JJs have provided electrostatic control of Josephson inductance and have opened up new possibilities such as superconducting qubits with proximitized $2$D electron gas ($2$DEG)~\cite{larsen_semiconductor-nanowire-based_2015,de_lange_realization_2015}, and parity-protected qubits~\cite{larsen_parity-protected_2020} among others. These developments illustrate how modifying the weak link material can dramatically expand the design space for Josephson devices. In practice, the weak link in a JJ can be of different types: an insulator, a normal metal, a ferromagnet, and a semiconductor, among others~\cite{Golubov_2004,geim_van_2013}. Depending on the nature of the weak link, the Cooper pair transfer process changes, adding novel functionalities with new tuning knobs. This naturally motivates the exploration of vdW crystalline materials as new weak links, where their unique properties can be directly harnessed, as discussed in the following subsections. Fig.~\ref{fig:roadmap}a shows the developmental roadmap for various vdW JJs over time and their role in advancing physics and quantum technology, see Fig.~\ref{fig:roadmap}b.

\subsection{Extensive Materials Library} 
Following the rise of graphene~\cite{geim_rise_2007}, various $2$D vdW materials have been discovered that exhibit nearly all phases of condensed matter, including magnets, superconductors, semiconductors, semi-metals, and band insulators, among others~\cite{geim_van_2013}. The interest in vdW materials has grown for both fundamental research and technological advancements~\cite{liu_two-dimensional_2020,montblanch_layered_2023,wang2024two}. With the recent development of $2$D moiré systems, there are new opportunities for quantum simulations~\cite{deshmukh_simple_nodate,andrei_marvels_2021}, enabling the testing of complex Hamiltonians such as the Hubbard model that describes many-body physics. Given the extensive library of $2$D materials, recent interest has also shifted towards integrating cQED devices with vdW materials. Inspired by the work on superconductor-semiconductor hybrids, which offer potential for qubits with topological protection and gate tunability~\cite{larsen_parity-protected_2020,larsen_semiconductor-nanowire-based_2015,de_lange_realization_2015,casparis_superconducting_2018}, there has been exploration of graphene-based JJs as we discuss extensively later in \hypersetup{linkcolor=black}\hyperref[sec:2.3]{section 2.3}. For cryogenic commercial applications, utilizing vdW high- T$_\mathrm{c}$ superconductors, various superconducting electronic devices have been proposed and demonstrated, such as Josephson diodes and qubits, among others, which can operate at 77 K~\cite{ghosh_high-temperature_2024,confalone_cuprate_2025}.

\subsection{Combining Materials beyond epitaxy} 
Growing abrupt interfaces between different materials, for example, superconductors and ferromagnets, is challenging because of mutual incompatibility and mixing of materials~\cite{gao_reaction_1987,buzdin_proximity_2005}. Here again, vdW materials provide a unique advantage because their assembly does not rely on lattice matching. The vdW heterostructures are typically stacked by combining flakes one by one using a transparent polymer film~\cite{onodera_assembly_2020}. The layer-by-layer assembling of a 2D heterostructure opens up exciting possibilities that are unavailable with epitaxial growth techniques such as chemical vapor deposition, molecular beam epitaxy, and atomic layer deposition. Here, the vdW materials and resulting devices offer the unique opportunity of combining crystalline ferromagnets and superconductors, among others, with remarkable interfacial integrity~\cite{ai_van_2021,idzuchi_unconventional_2021,kang_van_2022}.
Atomically thin interfaces have enabled the realization of vdW JJs and various interfacial phenomena, such as chiral superconducting order ~\cite{can_high-temperature_2021,zhao_time-reversal_2023}. As we advance, the identification of specific material combinations that lead to different weak links will be discussed in greater detail in the following sections.

\subsection{Gate Tunability of Superconducting Properties} 
\label{sec:2.3}
Experimental realizations of superconductor-normal-superconductor
(SNS) Junctions using graphene as the weak links provide a platform for creating gate-tunable JJs. These SNS JJs offer electrostatic control over switching currents or Josephson inductance in a field effect transistor (FET) geometry. This represents a significant shift from conventional oxide-barrier JJs, where tuning typically relies on magnetic flux using a SQUID-based geometry. Fig.~\ref{fig:grjj}a shows a schematic of planar graphene JJ and compares it with vertically formed S-I-S tunnel JJ. The graphene JJ uses a local gate, either at the top or bottom of the graphene channel, for electrostatic control of the switching current ($I_\mathrm{c}$) and consequently the junction inductance ($L_\mathrm{J}$). The gate tuning also enables control of the CPR, which in turn determines the degree of nonlinearity~\cite{haller_phase-dependent_2022}. Recently, there has been extensive use of graphene JJs in various quantum devices, which has led to state-of-the-art performance in the form of quantum noise-limited amplifiers~\cite{sarkar_quantum-noise-limited_2022} and single photon bolometric sensors~\cite{walsh_josephson_2021}. The electrostatic control is especially advantageous over magnetic flux control in cQED devices as it uses a very localized electric field, which minimizes cross-talk. 

\subsection{Easy to Interface with Existing Superconducting Device Architecture} 
The vdW JJs are integrable with the existing cQED device architecture because of their ease of fabrication and small footprints. This compatibility is crucial, since it ensures that advances enabled by vdW materials can be adopted without the need for an entirely new hardware platform. The JJs are made in a typical three-terminal FET geometry, with two leads for the JJ and an on-chip DC line for applying the local gate voltage. This leads to the usage of vdW JJs in various superconducting device architectures for optimal performance and with new tuning knobs. Importantly, the vdW approach is not limited to junctions alone but extends naturally to other circuit elements. In addition to junctions~\cite{lee_two-dimensional_2019,tian_josephson_2021,balgley_crystalline_2025,wang_two-dimensional_2024}, there is also a search for better dielectrics for making compact and low-loss capacitors in cQED devices. Studies have found hexagonal Boron Nitride (hBN) as a low-loss dielectric element for capacitors in transmon qubits that minimize the footprint of the device~\cite{antony_miniaturizing_2021,wang_hexagonal_2022}.

\subsection{Twist Angle to Engineer Wave Functions and Order Parameters} 
Beyond material choice and gating, another powerful knob offered by vdW systems is the relative twist angle between stacked layers. Engineering devices with a twist angle has opened up the possibility of exploring the physics of electronic interaction in the 2D or quasi-2D limit. The angle of rotation, or twist, between two stacked layers of a material plays a crucial role in determining the electronic hybridization between the two layers. In a twisted graphene system, a small twist angle modifies the Hamiltonian and gives rise to the emergent band structure of the combined system by hybridizing the electronic band of each graphene layer; this has given birth to a new field called moir\'e physics~\cite{doi:10.1073/pnas.1108174108,cao2018unconventional,andrei_marvels_2021, adak2024tunable}. For a particular value of the twist angle known as the'magic angle', this emergent band becomes very flat, resulting in superconductivity and other correlated phases. In the case of transition metal dichalcogenides (TMDC), a large angle of twist ($\sim$5\textdegree) between two flakes can give rise to superconductivity~\cite{xia_superconductivity_2025,guo_superconductivity_2025}. Besides graphene and TMDC systems, hybrid twisted high-$T_c$ superconducting systems have recently drawn the interest of the community. By introducing the twist between two layers, one can exploit the symmetry of the order parameter associated with the superconducting planes of these systems. Some theoretical work suggests that one can get an emergent time-reversal symmetry-broken topological superconducting order at the twisted interface~\cite{can_high-temperature_2021}; this is supported by recent experiments, an aspect we discuss later in detail in \hypersetup{linkcolor=black}\hyperref[sec:5.2]{section 5.2}. 

\subsection{Floquet Studies to Realize Synthetic Hamiltonians} 
In a quantum system, if the Hamiltonian is time-periodic, it leads to the appearance of quasi-energy levels; this physics is well described by Floquet theory and is a route to engineering non-equilibrium quantum states~\cite{rudner_band_2020}. In addition to structural control through twist, time-periodic driving provides another route to engineer novel superconducting states in vdW JJs. In particular, graphene offers an ideal platform for studying Floquet physics, as such low-dimensional electronic systems can be coupled to an external electric field without screening, but at the same time, it needs careful engineering, as ultrafast decoherence times of electrons hinder the generation of steady Floquet states\cite{Heide_2021,Merboldt_2025}. Graphene-based JJs yield larger Josephson current in comparison to traditional JJs, due to the unique electronic structure and high mobility of the electronic bands~\cite{heersche_bipolar_2007,lee_ultimately_2015} and transitions between mesoscopic regimes (e.g., short versus long junction behavior) can be achieved by tuning the Fermi energy of graphene~\cite{Bretheau_2017}. A key example of Floquet engineering, which uses periodic driving to generate new electronic states~\cite{Merboldt_2025}, was presented in Ref. \citeonline{Park_2022}, where irradiating the ABS in a graphene JJ with GHz radiation was claimed to produce steady-state Floquet-Andreev states stabilized by the presence of dissipative processes, resulting in peaks in differential conductance above and below the original ABS peaks as the microwave power was increased. While alternate propositions exist\cite{Haxell_2023}, there is debate within the community about the microscopic mechanism of the observation.
They underscore the need for better theoretical understanding, where non-equilibrium effects of the system and the probe can be treated within a unified framework, as well as further experiments.

\section{Josephson Devices Combining Graphene and Conventional Superconductors}
\label{sec:3}
The rapid development of vdW JJs has been propelled by a series of groundbreaking experiments and theoretical insights. While the idea of Josephson devices was proposed for tunneling architecture, later proximitized Josephson devices were realized. Graphene, as a semi-metal with tunable carrier density, offered a natural starting point for realizing SNS JJs with bipolar response (see \hypersetup{linkcolor=black}\hyperref[box1]{Theory Box} for introduction to SNS JJs). Graphene was also a fortunate choice for realizing transparent contacts, as its work function matches well with many conventional superconductors and it avoids the formation of large Schottky barriers, resulting in clean and highly transparent contacts.
In addition, with hBN-encapsulated graphene, the large mean free path led to explorations of ballistic Josephson junctions~\cite{dean_boron_2010}. The following subsections review the major advances using graphene JJs.

\subsection{Graphene-Based Bipolar S-N-S JJs} 
Graphene-based junctions have provided a fruitful platform, enabling detailed investigations of proximity-induced superconductivity. Besides gate-tunable bipolar supercurrent, graphene JJs can have large critical current densities and high contact transparency, motivating researchers to explore them for various applications. The first generation of graphene JJ FET devices (GrJoFET) was made by surface contact of the graphene flake with superconductors~\cite{heersche_bipolar_2007,du_josephson_2008,girit_tunable_2009,ojeda-aristizabal_tuning_2009,borzenets_phase_2011,komatsu_superconducting_2012,rickhaus_quantum_2012} with a gate and field-tunable supercurrent. Later, edge contact with the graphene flake with superconductors helped realize clean SNS junctions having transparent contacts~\cite{Calado_2015}. Fig.~\ref{fig:grjj}b shows an optical image of the graphene JJ samples. The microscopic mechanism of SNS JJs (see \hypersetup{linkcolor=black}\hyperref[box1]{Theory Box}) relies on the formation of Andreev bound states in the weak link region. The Andreev bound states, enabled by the transparent contacts, play a crucial role in graphene JJs, and these are depicted in Fig.~\ref{fig:grjj}c. Spectroscopy of these Andreev bound states has also been performed using a tunnel junction on top of the graphene weak link -- a unique advantage of the vdW architecture not possible in conventional SNS JJs (see Fig.~\ref{fig:grjj}d).  

Large supercurrent densities  (up to $\sim200$ nA $\mu$m$^{-1}$) were observed over a long channel length up to several $\mu$m of the JJ. See Fig.~\ref{fig:grjj}e for gate-tunable switching currents in these devices. The JJ showed ballistic transport of the carriers, evidenced by oscillations in both the switching current and normal-state resistance, due to phase-coherent interference caused by the formation of a Fabry–Pérot cavity along the length of the junction. Subsequently, several studies have been carried out to characterize supercurrent transport in graphene JJs having different architectures and geometries~\cite{mizuno_ballistic-like_2013,choi_complete_2013,borzenets_phonon_2013,lee_ultimately_2015,borzenets_ballistic_2016,ben_shalom_quantum_2016,allen_spatially_2016,zhu_edge_2017,nanda_current-phase_2017,park_short_2018,larson_zero_2020,kalantre_anomalous_2020,dou_microwave_2021,haller_phase-dependent_2022,huang_study_2022,schmidt_tuning_2023,huang_observation_2023,messelot_direct_2024,schmidt_anisotropic_2025}. These findings established that ballistic transport and high transparency could be reliably achieved in graphene devices, paving the way for integration into cQED devices. Fig.~\ref{fig:grjj}f lists the key advantages of graphene JJs over the conventional S-I-S tunnel JJs for practical device applications.

\subsection{Graphene JJ Operational in the Linear Regime as an Inductor} 
Following the demonstration of ballistic transport in GrJoFET, there has been a growing interest in integrating these elements in various cQED devices. As we discussed in the \hypersetup{linkcolor=black}\hyperref[box1]{Theory Box}, the JJs act as an inductor with AC drive. Fig.~\ref{fig:grJJqubit}a shows the scheme of using gate-tunable JJs as inductors in microwave resonators. These inductive elements are compatible with the pre-existing cQED device architectures. Schmidt et al.~\cite{schmidt_ballistic_2018} demonstrated a gate-tunable superconducting inductor element by coupling a graphene JJ to a quarter-wave resonator, see Fig.~\ref{fig:grJJqubit}b. The authors used a DC bias-compatible resonator for a simultaneous DC+microwave probe on the device. Their experiments shed light on the loss mechanisms in the GrJoFET. Although graphene JJs showed zero resistance in DC transport experiments, it was discovered that they exhibit losses when embedded into microwave circuits. Their systematic study helped to understand the loss mechanism in the JJ arising from the subgap quasiparticles in the $\mu$eV energy scales, which are inaccessible in DC transport experiments~\cite{schmidt_ballistic_2018}.

\subsection{Nonlinear Graphene JJ Enables Realization of a Qubit}
\begin{figure*}[h!]
    \centering
	\includegraphics[width=17.5cm]{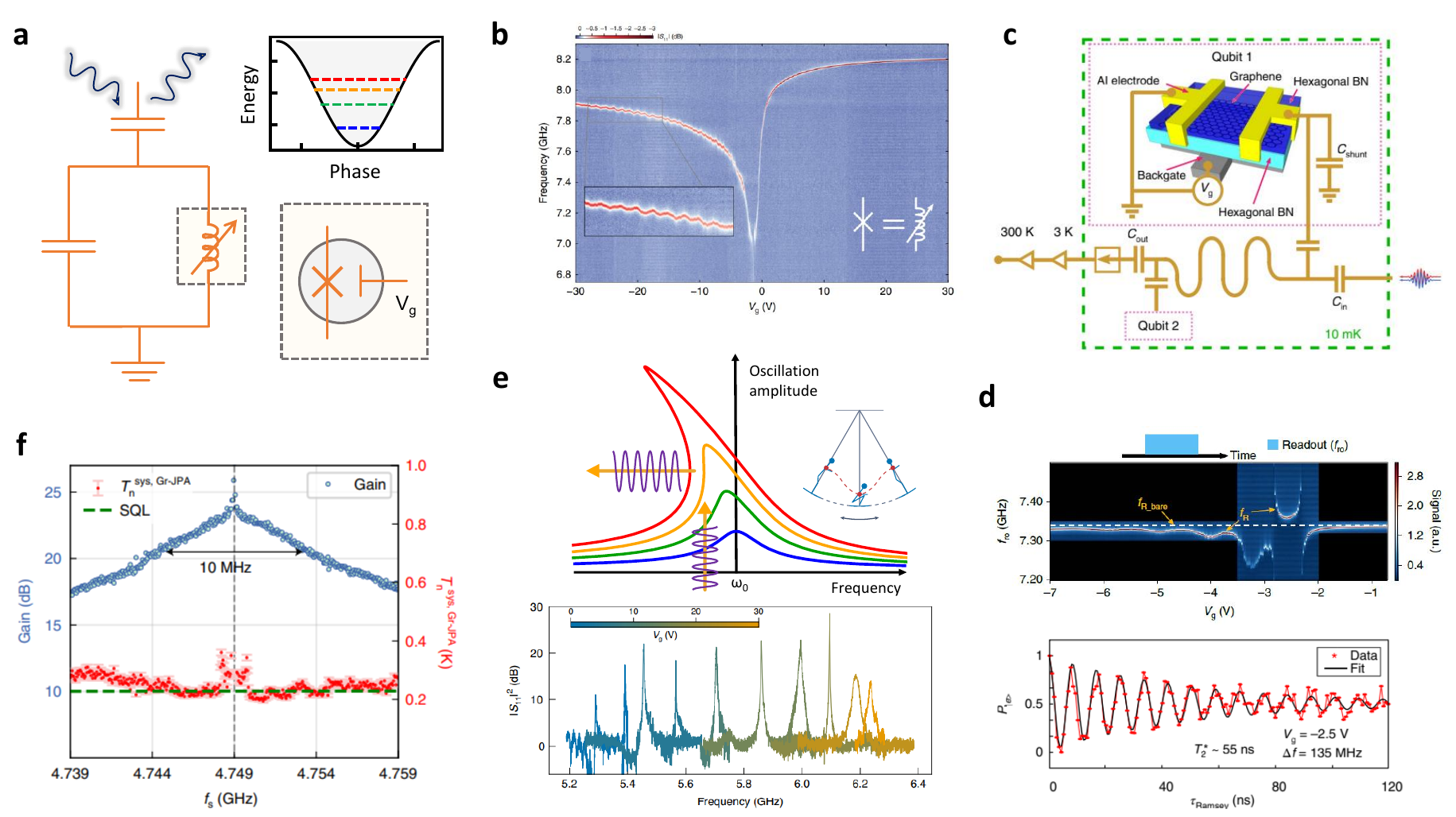}
    \caption{ \label{fig:grJJqubit} {\textbf{Graphene JJ as gate tunable inductor in cQED devices -- resonator, qubit, and amplifiers.} 
    \textbf{a,}~The schematic shows a capacitively coupled microwave 
    LC resonator in cQED architecture. The resonator consists of a tunable inductor element. The bottom inset shows a gate-tunable Josephson junction. The top inset shows the cosinusoidal potential of the "phase particle" in the RCSJ model, which in practice mimics an anharmonic LC oscillator's potential.   
    \textbf{b,}~Shows the resonance frequency shift of a tunable quarter-wave resonator as a function of gate voltage~\cite{schmidt_ballistic_2018}. The tunability comes from a graphene JJ where the Josephson inductance is tuned through electrostatic gating.
    \textbf{c,}~Shows the graphene JJ based planar superconducting qubit architecture~\cite{wang_coherent_2019}. The JJ is made on an hBN-graphene-hBN heterostructure and coupled to a resonator for readout. 
    \textbf{d,}~The top panel shows the qubit frequency tunability with gate voltage and the qubit-cavity hybridization, which is used for the dispersive readout scheme~\cite{wang_coherent_2019}. The bottom panel shows the Ramsey dephasing time ($T_2^*$) measured at a fixed gate voltage. 
    \textbf{e,}~Shows the amplitude response of a non-linear Duffing oscillator for increasing drive amplitudes~\cite{sarkar_quantum-noise-limited_2022}. As the drive increases, the nonlinear terms become active, causing the resonance peak to shift to lower frequencies. Beyond a certain drive, the amplitude response becomes very sharp (orange curve). With any further increase in drive power, the amplitude becomes a multivalued function of frequency (red curve), and the oscillator enters an unstable state. The orange curve corresponds to a bias point at which the device operates as a parametric amplifier, amplifying an input signal. The inset shows the classic example of a parametric amplification process in real life, where a person swinging on a rope modulates their center of mass to achieve a parametric resonance condition, which in turn increases the oscillation amplitude~\cite{fong_graphene_2022,aumentado_superconducting_2020}. The figure at the bottom shows the microwave signal gain traces of a gate-tunable graphene parametric amplifier measured at different gate voltages~\cite{butseraen_gate-tunable_2022}.
    \textbf{f,}~Shows the measured gain (left y-axis) and noise temperature (right y-axis) on another graphene JPA device~\cite{sarkar_quantum-noise-limited_2022}. The device shows $\sim20$ dB gain over $10$ MHz bandwidth and shows quantum noise-limited operation.}}
    
\end{figure*}

A natural evolution for GrJoFETs from tunable inductors was to realize transmon qubits. Gate-tunable transmons, also known as gatemon, are superconducting qubits where one can tune the qubit frequency with electrostatic gating~\cite{casparis_superconducting_2018}. This is in contrast to standard transmons that incorporate SQUID architecture and a magnetic field for tunability. The first step toward realizing a circuit-based qubit is to construct an anharmonic LC resonator -- whose lowest two energy levels define the qubit, see Fig.~\ref{fig:grJJqubit}a. Wang et al.~\cite{wang_coherent_2019} demonstrated a gatemon qubit realized in a planar architecture, see Fig.~\ref{fig:grJJqubit}c. This is the first demonstration where the authors could manipulate the qubit states and showed temporal coherence and control. The gate tunability helps to tune the qubit frequency and bring it close to the cavity frequency for hybridization to take place. The qubit states are probed through the cavity using the dispersive shift mechanism~\cite{wallraff_strong_2004}. The measured relaxation time and coherence times were modest, $T_1\sim36$ ns and $T_2^{*}\sim55$ ns, respectively, see Fig.~\ref{fig:grJJqubit}d. The graphene transmon is in its developmental stage compared to state-of-the-art transmons~\cite{bland_2d_2025}. While the improvement in coherence time requires systematic investigations, a major reason could be that the subgap energy states within the graphene JJ in the GHz energy scale lead to a lossy nature of the qubit. These experiments pave the way for the integration of 2D materials with cQED devices to explore the coherent time dynamics of individual quantum degrees of freedom -- an aspect that can be useful in quantum sensing applications. 

In addition to gate tunability, another key advantage of GrJoFETs is their resilience to magnetic fields, where conventional superconducting devices often fail. Kroll et al. incorporated a graphene JJ into a superconducting cavity to realize a gate-tunable transmon that operates at a large parallel magnetic field~\cite{kroll_magnetic_2018}. They see qubit-cavity hybridization as they tune the qubit frequency with gate voltage, and they perform two-tone energy level spectroscopy to find the qubit frequency. The qubit frequency remains nearly constant under an applied parallel magnetic field, enabling operation in fields up to $\sim1$ T, an order of magnitude higher than prior studies. This work demonstrates the potential of GrJoFET as field-resilient elements in topological and hybrid quantum computing schemes. 

\subsection{Nonlinear Graphene JJ Enables Quantum Noise Limited Amplification} 
Alongside qubit demonstrations, another vital application of GrJoFETs has been in quantum-limited amplification. In a typical cryogenic measurement setup, signals from a quantum device are read out through a chain of amplifiers, where each of them adds some amount of noise. While the overall gain is the product of the individual amplifier gains, the total noise is largely dominated by the first amplifier closest to the device. Therefore, it is crucial that the first amplifier introduces as little noise as possible. 

Josephson parametric amplifiers (JPA) are widely used devices in various cQED experiments for quantum signal processing~\cite{bergeal_phase-preserving_2010,aumentado_superconducting_2020}. To date, most commercial low-noise cryogenic amplifiers are based on high electron
mobility transistors (HEMT). However, as resistive components, they exhibit Johnson
noise ($\propto k_BT$) at temperature $T$. In contrast, Josephson parametric amplifiers (JPAs) are realized using lossless nonlinear inductors, which do not exhibit resistive fluctuations.
Instead, they are limited by quantum fluctuations ($\sim\hslash\omega/2$). The JPAs working at the quantum noise limit add very minimal noise, about a noise photon, and are commonly used in cryogenic experiments, with JPA at the $10$ mK stage, followed by the HEMT amplifier at $4$ K stage in a typical dilution refrigerator. Fig.~\ref{fig:grJJqubit}e shows the working scheme of a JPA from the viewpoint of a non-linear Duffing oscillator having a drive power-dependent amplitude response. The existing JPAs typically employ Al–AlO$_\mathrm{x}$–Al tunnel junctions configured in a SQUID geometry, where magnetic flux is used to tune the frequency~\cite{aumentado_superconducting_2020}. Recently, there has been interest in making JPAs using graphene JJs, which provide electrostatic control for frequency biasing of the device. It turns out that, while subgap losses in the graphene JJ limit its use in gatemon qubits, the losses are not detrimental to prohibit usage in a JPA. State-of-the-art JPAs made using graphene JJs that have gate-tunable linear resonance frequency have recently been demonstrated with $\sim20$ dB gain, having $10$ MHz bandwidth and quantum-limited noise performance, see Fig.~\ref{fig:grJJqubit}f, making it an attractive platform for highly sensitive signal processing~\cite{sarkar_quantum-noise-limited_2022,butseraen_gate-tunable_2022,fong_graphene_2022}. In addition, by leveraging the exceptional thermodynamic properties of graphene, there are opportunities to realize integrated quantum sensors operating at finite magnetic fields. In general, this work opens up exploration of scalable device architecture with gate-tunable properties such as Josephson inductance and nonlinearities.

\subsection{Graphene JJ as Sensors -- Routes for Dark Matter Search} 
Beyond their role in qubits and amplifiers, GrJoFETs also opened up exciting opportunities in quantum sensing owing to the exceptional electrical and thermal properties of vdW materials and the intrinsic sensitivity of the Andreev bound states in graphene JJs. Graphene's vanishingly small density of states at charge neutrality yields a reduced heat capacity and electron–phonon thermal conductance, both of which are highly favorable properties for bolometric applications~\cite{yan_dual-gated_2012}. Studies have shown that graphene absorbs photons across a wide range of frequencies, starting from GHz to near-infrared, due to the zero bandgap nature of monolayer graphene~\cite{cai_sensitive_2014,el_fatimy_epitaxial_2016,lee_graphene-based_2020,walsh_josephson_2021}. This broadband absorption, coupled with exceptional thermal properties, makes GrJoFETs natural candidates for bolometric sensors. By leveraging the sensitivity of GrJoFETs, a promising direction is to use them as quantum sensors for various degrees of freedom -- including photons, phonons, magnons, etc. Among these, the GrJoFET bolometer has made significant progress -- an aspect we discuss next.

\begin{figure*}[h!]
    \centering
	\includegraphics[width=18cm]{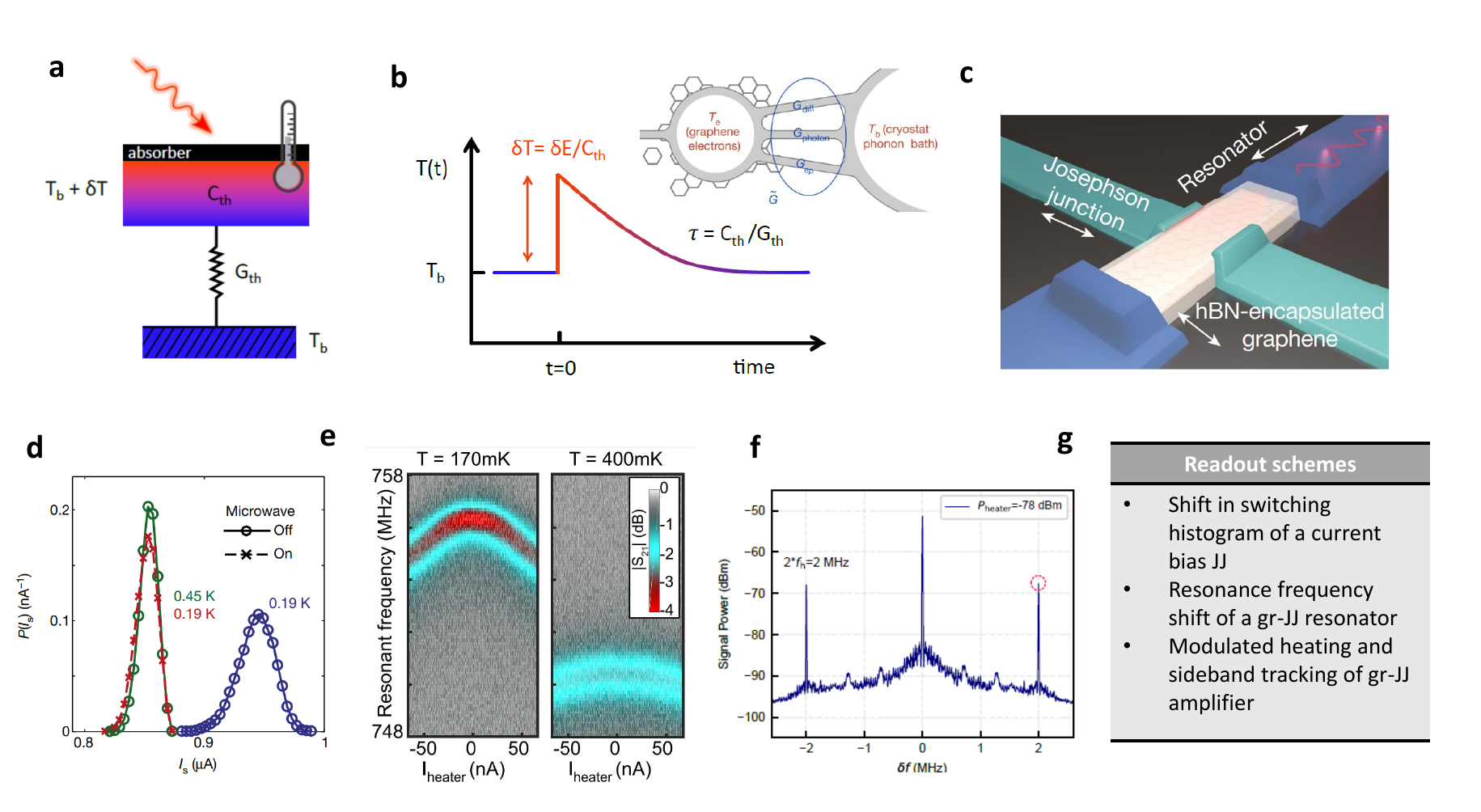}
    \caption{ \label{fig:grbolo} {\textbf{vdW JJ as bolometric sensors.} 
    \textbf{a,}~Shows the basic working principle of bolometer, which has a heat capacity $C_\mathrm{th}$, is in thermal equilibrium with a bath at temperature $T_\mathrm{b}$. The bolometer absorbs incident radiation, causing its temperature to increase by $\delta T$. It releases heat to the bath through a thermal link with a thermal conductance $G_\mathrm{th}$ and resets to sense subsequent photons. 
    \textbf{b,}~Shows the typical response of a linear power sensor. Upon absorbing heat energy of $\delta E$ instantaneously, its temperature rises by $\delta T$. The sensor then cools back to the bath temperature $T_\mathrm{b}$ with a characteristic time $\tau =C_\mathrm{th}/G_\mathrm{th}$. Different channels, all acting in parallel, release the heat from the absorber to the bath. Here, $G_\mathrm{ep}$, $G_\mathrm{diff}$, and $G_\mathrm{photon}$ refer to the heat removal through phononic heat transfer, electronic diffusion, and radiative cooling processes, respectively~\cite{kokkoniemi_bolometer_2020}. 
    \textbf{c,}~Shows schematic of a graphene JJ-based bolometer device architecture~\cite{lee_graphene-based_2020}. The device uses a half-wave superconducting resonator. The blue electrodes are the central conductor of the resonator. A graphene flake is coupled at the current antinode of the resonator. The authors use a "$+$" like device geometry. The JJ is formed in the orthogonal direction to the central line. The microwave heating signals are sent through the central conductor. The JJ is read out using switching measurements through the orthogonal probes. 
    \textbf{d,}~Shows data from the bolometer in (c), where the device consists of a DC current-biased graphene JJ as the absorber of heat. When a microwave heating signal is applied, the switching current of the JJ changes. By measuring the shift in the switching histogram, the authors extract the sensitivity of the bolometer~\cite{lee_graphene-based_2020}.
    \textbf{e,}~Shows data from a bolometer device realized by coupling a graphene JJ to a quarter-wave superconducting resonator~\cite{katti_hot_2023}. When a heating current is applied, the Joule heating increases the local temperature of the JJ and hence the device shows a shift in resonant frequency. By measuring the resonant frequency shift, the authors demonstrate inductance thermometry. 
    \textbf{f,}~Shows the readout of a graphene JPA bolometer device based on modulated heating and frequency up-conversion method~\cite{sarkar_kerr_2025}. The author uses a broadband JPA device designed at $\sim5$ GHz. A low-frequency AC signal at $1$ MHz is applied for heating the device, which generates sidebands. The JPA enhances the sideband signals and provides a sensitive detection of minute heating effects. 
    \textbf{g,}~Various readout schemes used to probe the graphene JJ bolometers.}}
    
\end{figure*}

Bolometers are radiation sensors that are at the heart of material science, thermal imaging, qubit readout, and most recently experiments towards dark matter search among others~\cite{paolucci_fully_2021,braine_extended_2020,opremcak_measurement_2018,gunyho_single-shot_2024}. Fig.~\ref{fig:grbolo}a,b shows the working scheme of a bolometer, where absorbed radiation generates heat that changes a local property in the device like resistance or inductance. Bolometers with different architectures have been demonstrated using GrJoFETs; see Fig.~\ref{fig:grbolo}c. Some of these devices use a DC-biased JJ to realize bolometers, which rely on changes in the switching histogram of a JJ upon photon absorption and have demonstrated microwave and near-infrared (NIR) photon sensing~\cite{walsh_graphene-based_2017,lee_graphene-based_2020,walsh_josephson_2021,huang_graphene_2024}. Although exquisite sensitivity $\sim0.7$ aW/$\mathrm{\sqrt{Hz}}$ has been reported in these devices~\cite{lee_graphene-based_2020}, the low frequency switching protocol makes the readout scheme slow. Whereas some devices have utilized GrJoFET coupled to microwave resonators, which rely on the resonance frequency shifts in the devices for heat sensing, this technique is comparatively fast for probing the bolometers~\cite{kokkoniemi_bolometer_2020,katti_hot_2023}. Another report shows the usage of an integrated JPA device for heat sensing, where the inherent nonlinearity of the device helps in enhancing the bolometer sensitivity~\cite{sarkar_kerr_2025}. See Fig.~\ref{fig:grbolo}d-g for various readout schemes used in the graphene JJ bolometer devices. The electrostatic tunability of the bias points of the bolometers and the potential for multiplexing open up the possibility of realizing next-generation integrated quantum sensor arrays similar to the microwave kinetic inductance detector (MKID) and the superconducting nanowire single-photon detector (SNSPD)-based imaging devices~\cite{walter_mkid_2020,oripov_superconducting_2023}.

\section{JJs Made of Emergent Moir\'e Superconductors} \label{sec:4}   
The formation of a moir\'e pattern occurs when 2D sheets of any lattice (here, we focus on the honeycomb lattice of graphene) are stacked with a relative twist angle. While graphene JJs have already proven versatile, introducing moiré physics through twist adds yet another dimension of tunability. In the previous section, we have seen JJs that utilize graphene as a normal metal coupled to conventional superconductors. In contrast, moir\'e JJs allow the realization of an all-carbon JJ without having to use conventional superconductors. Often used to understand the unconventional nature of the moir\'e superconductors, they can also be used to realize interesting devices with unique technological applications.

\begin{figure*}[h!]
       \centering
	\includegraphics[width=16.5 cm]{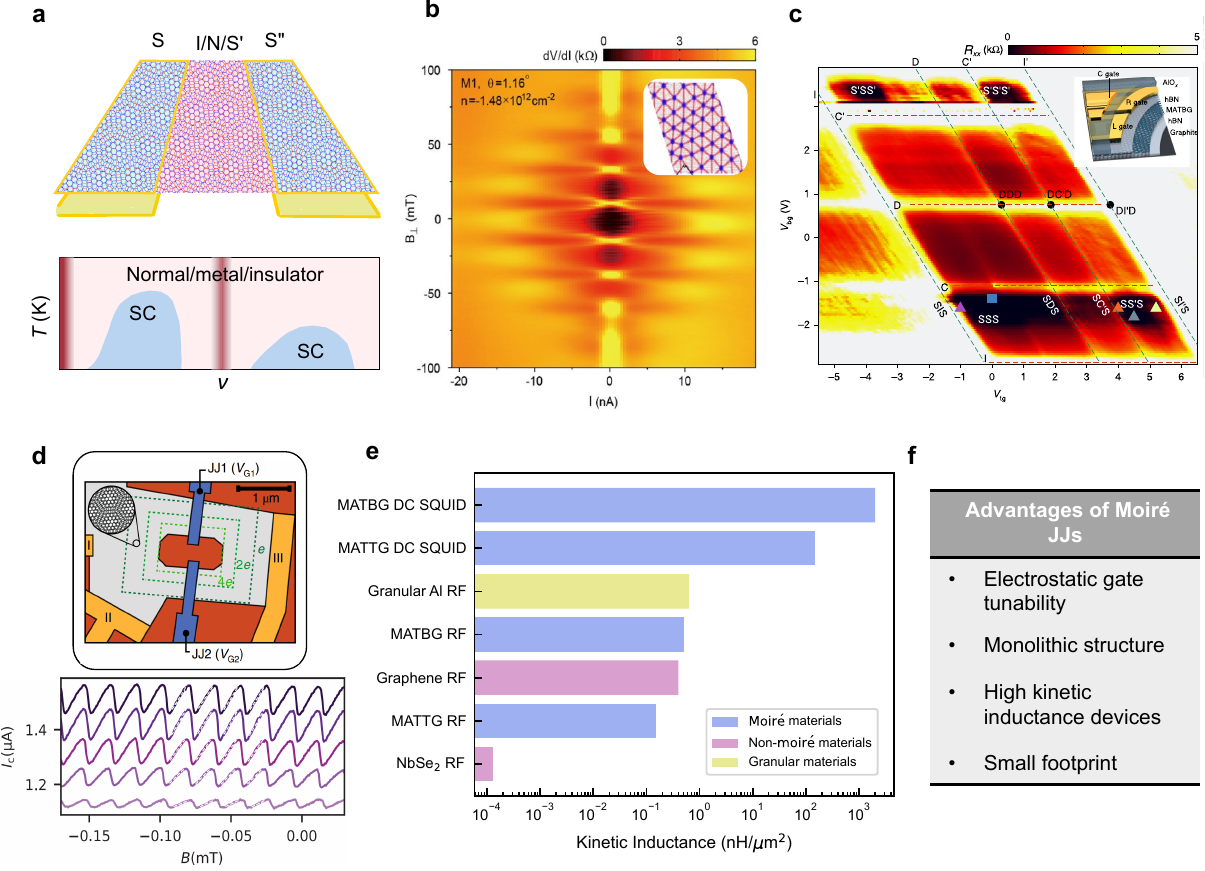}
    \caption{ \label{fig:moirejj} {\textbf{All-carbon Josephson junctions utilizing emergent moir\'e superconductors} 
     \textbf{a,}~Schematic of a moir\'e JJ along with a representative phase diagram of moir\'e materials, including a superconducting (S/S'/S") phase, and a normal metal (N)/insulator (I) phase. The moir\'e superconductor can be tuned by changing the doping $\nu$, temperature $T$, and also a perpendicular electric field.
     \textbf{b,}~Intrinsic JJs are formed due to moiré inhomogeneities. The Fraunhofer interference pattern shows the presence of phase coherence within the moir\'e sample \cite{cao2018unconventional}. \textbf{Inset,}~Schematic showing moir\'e inhomogeneities in a twisted graphene sample. The dark colored regions show the percolating non-superconducting paths \cite{mukherjee2025superconducting}.
     \textbf{c}~Moir\'e JJs are made using gate-defined scheme where each part of the JJ is made from a phase in the phase-diagram of the moir\'e material and is biased to different phases using the electrostatic gates \cite{rodan-legrain_highly_2021}. Different phases are denoted as S/S', superconducting; C, correlated insulator at half-filling; D, Dirac (charge-neutral) point; I, full-filling band-insulating state.
     \textbf{Inset,}~Schematic of gate defined JJ architecture \cite{de2021gate}.    \textbf{d,}~Superconducting quantum interference devices (SQUIDs) utilising two moir\'e JJs in parallel \cite{portoles2022tunable}. The current-phase relationship of the SQUID is tuned with gate voltage (curves of different color), revealing a tunable kinetic inductance of the device \cite{jha2025large}. 
    \textbf{e,}~Plot showing kinetic inductances in various materials, including moir\'e materials, which have high inductances and can be used in low-noise quantum circuits as inductors. Ref~\cite{portoles2022tunable,jha2025large,valenti_interplay_2019,tanaka2025superfluid,schmidt_ballistic_2018,banerjee2025superfluid,kreidel_measuring_2024}.
    \textbf{f,}~Key advantages offered by moir\'e JJs to integrate moir\'e materials into different quantum device architectures.
    }}
\end{figure*}

Fig.~\ref{fig:moirejj}a shows a typical phase diagram for a graphene-based moir\'e device hosting multiple phases, accessible by tuning the doping $\nu$ in the system with electrostatic gating. The twist between successive layers results in moir\'e inhomogeneities at the mesoscopic scale primarily due to lattice relaxation \cite{nam2017lattice,turkel2022orderly, mukherjee2025superconducting}. These inhomogeneities, while sometimes seen as a challenge, naturally give rise to JJ behavior embedded within the moiré system itself. In magic-angle twisted bilayer graphene (MATBG) and twisted trilayer graphene (MATTG), these moir\'e inhomogeneities host percolating non-superconducting pathways even when the sample is doped to the superconducting phase. This phenomenon allows for the formation of intrinsic JJs and the consequential observation of a Fraunhofer pattern in both MATBG and MATTG \cite{cao2018unconventional,yankowitz2019tuning,park2021tunable,hao2021electric} shown in Fig.~\ref{fig:moirejj}b. 

Moir\'e engineering enables the coexistence of multiple phases—such as superconductivity, correlated insulators, and metallic states—within the same sample, which can be tuned using applied gate voltages.  Figure \ref{fig:moirejj}c illustrates the gate tunability in creating JJs by electrostatically doping specific device areas. As a result, multiple Josephson device architectures have been realized with MATBG - the first member in the moiré superconductor family. One single device of the MATBG JJ can be used to realise JJ, tunelling barriers and single electron transistors - all with the same gate architecture on a single MATBG sample \cite{rodan-legrain_highly_2021}. The MATBG JJ can also successfully couple to microwave voltages, and has been demonstrated with the help of the ac Josephson effect measurement of Shapiro steps \cite{de2021gate}. Interestingly, such microwave measurements were less sensitive to the angle inhomogeneities ubiquitous in twisted devices. Although, the MATBG JJ showed a topologically trivial character, subsequent work has broadened the scope of moiré JJs by using them to probe valley polarization, orbital magnetization, and quantum geometric effects \cite{diez2023symmetry,diez2025probing}. Thus, moiré JJs are valuable for both technological applications and as probes that can be used to explore unresolved questions in the field of moir\'e devices. We discuss next, two unique applications of JJs made with moir\'e  superconductors and probing the role of the quantum geometry in such devices.

\subsection{Quantum Interference Probes Moir\'e Superconductivity} 
\label{sec:4.1}
One natural extension of moiré JJs is their incorporation into superconducting quantum interference devices (SQUIDs). A SQUID is a device where two or more JJs form a loop, enabling quantum interference of their macroscopic phases—analogous to Young’s double-slit experiment. It measures phase drops across the junctions, including contributions from the Aharonov–Bohm flux threaded through the loop. Neglecting lead inductance, the interference condition is $\phi_a -\phi_b = 2\pi \frac{\Phi}{\Phi_0}$, where $\phi_a$ and $\phi_b$ are the phase drops across the two SQUID arms, $\Phi$ is the magnetic flux threaded through the loop, and $\Phi_0$ is the flux quantum. Owing to its extreme flux sensitivity, the SQUID functions as one of the most precise magnetometers \cite{weinstock_squid_1996}.

SQUIDs formed with moiré devices offer a two-fold advantage: electrostatic tunability and a single-device architecture. Recently, SQUIDs using MATBG and MATTG have been studied. MATBG SQUID architecture is a monolithic gate-defined device, where the loop of the SQUID is formed by etching the graphene heterostructure \cite{portoles2022tunable}, see Fig.~\ref{fig:moirejj}d. Each JJ arm in the SQUID can be controlled by different gates and can not only be switched on and off, but also be tuned to host different critical currents. Independent control over individual JJs can morph the SQUID from an asymmetric to a symmetric SQUID allowing for studying a host of different phenomena in the same device. The SQUID configuration allows study of the CPR of the weak links (see the \hypersetup{linkcolor=black}\hyperref[box1]{Theory Box}). This flexibility allows SQUIDs to serve as precision probes of the CPR and kinetic inductance in moiré systems. The slope of the CPR can be used to infer the kinetic inductance of the SQUID arms and weak links. Ref. \citeonline{portoles2022tunable} reports MATBG SQUID to host a high inductance of 2 $\mathrm{\mu}$H at low temperature, however this was notably an order of magnitude higher than succeeding studies. SQUIDs made out of MATTG are reported to host inductances of around 150 nH/$\square$\cite{jha2025large} and is comparable to the estimates from recent RF studies of kinetic inductances of moir\'e superconductors~\cite{tanaka2025superfluid,banerjee2025superfluid}. While in the MATBG SQUID study \cite{portoles2022tunable}, the arms are made of graphene-based moir\'e superconductor, the MATTG SQUID study~\cite{jha2025large} uses MoRe arms and only uses the MATTG to form the weak links, showcasing multiple architecture choices. Additional SQUID geometries are discussed later, where similar CPR-based approaches are employed to study the disparate physics of either magnetic states (\hypersetup{linkcolor=black}\hyperref[sec:5.4]{section 5.4}) or to reveal the topological characteristics of the weak links (\hypersetup{linkcolor=black}\hyperref[sec:6.2]{section 6.2}).

\subsection{High Kinetic Inductance of Moir\'e Superconductors} 
High inductance elements are used routinely in superconducting qubit circuits to enhance circuit performance and control. Typical materials used for this purpose include NbN, NbTiN, TiN, and granular Al~\cite{glezer_moshe_granular_2020}. 
In superconductors, kinetic inductance arises from the finite phase stiffness of Cooper pairs -- lower phase stiffness (or superfluid density) leads to a larger kinetic inductance ($L_{k}$) that resists changes in current. Moir\'e materials that exhibit flat electronic bands are expected to have a high kinetic inductance. This is a consequence of their combination of a large effective mass of charge carriers and a low carrier density ($\sim 10^{16}\ \mathrm{m}^{-2}$) in the superconducting phase, which together significantly enhances the kinetic inductance \cite{jha2025large}. Tunability of the superconducting phase is also inherited to tune the kinetic inductance of the system, with temperature, magnetic field, and, more importantly, electrostatic gate voltages. Studies investigating the kinetic inductance of moir\'e materials also seek to understand the origin of superconductivity in such materials. In 2025, two groups incorporated moir\'e superconductors, namely twisted bilayer and twisted trilayer graphene, into RF resonant circuits to measure the kinetic inductance of such devices \cite{tanaka2025superfluid,banerjee2025superfluid}. Kinetic inductance of moir\'e devices can be measured through multiple ways, including a SQUID architecture \cite{jha2025large}, and also from Berezinskii-Kosterlitz-Thouless relations~\cite{mukherjee2025superconducting}. 
The kinetic inductance values for moir\'e materials are summarized and compared with some non-moir\'e materials in Fig.~\ref{fig:moirejj}e, showcasing how moir\'e materials have comparable $L_k$ to commonly used high inductance non-moir\'e and granular materials. Fig.~\ref{fig:moirejj}f lists some advantages of moir\'e JJs. If the challenge of scalability and angle inhomogeneity is overcome, then moir\'e superconductors could play an important role in superconducting technology.

\subsection{Quantum Geometry Governs JJs in Flat Band Systems} 
Flat-band JJs are a recent development in which a flat-band (FB) material, such as twisted bilayer graphene, Lieb lattice, or kagome lattice, is used as the weak link. In these materials, the Fermi velocity vanishes, leading to the theoretical prediction of a vanishing superfluid velocity. According to conventional wisdom, such systems should not support supercurrents, even in very short junctions. These systems are characterized by a band geometric quantity known as the \textit{quantum metric} \cite{Torma2022,Peotta_2015}, given by
$
    g_{n;\alpha\beta}(\mathbf{k}) = \Re\!\left[ \langle \partial_\alpha u_n(\mathbf{k}) \,|\, \big( 1 - |u_n(\mathbf{k})\rangle \langle u_n(\mathbf{k})| \big) \,|\, \partial_\beta u_n(\mathbf{k}) \rangle \right],
$
where $|u_n(\mathbf{k})\rangle$ denotes the Bloch state of band $n$, and $\alpha,\beta$ label spatial coordinates. This quantity encapsulates the geometric properties of the Bloch state. It has been theoretically shown in a one-dimensional S/FB/S JJ \cite{PhysRevResearch.7.023273} that the quantum geometry introduces a characteristic length scale, \( \xi_{\text{QM}} \), which acts as an effective coherence length \( \xi \). If \( \xi_{\text{QM}} \geq L \), where \( L \) is the junction length, the system can exhibit a finite critical current and a proximity effect despite the vanishing Fermi velocity.

A recent experiment on an NbTiN/MATBG/NbTiN JJ was carried out to test the theoretical predictions for flat band JJs \cite{diez2025probing}. 
A strong proximity effect was observed in the flat-band regime, comparable in magnitude to that of the higher-energy dispersive bands. Furthermore, the conventional relationship \( I_c \propto 1/R_N \), where \( R_N \) is the resistance of the normal state, does not hold in this system. 
Whether this phenomenon is generic or system-specific remains an open question. More research is needed in a broader class of flat-band systems in both one and two dimensions \cite{virtanen2024superconducting}.
Future theoretical work should account for strong electronic interactions in modeling the mechanism of induced superconductivity.

\section{vdW based S-I-S JJs} 
\label{sec:5}
So far, in this review, we have discussed the physics and implementation of SNS JJs; however, as discussed in the \hypersetup{linkcolor=black}\hyperref[box1]{Theory Box}, there is another kind of JJ based on tunneling phenomena that have played an important role in the evolution of Josephson physics. 

Following the discussion of graphene-based S-N-S JJs based on proximity effect, we now focus on diverse S–I–S junctions realized across distinct vdW material platforms. vdW JJs incorporating tunnel barriers serve as key building blocks for exploring novel quantum physics and device functionalities. In this section, we will discuss the physics and device architecture of three kinds of S-I-S JJs -- in the first kind, the vdW gap formed by layering two superconducting flakes serves as the tunnel barrier, referred to as vdW gap JJs. In the second kind, an internal crystal structure of the vdW material provides a tunneling barrier, as exemplified by high Tc cuprate JJs. In the third category, an explicit tunnel barrier is used, as demonstrated in magnetic JJs. The unique physical properties of each individual vdW material can significantly influence the characteristics of the junction and can add novel control knobs. In Fig.~\ref{fig:JJ_squid}a, we demonstrate a schematic illustrating the vdW nature of the JJ.

\subsection{vdW Gap JJ} 

In a vdW gap JJ, Cooper pair tunneling occurs across the vdW gap between two superconducting flakes. The vdW gap occurs when two different flakes are stacked on top of each other without the need for a tunnel barrier layer between them. This mechanism contrasts with conventional tunnel JJs, as the insulating barrier here originates naturally from the weak vdW bonding rather than an engineered insulating layer. In Fig.~\ref{fig:JJ_squid}b we show a schematic of the vdW gap JJ, where the vdW gap that serves as an insulator between two superconducting layers has to be thinner than the coherence length of the superconducting flake (see \hypersetup{linkcolor=black}\hyperref[box1]{Theory Box}). Yabuki et al. fabricated  JJ devices from NbSe$_2$ flakes \cite{yabuki_supercurrent_2016}, confirming a vdW gap between the top and bottom flakes using transmission electron microscopy (TEM).
Furthermore, from transport measurement, the observed Fraunhofer interference pattern as seen in Fig.~\ref{fig:JJ_squid}b confirms the Josephson coupling and the magnitude of the vdW gap between the two flakes \cite{yabuki_supercurrent_2016}. 

These first experiments with vdW Josephson junctions used superconductors with isotropic superconducting order parameter, \textit{s}-wave, superconductors. However, the vdW device architecture opened up the possibility of making JJs using superconductors of anisotropic order parameters such as \textit{d}-wave superconductors. The high-T$\text{c}$ cuprate superconductors have an anisotropic order parameter, and some of them are exfoliable to realize JJs, an aspect we discuss next.

\subsection{JJs Using \textit{d}-wave Superconductors and Resulting Non-reciprocity}
\label{sec:5.2}
The superconductivity in high-$T_c$ cuprates arises from the Cu$\mathrm{O}_2$ planes, where the order parameter exhibits
$d$-wave pairing symmetry. These planes are stacked with alternating charge reservoir layers, forming a natural superlattice along the $c$-axis. As a result, JJs are intrinsically present along $c$-axis of the layered cuprate crystal structure. 
Bi$_2$Sr$_2$CaCu$_2$O$_{8+\delta}$ (Bi-2212 or BSCCO) is a prototypical vdW cuprate, in which superconducting Cu$\mathrm{O}_2$ planes are separated by insulating BiO–SrO layers. The weak interlayer vdW coupling enables BSCCO flakes to be readily cleaved along these planes, allowing the fabrication of heterostructures~\cite{zhao_time-reversal_2023,zhu_presence_2021,lee_encapsulating_2023,ghosh_high-temperature_2024,patil_pick-up_2024}. 
Introducing a twist angle between the two \textit{d}-wave superconducting flakes misaligns the order parameter of the top and bottom layers. This misalignment modifies the overlap matrix element between the macroscopic wave functions of the twisted cuprate layers. Such an approach to realizing JJs can allow for engineering of new superconducting interfacial orders -- an aspect unique to the vdW materials and their assembly. However, to realize these experiments, it is crucial to maintain the integrity of the interfaces. Techniques that exfoliate the cuprate at cryogenic temperatures preserve interfacial stoichiometry and offer a promising way forward \cite{zhao_time-reversal_2023, ghosh_high-temperature_2024, patil_pick-up_2024, wang_prominent_2023,lee_twisted_2021}. 

Fig.~\ref{fig:JJ_squid}c shows the schematic and optical image of the twisted BSCCO heterostructure. Zhao et al.\cite{zhao_time-reversal_2023} in their pioneering work fabricated twisted BSCCO devices of different twist angles and demonstrated the nature of interfacial time-reversal symmetry-broken superconducting order parameter. The twist angle between the two layers is crucial for tuning the Josephson coupling, due to the $d$-wave order parameter \cite{zhao_time-reversal_2023,ghosh_high-temperature_2024,can_high-temperature_2021}. 
These JJs are well suited for application at liquid N$_2$ temperature in quantum circuits\cite{confalone_cuprate_2025}.
Moreover, recent studies have demonstrated high $T_c$ cuprate-based Josephson diodes showing non-reciprocal transport -- an aspect we discuss next.

The semiconducting diode effect is a classic example of non-reciprocal transport that arises from the breaking of inversion symmetry. The non-reciprocity of \textit{pn}-junction diode is ubiquitous in electronic circuits.  In an analogous manner diode response in the superconducting systems is important both for potential practical applications and fundamental physics. Diode response in the superconducting system implies that the maximum dissipationless current flowing for two signs of bias are unequal. The Superconducting Diode Effect (SDE) refers to the emergence of non-reciprocal current–voltage ($I$–$V$) characteristics. Depending on the material platform, the SDE can arise from different physical mechanisms and can be controlled through various tuning parameters such as electrostatic gating, magnetic fields, or twist angle \cite{nadeem_superconducting_2023}.  

 In twisted high T$_\mathrm{c}$ JJs, both zero field\cite{zhao_time-reversal_2023} and field-induced diode effect\cite{ghosh_high-temperature_2024} have been observed. The physics behind the two cases, although connected, is fundamentally different. Experimental\cite{zhao_time-reversal_2023} and theoretical studies\cite{can_high-temperature_2021} suggest that twisting two flakes by 45 degrees allows for the stabilization of time reversal symmetry broken \textit{d$\pm i$d} interfacial order. Breaking of the time reversal and inversion symmetries is understood to be the reason for the observed non-reciprocal transport in the twisted BSCCO JJs. The field-induced diode effect in twisted BSCCO systems, as shown in Fig.~\ref{fig:JJ_squid}c, emerges from the complex interaction between Abrikosov and Josephson vortices. 
 
 Cuprate twisted JJs present an interesting platform to stabilize unconventional orders in high T$_c$ superconductor interfaces~\cite{zhao_time-reversal_2023}, and open interesting device possibilities for superconducting circuits that function at liquid nitrogen temperature\cite{ghosh_high-temperature_2024}. Further studies are needed to gain a deeper understanding of these aspects.

\subsection{Magnetic JJ and Route to Zero-Field Josephson Diode}

Zero-field Josephson diode effects discussed earlier can also be realized in different vdW JJs by incorporating known crystalline magnetic materials as the tunnel barrier -- we discuss this next.

The tunnel barrier in a JJ that serves as a weak link (see \hypersetup{linkcolor=black}\hyperref[box1]{Theory Box}) can also be a magnetic insulator, an idea that has been explored with conventional superconducting and magnetic systems.
Magnetic JJs are demonstrated for heterostructures like NbSe$_2$/Cr$_2$Ge$_2$Te$_6$(CGT)/NbSe$_2$\cite{huang_manipulation_2024,ai_van_2021,idzuchi_unconventional_2021} and NbSe$_2$/Nb$_3$Br$_8$/NbSe$_2$\cite{wu_field-free_2022}. In Fig.~\ref{fig:JJ_squid}d, we show the schematic of the NbSe$_2$/Nb$_3$Br$_8$/NbSe$_2$ heterostructure, where Nb$_3$Br$_8$ is the magnetic insulator. An interesting aspect of magnetic JJs is that $I-V$ characteristics can be intrinsically non-reciprocal without applying an external magnetic field\cite{wu_field-free_2022} as time reversal symmetry is broken by the exchange interaction and inversion symmetry is broken by the twisted architecture; a phenomenon known as the zero-field diode effect \cite{nadeem_superconducting_2023}. 

In addition to the diode effect, magnetic JJs offer a natural route to realizing $\pi$-junction behavior due to the coexistence of two competing physical phenomena -- superconductivity and magnetism. A $\pi$-junction differs fundamentally from a conventional Josephson junction, as its current–phase relation (CPR) is shifted by phase of $\pi$. These junctions are particularly intriguing because magnetic interactions lift spin degeneracy, enabling spinless or effectively $p$-wave pairing of electrons, which offers a promising platform for exploring Majorana zero modes. We now try to understand the underlying microscopic origin of $\pi$-JJ formation, known as
Fulde-Ferrell-Larkin-Ovchinnikov (FFLO) mechanism, in which Cooper pairs acquire a finite center-of-mass momentum. In a superconductor-ferromagnet-superconductor (S-F-S) junction, the exchange field of the ferromagnetic layer induces Zeeman splitting of the Fermi surface, imparting finite momentum to the center of mass of the Cooper pairs and thereby realizing an FFLO-like state\cite{buzdin_proximity_2005}. Unlike the conventional picture, the superconducting order is not spatially uniform rather it has a functional form $\Delta_{\mathrm{FFLO}}(\vec{r})=\Delta_0 e^{i\vec{q}.\vec{r}}$, where $\vec{q}$ is the finite momentum of the Cooper pair and $\Delta_0$ is the amplitude of the superconducting order parameter. It implies the spatial modulation of the superconducting order across the length of the junction. The thickness of the ferromagnetic barrier can lead to a superconducting order parameter with opposite signs at the two S/F interfaces, thereby realizing a $\pi$-junction. The thickness-dependent $0$-$\pi$ transition has been observed in NbSe$_2$/CGT/NbSe$_2$ JJs\cite{kang_van_2022}. Thus, magnetic JJs not only enable zero-field diode effects but also provide a natural system for studying unconventional superconducting states. From this platform, we now turn to SQUID architectures, where vdW JJs are integrated into loops to probe phase coherence and quantum interference.

 \subsection{Quantum Interference in vdW JJs} 
\label{sec:5.4}
SQUIDs based on vdW JJs are important both for probing fundamental physics of superconductivity and their potential application as highly sensitive magnetic-field and flux sensors. The vdW junctions can be integrated into two distinct SQUID architectures -- DC and RF -- each offering a unique route to accessing the junction's CPR. The CPR of a JJ is more than a mathematical expression that connects the supercurrent and the phase difference across the JJ; it provides deep insight into the microscopic mechanisms of Cooper-pair tunneling\cite{golubov_current-phase_2004,endres_currentphase_2023} and can reveal the topological nature \cite{endres_currentphase_2023, kudriashov_non-reciprocal_2025} of the junction. In an RF-SQUID configuration, a single vdW Josephson junction is embedded in a superconducting loop that is inductively coupled to a resonant tank circuit, enabling direct extraction of the junction’s CPR. Our discussion will focus on the other architecture, namely, DC SQUID. A DC SQUID is an architecture where there are two JJs embedded in a superconducting loop, where threading a magnetic field controls the phase of the JJs relative to each other. A DC SQUID, in which the Josephson inductances of the two arms differ substantially, provides an effective platform for extracting the CPR of the junction; thus, it can reveal distinct physics across different vdW material platforms, as discussed in the case of moiré JJs in \hypersetup{linkcolor=black}\hyperref[sec:4.1]{section 4.1}.

\begin{figure*}[h!]
    \centering
	\includegraphics[width=16.5cm]{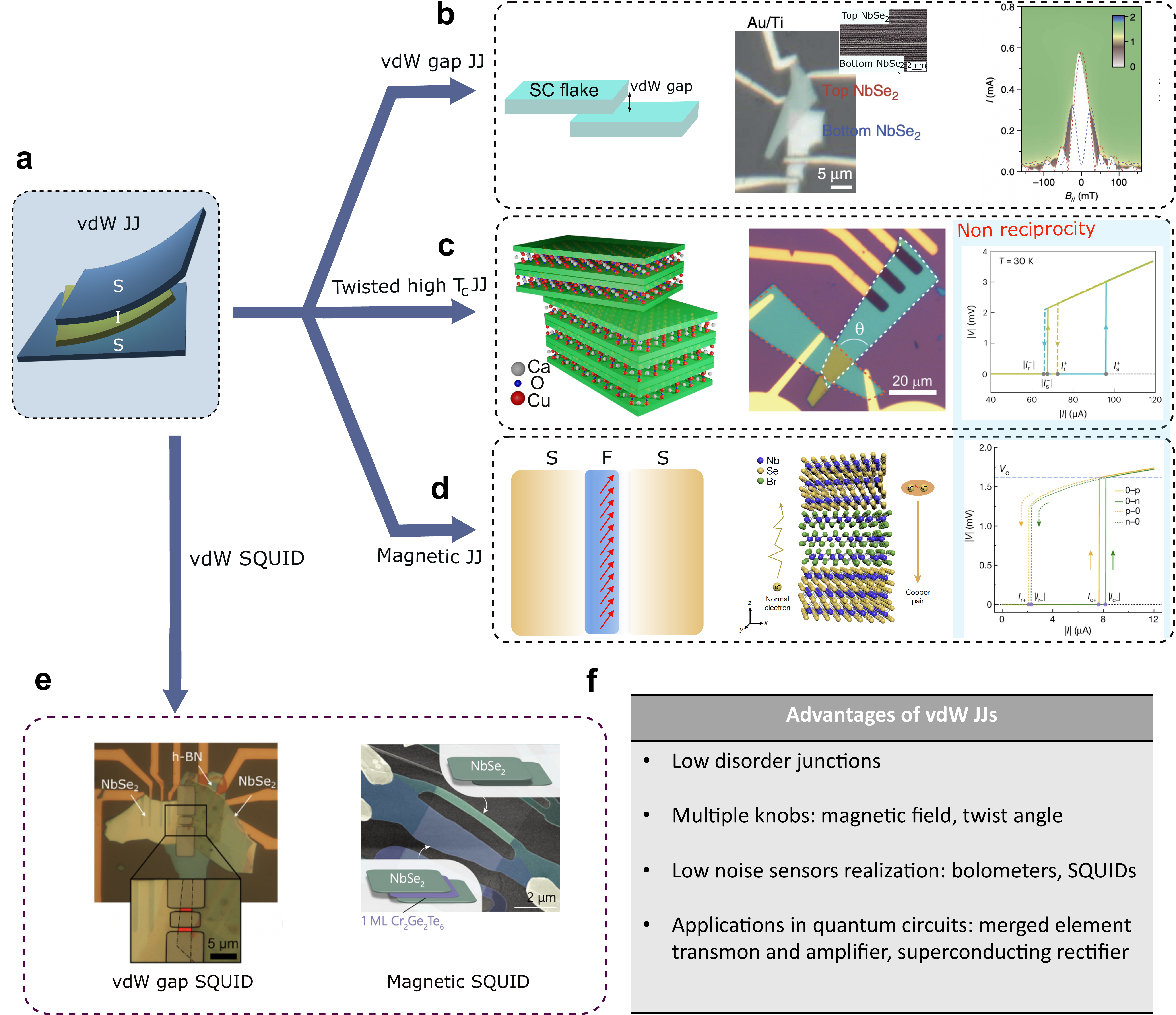}
    \caption{ \label{fig:JJ_squid} {\textbf{S-I-S JJs beyond graphene-based weak links.} 
   \textbf{a,} A schematic representation of vdW S-I-S JJ. 
   \textbf{b,} vdW gap-based JJ fabricated from NbSe$_2$ flakes \cite{yabuki_supercurrent_2016}. The observed Fraunhofer pattern is the signature of the JJ, formed between two superconducting flakes.
   \textbf{c,} Twisted high T$_\mathrm{c}$ cuprate JJ fabricated using BSCCO\cite{zhao_time-reversal_2023}. In the left schematic, the green, semitransparent blocks represent the insulating BiO-SrO layers. Non-reciprocal $I-V$ characteristics suggest the broken inversion and time reversal symmetry of the system\cite{ghosh_high-temperature_2024}. 
   \textbf{d,} Magnetic JJ: NbSe$_2$ is the superconducting flake and Nb$_3$Br$_8$ is the magnetic insulator\cite{wu_field-free_2022}. The zero-field diode effect demonstrates time-reversal symmetry breaking due to the presence of the magnetic insulating layer.
   \textbf{e,} SQUIDs based on vdW heterostructure: \textbf{left}, SQUID fabricated using NbSe$_2$\cite{farrar_superconducting_2021} vdW gap Josephson junction; \textbf{right}, SQUID fabricated using magnetic Josephson junction of  NbSe$_2$/CGT/NbSe$_2$\cite{idzuchi_unconventional_2021}.
   \textbf{f,} The advantages of vdW Josephson junctions from both a physics perspective and an applications standpoint.
    }}
    
\end{figure*}

Asymmetric DC SQUIDs can be realized in two types of architectures. In the first approach, the entire SQUID loop is patterned directly onto a vdW heterostructure. A notable example is a SQUID incorporating a magnetic JJ, where one arm contains a magnetic tunnel barrier forming the JJ, while the reference arm consists of a conventional vdW gap junction\cite{ai_van_2021,idzuchi_unconventional_2021}. This configuration enables direct access to the doubly degenerate ground state intrinsic to magnetic JJs. In such junctions, the RCSJ potential exhibits two phase minima, which manifest in the magnetic-field interference pattern of the SQUID as a shift of the critical-current maxima away from zero magnetic field\cite{ai_van_2021}. Fig.\ref{fig:JJ_squid}e, shows the device image of the magnetic JJ SQUID fabricated using NbSe$_2$/CGT/NbSe$_2$\cite{idzuchi_unconventional_2021}. Similar SQUID architecture is also demonstrated in other vdW material platforms, such as NbSe$_2$/NbSe$_2$ JJ\cite{farrar_superconducting_2021} and NbSe$_2$/Bi$_2$Se$_3$/NbSe$_2$ JJ\cite{kudriashov_non-reciprocal_2025}.
Conventional NbSe$_2$-based vdW gap SQUIDs\cite{farrar_superconducting_2021}, shown in Fig.~\ref{fig:JJ_squid}e, are composed of $s$-wave superconductors. These SQUIDs exhibit a large modulation depth in the voltage–flux characteristics\cite{farrar_superconducting_2021}, making them highly sensitive flux-noise sensors. Whereas SQUIDs incorporating a topological insulator (TI) barrier, such as NbSe$_2$/Bi$_2$Se$_3$/NbSe$_2$\cite{kudriashov_non-reciprocal_2025}, provides valuable insight about the topological nature of the S/TI interface.

In the second architecture, the SQUID loop is fabricated from a conventional superconductor, and the vdW material is incorporated into one of its arms. Such hybrid SQUIDs have been realized using WTe$_2$\cite{endres_currentphase_2023}, as we will discuss in the next section.

In Fig.~\ref{fig:JJ_squid}f, we highlight key advantages of vdW JJs. Owing to the minimal lithographic steps during fabrication, vdW JJs exhibit lower disorder than sputtered heterostructures. Their properties can be tuned via a magnetic field or the twist angle between layers. From an application standpoint, vdW JJs are well-suited for low-noise sensors such as bolometers\cite{sarkar_kerr_2025} and SQUIDs\cite{farrar_superconducting_2021, idzuchi_unconventional_2021, ai_van_2021}, and have shown promise in applications in quantum circuits such as Josephson diodes\cite{zhao_time-reversal_2023,ghosh_high-temperature_2024}. Recent studies also show possibilities of using the vdW JJ as merged element transmon qubits~\cite{zhao_merged-element_2020, mamin_merged-element_2021} and amplifiers~\cite{sun_broadband_2025}, an aspect we discuss later in \hypersetup{linkcolor=black}\hyperref[sec:8.1]{section 8.1}. Collectively, vdW SQUIDs leverage the versatility of these junctions to create powerful devices that bridge fundamental studies of quantum phase dynamics with emerging applications in precision sensing and circuit integration.

Taken together, these examples illustrate the complex physics of vdW-based tunnel barrier JJs and their applications, where material choice, twist angle, and magnetic functionality provide powerful tuning knobs. Beyond such material and architecture-driven approaches, an exciting frontier lies in flat-band systems, where the geometry of electronic states, rather than conventional band dispersion, dictates Josephson behavior.

\section{vdW Topological JJs} 
\label{sec:6}
Building on the discussion in previous sections, which highlight how structural engineering can yield exotic superconducting states, we now shift focus to topological systems. Topological Josephson junctions represent a frontier in condensed matter physics, where the Josephson effect is combined with topologically non-trivial electronic states to realize emergent quasiparticles and unconventional superconducting phenomena. Unlike conventional JJs, where Cooper pair transfers through a trivial metallic or semiconducting weak link, topological JJs leverage systems with protected edge or surface states, such as quantum Hall edges, topological insulators, or multiterminal networks invoking synthetic topology, to engineer robust, phase-coherent modes that are resilient to disorder and dephasing.

\subsection{JJs Realized Using Quantum Hall States} 
\begin{figure*}[h!]
       \centering
	\includegraphics[width=17 cm]{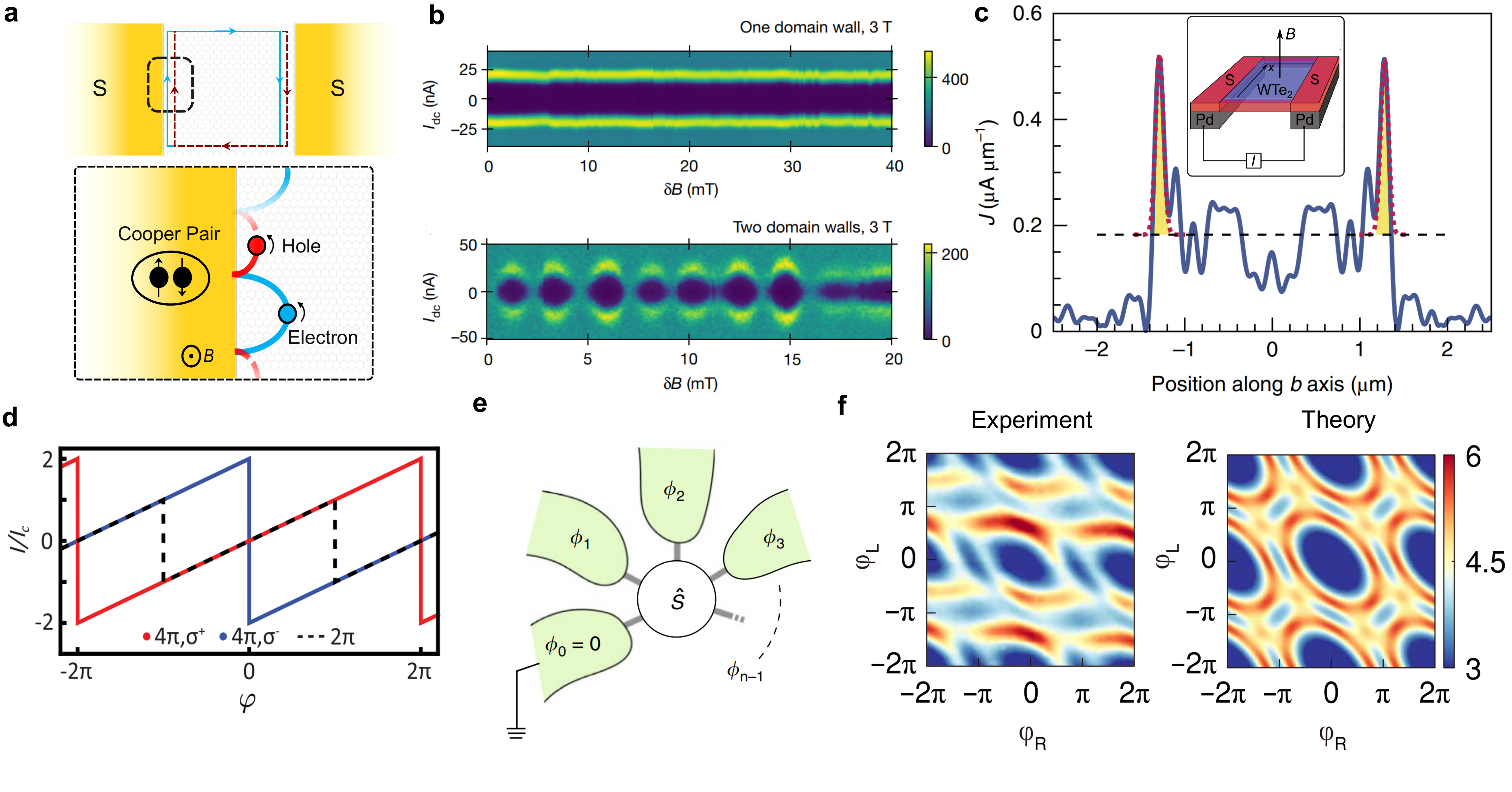}
    \caption{ \label{fig:topojj} {\textbf{JJs exhibiting topological character via quantum Hall states, topological insulators and multiterminal architecture.} 
     \textbf{a,}~Schematic of a topological Josephson junction in the quantum Hall regime with edge states. Inset, schematic of successive Andreev reflections of the quantum Hall edge channel carriers (here, a quasi-electron is reflected as a quasi-hole) leading to alternating skipping orbits at the interface of a topological material supporting edge states with a superconductor, in the presence of magnetic field. The electrons and holes are shown to follow the semi-classical cyclotron trajectory.
     \textbf{b,}~ Differential resistance maps of Josephson junctions with one (upper panel) and two (lower panel) AB/BA domain walls in a minimally twisted bilayer graphene sample. The domain walls act as the weak link in the quantum Hall regime~\cite{Barrier_2024}.
     \textbf{c,}~ Extracted Fourier transform of the magnetic field interference pattern of the critical current of the WTe$_2$ Josephson junction; revealing a high Josephson current density at the edge of the sample, highlighted by the yellow shaded region. The dashed line represents the bulk contribution \cite{choi2020evidence}. Inset, Josephson junction with WTe$_2$ as the weak link and Pd as the superconductor. Due to the topological character of WTe$_2$, it is predicted to host edge channels \cite{kononov2020one}. 
     \textbf{d,}~Schematic for theoretically expected normalized current ($I/I_{c}$)-phase ($\phi$) relation (CPR) for a topological Josephson junction. The parity-protected Andreev bound states of opposite parity ($\sigma^+$and $\sigma^-$) result in a 4$\pi$-periodic supercurrent through the SQUID. Due to spin-momentum locking of the helical states, the CPR is multivalued and of a characteristic saw-tooth shape. A 2$\pi$ periodic CPR of a topologically trivial junction is shown with a dashed line\cite{endres_currentphase_2023}. 
    \textbf{e,}~ Schematic of a multiterminal Josephson junction with multiple superconducting leads with phases $\phi_1$, $\phi_2$, ..$\phi_n$ connected to a single scattering region described by a scattering matrix $\hat{S}$. Due to gauge dependence, only $n-1$ of these leads have independent phases, that act as synthetic dimensions. Multiterminal JJs can give rise to topological features such as Weyl singularities in the synthetic spectrum~\cite{Riwar_2016}.  \textbf{f,}~ Experimental (left) and theoretical (right) differential tunneling conductance as a function of phase differences $\varphi\mathrm{_{R}}$ and $\varphi\mathrm{_{L}}$ between two terminals of a three-terminal Josephson junction. The color scale represents ($dI/dV$) in $\mu S$ units for the experiment panel, and in arbitrary units for the theory panel. In this synthetic two-dimensional space, various transitions between gapless (red) and gapped (blue) regions can be identified, showcasing the rich Andreev spectrum of the system~\cite{Jung_2025}.
    }}
\end{figure*}

The quantum Hall effect arises when a two-dimensional electron gas, subjected to a strong perpendicular magnetic field and cooled to low temperatures, exhibits Landau quantization of electronic states, leading to dissipationless edge transport and quantized Hall conductance in integer or fractional units of $e^2/h$ reflecting the underlying topological order of the system \cite{Stone_1992,Prange_1990,Goerbig_2011a,Goerbig_2011b}. In a conventional JJ, the critical current follows a Fraunhofer interference pattern with a periodicity of $\Phi_0=h/2e$ arising from the interference of Cooper pairs carrying charge $2e$ in the presence of magnetic flux. In contrast, a quantum Hall JJ supports a supercurrent mediated by charge $e$ quasiparticles (electrons and holes) traveling on spatially separated chiral edge states, shown in schematic Fig.~\ref{fig:topojj}a. Their interference, governed by the enclosed magnetic flux, results in a $2\Phi_0=h/e$ periodicity. The interplay between the quantum Hall effect and superconductivity is predicted to yield excitations with non-trivial braiding statistics, most notably Majorana fermions and non-Abelian anyons~\cite{Stern_2013,Alicea_2015,Lindner_2012,Clarke_2013,Mong_2014,San_Jose_2015}. To probe these phenomena, several groups have fabricated JJs in quantum Hall systems by coupling superconducting leads to a quantum Hall bar.

Amet \textit{et al.} investigated a graphene-based JJ with MoRe superconducting electrodes~\cite{amet_supercurrent_2016}. Although the junction geometry could, in principle, support chiral edge-state supercurrents in the quantum Hall regime, the expected $2\Phi_0$ flux periodicity was not observed. This absence was later attributed to additional conduction channels arising from charge accumulation along etched graphene edges, effectively forming two trivial JJs, one at each edge, which give rise to the observed $\Phi_0$ oscillations~\cite{Seredinski_2019}. Similarly, a study on bilayer graphene with MoRe contacts in the quantum Hall regime also found no evidence of $2\Phi_0$ periodicity~\cite{Indolese_2020}.
A subsequent generation of graphene JJs employed MoGe contacts and ultra-narrow junctions (125--330~nm)~\cite{Vignaud_2023}. In these devices, clear supercurrent oscillations with period $2\Phi_0$ were detected, providing evidence for chiral supercurrents along the quantum Hall edges. However, the critical currents remained extremely small ($\sim \SI{1}{\nano\ampere}$ at mK temperatures) and poorly reproducible, even among nominally identical junctions.

More recently, JJs based on minimally twisted bilayer graphene have exploited the one-dimensional domain walls that form between AB and BA stacking regions~\cite{Barrier_2024}. Devices with NbTi leads and domain-wall widths of $\sim10~\mathrm{nm}$ and lengths of 100--200~nm exhibit robust, non-oscillatory supercurrents of 10--20~nA per domain wall, shown in Fig~\ref{fig:topojj}b, essentially limited by the quantum conductance of strictly one-dimensional ballistic channels. These modes represent a promising avenue for achieving chiral superconductivity and, consequently, non-Abelian anyons, a prospect requiring further theoretical and experimental exploration. The fate of these modes in proximity to bulk domains where symmetry is broken, such as in magic-angle helical trilayer graphene~\cite{devakul2023magic}, also remains an open question.

\subsection{JJs Realised Using Topological Insulators}
\label{sec:6.2}
Topological insulators (TIs) are materials distinguished by an insulating bulk and topologically protected gapless states localized on surfaces or hinges, often protected by symmetries. In $d$ dimensions, an $n$-th order topological insulator $(d\geq n)$ hosts $(d-n)$ dimensional topologically protected conducting channels. These materials have gained significant attention for their potential to facilitate the exploration and realization of zero-energy Majorana modes that reside on the surface of topological superconductors~\cite{hasan_colloquium_2010,qi_topological_2011,breunig_opportunities_2021}. Topological superconductivity can be induced via the proximity effect between conventional superconductors and TIs or arise intrinsically in hybrid topological materials exhibiting a bulk superconducting state. JJs that incorporate such topological materials are predicted to host 4$\pi$-periodic Andreev bound states, interpreted as arising from the hybridization of Majorana bound states localized at each junction end. Fig.~\ref{fig:topojj}d shows a schematic 4$\pi$-periodic CPR. These topological JJs are also predicted to exhibit characteristics such as a fractional AC Josephson effect occurring at half the standard Josephson frequency. In the presence of an AC bias, only even Shapiro steps are predicted to be observable. The properties of topological JJs hold implications for the development of fault-tolerant qubits in quantum computing~\cite{das_sarma_topological_2006,nayak_non-abelian_2008,google_quantum_ai_and_collaborators_non-abelian_2023}.

Recent experimental progress involves JJs based on $\mathrm{WTe_{2}}$, a Weyl semimetal predicted in the monolayer limit (alongside  $\mathrm{MoTe_{2}}$) to host helical one-dimensional hinge states. Detection of these states via standard transport measurements is complicated by the conducting bulk due to its semimetallic nature. In 2020, three independent studies reported evidence of one-dimensional conducting states that carry supercurrent at the edges of $\mathrm{WTe_{2}}$ JJs contacted with Nb or Pd\cite{huang2020edge, kononov2020one, choi2020evidence}. These investigations used a methodology analogous to hinge state detection in bismuth-based TIs, utilizing the inverse Fourier transform of Fraunhofer patterns $I_{c}(B)$ to resolve the spatial current density profile within the junction as shown in Fig.~\ref{fig:topojj}c. Candidate interpretations for these states include helical hinge states, one-dimensional edge states, and two-dimensional Fermi-arc surface states. The studies included monolayer, few-layer, and bulk $\mathrm{WTe_{2}}$ junctions, ranging from short ballistic to long devices. In particular, Pd contacts on $\mathrm{WTe_{2}}$ form superconducting $\mathrm{PdTe_{x}}$ \cite{endres2022transparent}, and such Pd-contacted $\mathrm{WTe_{2}}$ JJs have recently been integrated into asymmetric SQUIDs \cite{endres_currentphase_2023}, revealing a multivalued 4$\pi$-periodic CPR. In the community, there is debate whether this 4$\pi$-periodic CPR supports the topological nature of the JJ. However, observation of the characteristic odd-even Shapiro step modulation in these topological vdW JJs remains an open experimental question. Similarly, Nb-contacted $\mathrm{MoTe_{2}}$ JJs also demonstrate edge supercurrents but are distinct in that they exhibit spatial asymmetry at the junction\cite{chen2024asymmetric}.

\subsection{Multiterminal JJ with Synthetic Topological Architecture} 

Beyond two-terminal devices, Josephson physics can be enriched even further by using multiterminal architectures. Multiterminal Josephson junctions (MTJJs) with $n\geq3$ superconducting leads enable the engineering of synthetic topological matter, where the $n-1$ superconducting phase differences gives rise to synthetic dimensions; here the phase differences are analogous to the wave vectors $\vec{k}$ in solid state systems. Fig.~\ref{fig:topojj}e shows the schematic of an MTJJ. In such architectures, the Andreev bound state (ABS) spectrum acts as an effective "band structure", which can be further tuned by electrostatic gating or magnetic flux. 
Topologically non-trivial structure emerges when the ABS bands develop isolated gap closings at specific phase configurations, which endows the synthetic bands with Berry curvature and nonzero Chern numbers~\cite{Riwar_2016,meyer_nontrivial_2017,Coraiola_2023}. MTJJs can also give rise to new phenomena, such as `multiplet supercurrent' flow, where a supercurrent is carried not by the conventional Cooper pairs, but by a coherent process involving more than two superconducting terminals, where multiple Cooper pairs are transferred simultaneously in a correlated way~\cite{arnault_multiplet_2025}.

Graphene is routinely used in MTJJs \citeonline{Draelos_2019,Arnault_2021} due to its supporting properties, such as gate tunability and high-transparency contact engineering \cite{Calado_2015}. Various three-terminal JJs have been constructed to engineer such 2D band structures of ABS states; see Fig.~\ref{fig:topojj}f. In ref.~\cite{huang_evidence_2022} MTJJs based on graphene provide some evidence of higher-order coherent transport in the form of “quartets” where two Cooper pairs move together. When a finite bias is applied, the Andreev spectrum becomes a driven (Floquet) one, and in this picture, the quartet response arises due to transitions between Floquet-ABSs, which explains the characteristic bias-dependent features.  
A three-terminal graphene-based JJ study
~\cite{Jung_2025} predicts signatures of topological transitions of these ABS bands. Nevertheless, unambiguous experimental evidence of nontrivial topology and coherent mixing between Andreev states remains elusive \cite{Prosko_2024}, underscoring the need for further experimental investigation.

\section{Challenges With vdW JJs}
\label{sec:7}

In general, the performance of a JJ is determined not only by the intrinsic properties of the chosen materials but also by the quality of fabrication steps such as exfoliation, stacking, contacting, and encapsulation. The following subsections outline key experimental challenges, including contact transparency, environmental sensitivity, and interface quality, and highlight methods developed within the community to address them.

\subsection{Fabrication Difficulties} 
One of the challenges in graphene JJ fabrication is the ability to make transparent superconducting contacts. For SNS junctions, the transparency parameter $D$ defined in the \hypersetup{linkcolor=black}\hyperref[box1]{Theory Box} should be close to $1$ for a clean, highly transparent junction. There is a choice of superconducting metals such as Al, MoRe, Nb, and NbTiN, among others, which have been found to create good superconducting contacts with the graphene~\cite{huang_study_2022,jung_engineering_2025}. It is observed in the devices that the contact metals electron-dope the graphene, leading to an asymmetric response in switching current for electron and hole doping. Often, the switching currents are very small, and the junction is resistive on the hole-doped side, resulting in weak proximity. Different fabrication strategies have been used to enhance the switching current on the hole side by 2D contacts~\cite{jang_engineering_2024}. Sometimes, the reason for poor transparency of the contacts is trapped chemical residues coming from the fabrication processes. In such cases, a well-followed strategy is to perform vacuum annealing, which helps in improving the transparency~\cite{sarkar_kerr_2025}. Additionally, residual metal contamination in the etching chambers and impurities in the process gases can degrade the contact transparency. For other kinds of vdW JJ fabrication, also achieving a clean interface is very critical. For example, interfacial vdW JJ incorporating ambient-sensitive materials such as BSCCO and NbSe$_2$ faces significant fabrication challenges. These materials are vulnerable to degradation from moisture and oxygen, making it critical to avoid prolonged atmospheric exposure. Moreover, their fragile nature renders conventional chemical treatments incompatible. To preserve material integrity and mitigate contact resistance, a dry cryogenic exfoliation and stacking process\cite{zhao_time-reversal_2023,patil_pick-up_2024,lee_encapsulating_2023} is typically performed inside an inert argon-filled glovebox environment. Gold contacts are then deposited using a SiN stencil-based shadow-mask evaporation technique\cite{PhysRevLett.122.247001,deshmukh_nanofabrication_1999}.

\subsection{Loss Mechanisms – Subgap Resistance} 
In microwave resonators involving GrJoFETs, it has been observed that microwave loss arises from the JJ, which is related to the low transparency of the JJs or subgap states. The resonator's internal quality factor is quantified in terms of a resistance $R_\mathrm{sg}$, which comes as the parallel resistance along with the junction, in the RCSJ picture. The higher the value of the subgap resistance $R_\mathrm{sg}$, the better the quality of the junction. This quantity, which ultimately determines the junction performance in microwave circuits, is only accessible through $\mu$eV or GHz excitations and is hard to predict from low-frequency transport measurements~\cite{dou_microwave_2021}. Another possible source of losses in the JJ is energy relaxation to atomic defects. It is known that when graphene is etched to form the 1D edge contacts, this procedure leads to the formation of atomic defects along the edges of the sample~\cite{halbertal_imaging_2017,huang_study_2022}. Future studies with vdW contacts to graphene, such as NbSe$_2$ contacting graphene, can help to eliminate atomic defects and help in understanding the loss mechanism~\cite{dvir_planar_2021,kim_strong_2017}.

\section{Perspective – Things to Look Forward To} 
\label{sec:8}
Having examined the experimental challenges that currently limit the scalability and performance of vdW JJs, it is equally important to look ahead at how this platform may shape future directions in quantum science and technology. In the following subsections, we outline several forward-looking perspectives, ranging from qubits and non-reciprocal devices to flat-band systems -- that highlight both the technological opportunities and the open questions awaiting exploration.

\subsection{Qubits Using vdW JJ} 
\label{sec:8.1}
Among various superconducting qubit designs, transmon is perhaps the simplest and most successful architecture that has been well utilized in quantum processors~\cite{koch_charge-insensitive_2007}. However, transmons typically have a large footprint because of the large capacitor pads shunted across the junction, which host various surface oxides arising from the device fabrication steps. The distributed electric field across the capacitor pads couples to quantum two-level systems (TLS) in the lossy surface oxides, resulting in losses in the qubit. To mitigate these issues, a new kind of transmon called the merged-element transmon or "mergemon" has been proposed and demonstrated~\cite{zhao_merged-element_2020,mamin_merged-element_2021}. It replaces the large capacitor pads of conventional transmons and relies on the intrinsic capacitance of the JJ, thus achieving nearly $\sim100$ times reduction in the qubit footprint~\cite{zhao_merged-element_2020}. The development of mergemon is at an initial stage. The ideas have been implemented in thin-film-based JJs, like Al-AlO$_x$-Al JJs or Nb-amorphous Si-Nb. However, the qubit performance, such as relaxation time, decoherence time, and the qubit quality factor, requires improvement. In this regard, vdW JJs with crystalline materials and pristine interfaces are expected to be an ideal platform for realizing mergemon. There have been efforts to make vertically stacked JJs using crystalline vdW tunnel barriers such as MoS$_2$ and WSe$_2$~\cite{lee_two-dimensional_2019,balgley_crystalline_2025} for superconducting qubits. In addition, epitaxial growth enables the scalability of mergemons by allowing a reduced footprint and introducing new control parameters~\cite{antony_miniaturizing_2021,wang_hexagonal_2022}.

\subsection{Non-reciprocal Cryo Electronic Devices} 
Developing compact, non-reciprocal devices is crucial for advancing the performance and integration of quantum hardware components. This includes cryogenic couplers, circulators, switches, and other elements. Currently, the circulators used in RF electronics are typically large and rely on magnetic materials, which can interfere with the operation of other nearby components. Moreover, these nonreciprocal devices lack reconfigurability. Josephson diode-based devices may offer potential solutions in this area~\cite{nadeem_superconducting_2023,ingla-aynes_efficient_2025,castellani_superconducting_2025}. vdW Josephson diodes provide a new way of fabricating on-chip rectifiers that can be operated even at liquid nitrogen temperature~\cite{ghosh_high-temperature_2024}. Additionally, one can design RF power splitters using such diodes with opposite polarity that can be operated in the GHz frequency range at liquid nitrogen temperatures, and their biasing conditions can be tuned by employing a local flux line. Overall, these cryogenic non-reciprocal devices promise advancements in terms of energy efficiency, miniaturization, and tunability compared to existing electronics.

\subsection{Flatband Josephson Devices with High Kinetic Inductance}

Finally, flat-band superconductors provide a frontier where vdW JJs can serve not only as circuit elements but also as powerful probes of unconventional superconductivity and quantum geometry. The ability to extract intrinsic superfluid density via kinetic inductance measurements in JJs incorporating moiré superconductors opens powerful avenues for both technological applications and fundamental investigations. Flat-band systems, such as magic-angle twisted bilayer \cite{cao2018unconventional, portoles2022tunable}, trilayer graphene \cite{jha2025large}, and multilayer rhombohedral systems \cite{zhou_superconductivity_2021,han_signatures_2025} provide a unique platform where strong correlations, topology, and superconductivity converge. 

The ability to fabricate JJs using rhombohedral multilayers offers a unique opportunity to realize unconventional JJs, potentially exhibiting topological properties associated with chiral superconductors~\cite{han_signatures_2025} and fractional quantum anomalous Hall states~\cite{lu_fractional_2024}. As these unusual correlated states manifest themselves in the same material, one expects the usual challenges of contact transparency to be less pronounced. The fractional quantum anomalous Hall states in the rhombohedral graphene system have an advantage over the transition-metal dichalcogenide platform~\cite{park_observation_2023}, where the fractional quantum Hall state is challenging to couple to superconductors because of contact issues.

The observation of large kinetic inductance in moiré superconducting systems, despite the vanishingly small conventional stiffness expected in a flat band, challenges established paradigms rooted in Fermi liquid theory and points toward a fundamentally quantum mechanical origin of superconductivity\cite{Torma2022}. 
Kinetic inductance offers a direct probe of the temperature dependence of the superfluid density, which is sensitive to the underlying pairing symmetry and gap structure. Theoretical study taking into account interaction in the flat bands and the origin of high-kinetic inductance as a probe of the nature of superconductivity remains a challenge. 
Furthermore, they may provide a means to experimentally access the role of the quantum metric in enhancing superfluid stiffness in nominally dispersionless bands. This aspect remains largely unexplored in current experiments. On the practical side, the large kinetic inductance intrinsic to flat-band moiré systems holds great promise for the development of next-generation superconducting circuits. This includes highly nonlinear circuit elements for quantum computation, enhanced qubit–photon coupling regimes, and ultra-sensitive photon detectors.

In summary, vdW JJs have evolved from simple experimental curiosities into a versatile platform that bridges materials science, condensed matter physics, and quantum engineering.
Through this review, we have attempted to capture the burgeoning field of vdW JJs and the exciting possibilities they hold for the near future. Despite the challenges outlined here, the ability to control the symmetry of JJs using vdW architecture is a key achievement for science and technology.   
Looking forward, one may expect the field to move rapidly toward scalable device architectures with the vdW JJs serving as the building blocks for next-generation quantum science and engineering.

\bigskip\bigskip
\begin{center}
  \rule{0.05\textwidth}{0.2pt}\par\vspace{-11pt}
  \rule{0.15\textwidth}{0.2pt}\par\vspace{-11pt}
  \rule{0.25\textwidth}{0.2pt}\par\vspace{-11pt}
  \rule{0.35\textwidth}{0.2pt}\par\vspace{-11pt}
  \rule{0.25\textwidth}{0.2pt}\par\vspace{-11pt}
  \rule{0.15\textwidth}{0.2pt}\par\vspace{-11pt}
  \rule{0.05\textwidth}{0.2pt}
\end{center}

\clearpage

\begin{figure*}[h!]
    \centering
	\includegraphics[width=17.5cm]{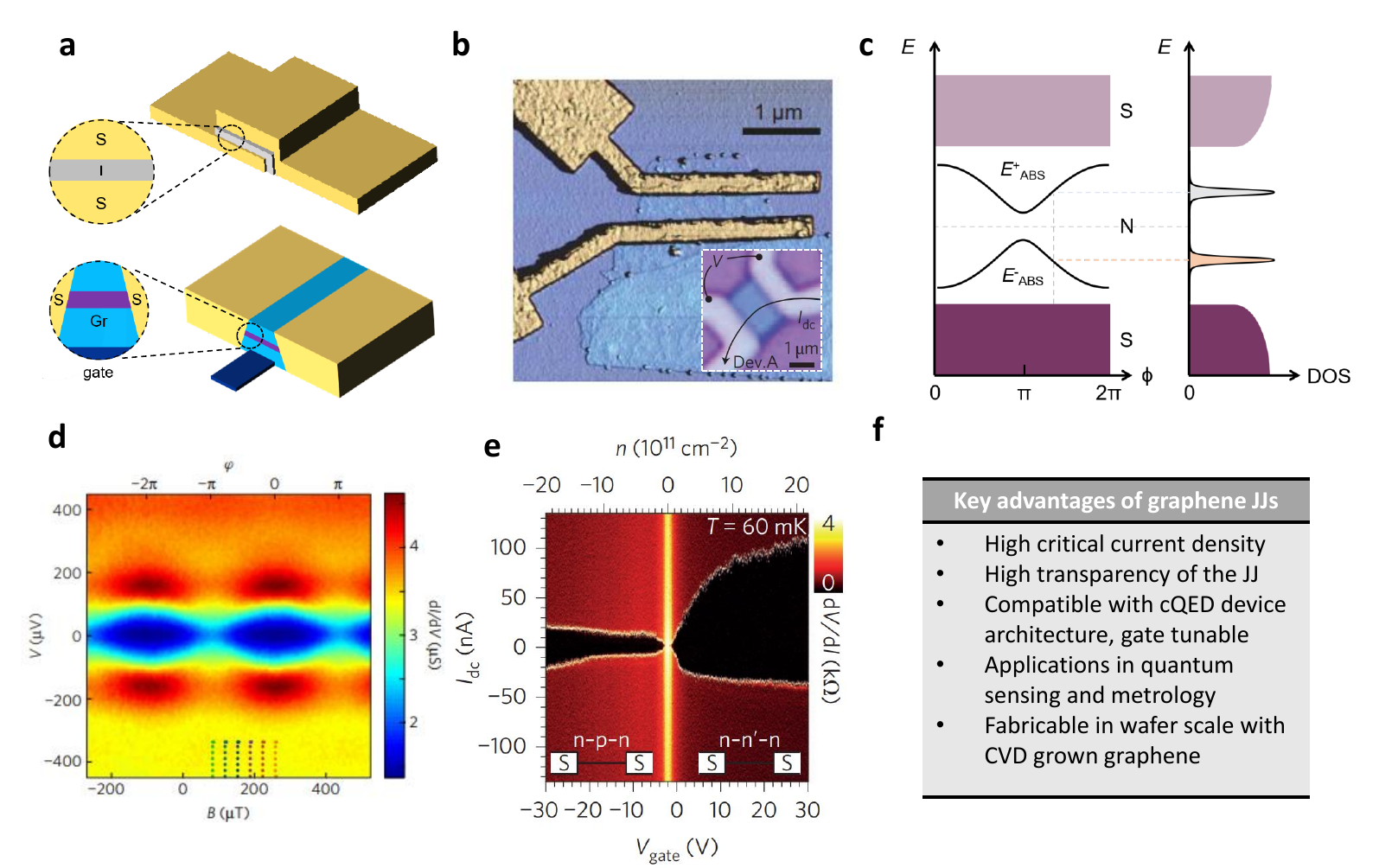}
    \caption{ \label{fig:grjj} {\textbf{Graphene based bipolar SNS Josephson junctions.} 
    \textbf{a,}~Shows the schematic of vertically formed S-I-S tunnel JJ and planar graphene JJ with superconducting contacts and bottom gate.
    \textbf{b,}~Shows the first generation graphene JJ realized with superconducting surface contacts on graphene flake~\cite{heersche_bipolar_2007}, and the inset shows the subsequent graphene JJ devices realized with superconducting edge contacts~\cite{Calado_2015}.
    \textbf{c,}~Shows the Andreev energy levels as a function of phase difference and the density of states as a function of energy in graphene JJ~\cite{Golubov_2004}. 
    \textbf{d,}~Shows the tunneling spectroscopy revealing the Andreev bands in graphene JJ devices~\cite{Bretheau_2017}. 
    \textbf{e,}~Shows the electrostatic gate tunable switching currents in graphene JJ devices~\cite{Calado_2015}. The switching current is asymmetric between the electron and hole sides due to electron-doping effects from the superconducting contacts, which is observed in all graphene JJ devices. 
    \textbf{f,}~Advantages of graphene JJs over the conventional S-I-S tunnel JJs.}}
    
\end{figure*}

\begin{naturebox}{Theory Box}
\begin{multicols}{2}
\phantomsection
\label{box1}
\begin{center}
    \includegraphics[width=0.43\textwidth]{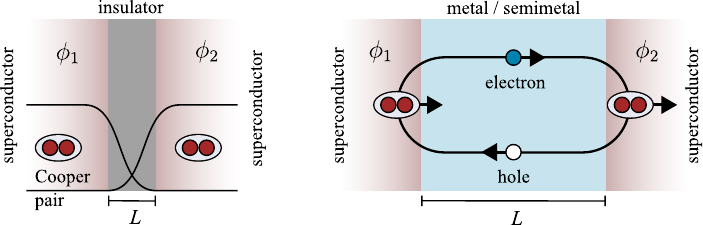}
    \label{fig:JJ-schematic}
\end{center}
\scriptsize
\subsection*{JJ basics}
A Josephson junction is a quantum device that enables the flow of supercurrent between two superconductors through a weak link or insulating barrier, governed by the macroscopic phase difference ($\phi=\phi_2-\phi_1$) of their superconducting order parameters. There are two primary types of Josephson junctions, distinguished by the nature of their barriers: \emph{Tunnel} Junctions (S-I-S), where a thin insulating barrier enables quantum tunneling of Cooper pairs due to a finite overlap of the macroscopic wavefunctions (left panel of the above figure), and \emph{weak-link} Junctions (right panel of the above figure), where the superconductors are coupled through a normal metallic region (S--N--S) or a through a narrow constriction also known as a point contact (S--c--S). All Josephson junctions are characterized by the transfer of Cooper pairs across a junction accompanied by a phase drop.

In the DC Josephson effect, a supercurrent flows across the junction without any voltage drop, provided the current remains below a critical value $I_c$. The AC Josephson effect arises when a voltage $V$ is applied across the two superconductors, causing the phase difference to evolve dynamically as $d\phi/dt=2eV/\hslash$. This time-dependent phase generates oscillations in the supercurrent at the Josephson frequency $\omega_j = {2eV}/{\hslash}$.

\subsection*{Classification of JJs}
Tunnel junctions consist of two superconducting electrodes separated by a thin insulating barrier (e.g., Al$_2$O$_3$), typically on the scale of nanometers ($L \ll l \ll \xi$, where $L$ is the separation between the superconducting electrodes, $l$ is the electron mean free path, and $\xi$ is the coherence length). The transport mechanism involves quantum tunneling of Cooper pairs through the barrier, described theoretically using the Ginzburg-Landau free energy framework, incorporating contributions from the individual superconductors ($f_1$ and $f_2$) and a coupling term $\gamma$ proportional to the overlap of their order parameters, $\psi_1 = \abs{\psi_1}e^{i\phi_1}$ and $\psi_2=\abs{\psi_2}e^{i\phi_2}$ with $\abs{\psi_{1,2}}\propto\sqrt{n_{1,2}}$ where $n_{1,2}$ is the Cooper pair density in the respective superconductors such that
\(
F = f_1(|\psi_1|) + f_2(|\psi_2|) - \gamma(\psi_1^*\psi_2 + \text{h.c.})
\). Minimizing the free energy with respect to the phase difference yields the sinusoidal CPR $I_J = I_c \sin\phi$, where $I_c$ depends on the Cooper pair densities and the tunneling strength $\gamma$. Near the critical temperature $T_c$, the critical current follows the Ambegaokar-Baratoff formula \cite{Ambegaokar_1963}:
\(
I_c \approx \frac{\pi \Delta}{2eR_N}\tanh\left(\frac{\Delta}{2k_BT}\right),
\)
with $\Delta$ being the superconducting gap and $R_N$ the normal-state resistance. A key parameter distinguishing tunnel junctions from weak links is their transparency—the probability that an electron (or hole) traverses the junction without scattering. High transparency ($D \approx 1$) corresponds to ballistic transport, characteristic of weak links, whereas low transparency ($D \ll 1$) indicates tunneling or diffusive transport, as in conventional tunnel junctions. We note that this is a simplified picture for ease of understanding, but real devices can be in between these two regimes. The reduced transparency of tunnel junctions leads to significantly lower critical current densities ($< 10^3$ A/cm$^2$). Their predictable sinusoidal response makes them ideal for metrological applications, such as voltage standards and superconducting qubits requiring precise phase control.

Weak link junctions utilize a normal metal region or a geometric constriction (rather than an insulator) to mediate the supercurrent. These junctions are classified as \textit{short} ($L \ll \xi$) or \textit{long} ($L \gg \xi$) based on the coherence length $\xi$, with transport properties further influenced by the electron mean free path $l$ (dirty vs. clean regimes). In such systems, Cooper pairs split into electron-hole pairs via repeated Andreev reflections, forming discrete Andreev-bound states (ABSs) within the weak link. These ABS energies depend on the phase difference $\phi$, enabling supercurrent flow even in the presence of a superconducting gap. Fig.~\ref{fig:grjj}c shows the Andreev energy levels as a function of phase difference and the density of states as a function of energy in graphene JJ. There have been tunneling spectroscopy studies~\cite{pillet_andreev_2010,Bretheau_2017,Park_2022,park_controllable_2024,jung_tunneling_2025} probing the Andreev bands in graphene JJ devices, see Fig.~\ref{fig:grjj}d. The CPR in weak links often deviates from the sinusoidal form, becoming skewed in ballistic or high-transparency regimes. 
Weak links exhibit higher critical currents ($10^3$--$10^6$ A/cm$^2$) and transparency ($D \approx 1$) compared to tunnel junctions, making them suitable for tunable superconducting qubits, phase-slip devices, and studies of non-equilibrium quantum transport. Below, we present a table comparing various properties of tunnel junctions and weak link junctions.
\begin{center}
\begin{tabular}{|p{2cm}|p{2.5cm}|p{2.5cm}|}
\hline
\textbf{Feature} & \textbf{Tunnel Junctions (S-I-S)} & \textbf{Weak Links (S-N-S/S-c-S)} \\
\hline
\textbf{Barrier Type} & 
Thin insulator (e.g., Al$_2$O$_3$, hBN)~\cite{Likharev_1979, nature2024abs,tian_josephson_2021} & 
Normal metal/constriction~\cite{Likharev_1979, nature2024abs} \\
\hline
\textbf{Length Scale} & 
$10^{-7}$ cm ($L \ll l \ll \xi$)~\cite{Likharev_1979} & 
0.1-10 $\mu$m ($L \lessgtr \xi$)~\cite{Likharev_1979} \\
\hline
\textbf{Transparency} & 
Low ($D \ll 1$)~\cite{royalsociety2018, nature2024abs} & 
High ($D \approx 1$)~\cite{royalsociety2018, nature2024abs} \\
\hline
\textbf{Critical Current} & 
$<10^3$ A/cm$^2$~\cite{Likharev_1979, pearson2011superconductivity} & 
$10^3$-$10^6$ A/cm$^2$~\cite{Likharev_1979, pearson2011superconductivity} \\
\hline
\textbf{Transport} & 
Direct tunneling~\cite{royalsociety2018, royalsociety2015, pearson2011superconductivity} & 
ABS formation~\cite{nature2024abs, royalsociety2018, royalsociety2015} \\
\hline
\textbf{Current-Phase} & 
$\sin\phi$~\cite{Golubov_2004, pearson2011superconductivity} & 
Non-sinusoidal~\cite{Golubov_2004, nature2024abs} \\
\hline
\textbf{Applications} & 
SQUIDs, Qubits~\cite{pearson2011superconductivity,Kjaergaard_2020} & 
Quantum sensors~\cite{lee_graphene-based_2020,braine_extended_2020} \\
\hline
\end{tabular}

\end{center}

\subsection*{Josephson Inductance}

When a time-dependent current flows through a JJ, it exhibits a finite inductance and a corresponding voltage drop. This behavior is analogous to Lenz’s law in classical electromagnetism, where a time-varying current induces an EMF voltage across an inductor. Drawing from this analogy, we can derive an expression for the effective inductance of a Josephson junction.

For a current-biased tunnel JJ, the current-phase relation is: \(I = I_c \sin\phi\),
where \( I_c \) is the switching current of the JJ. From the second Josephson relation, the voltage across the junction is given by:
$d\phi/dt=2eV/\hslash$. Now using these two Josephson relations and Lenz's law analogy: $V = L_J \dv{I}{t}$, one can derive an effective inductance for the JJ as $L_J = \frac{\Phi_0}{2\pi I_c \cos\phi}$, where \( \Phi_0 = h / 2 e \) is the flux quantum. We can identify \( L_J \) as the nonlinear Josephson inductance, which depends on the instantaneous phase drop \( \phi \). Alternatively, the Josephson inductance is obtained from the slope of the current phase relation as $L_J = \frac{\Phi_0}{2\pi} \left( \frac{dI}{d\phi} \right)^{-1}
$, see \hypersetup{linkcolor=black}\hyperref[box1]{Theory Box} second Fig. a. In the limit of small drive currents \( I_d \ll I_c \), we can express the Josephson inductance in terms of a linearized inductance \( L_J^{\text{lin}} = \Phi_0/2\pi I_c \), leading to $L_J^{\mathrm{non-lin}} = {L_J^{\text{lin}}}/{\sqrt{1 - (I_d/I_c)^2}}$. This expression highlights the nonlinear dependence of inductance on the drive current and is particularly important in the design of superconducting circuits such as qubits and parametric amplifiers.

\begin{center}
	\centering
	\includegraphics[width=0.5\textwidth,height=2.6 cm]{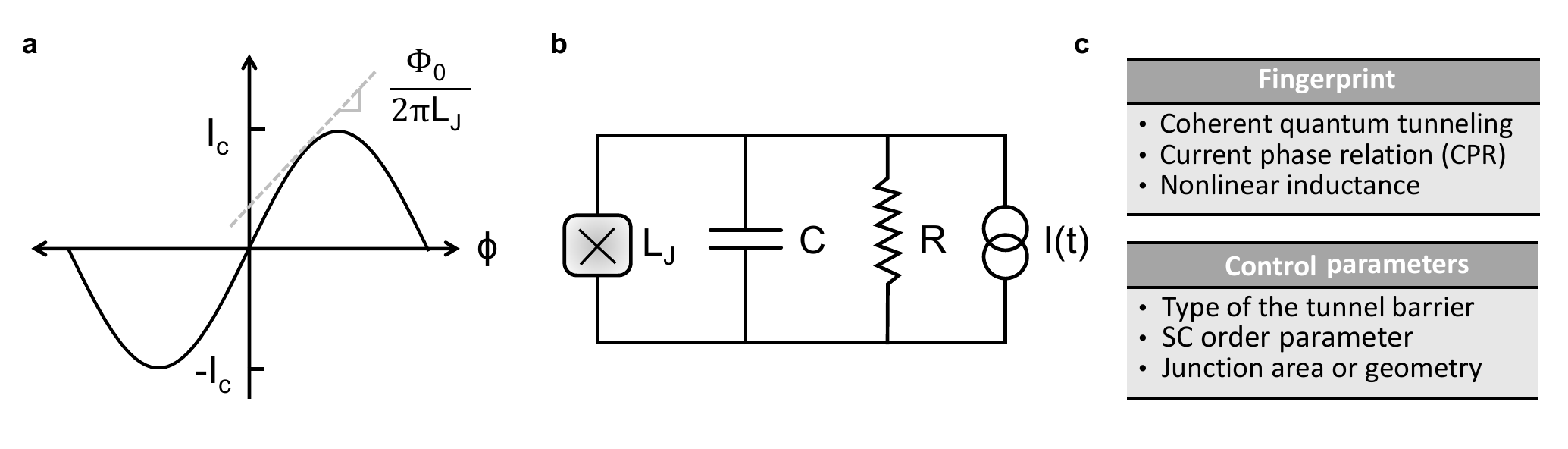}

\label{fig:rcsj}
\end{center}

\subsection*{RCSJ model}\label{sec:JJintro2}
The Resistively and Capacitively Shunted Junction (RCSJ) model is a widely adopted framework for describing the dynamics of JJs. In this model, the junction is represented by a nonlinear Josephson inductance shunted by a resistor \( R \) and a capacitor \( C \), both of which may include parasitic contributions beyond those intrinsic to the junction. A schematic of the RCSJ circuit is shown in \hypersetup{linkcolor=black}\hyperref[box1]{Theory Box} second Fig. b. Applying Kirchhoff's current law to the three branches of the circuit, and assuming a tunnel junction with a sinusoidal current-phase relation, the total current through the junction is given by $I(t) = I_c \sin \phi + \frac{V}{R} + C \dv{V}{t}= I_c \sin \phi + \frac{\Phi_0}{2\pi R} \dv{\phi}{t}+ \frac{\Phi_0 C}{2\pi} \dv[2]{\phi}{t}$, which can be rearranged as a second-order nonlinear differential equation:
\begin{align}
    \dv[2]{\phi}{t} + 2\Gamma \dv{\phi}{t} + \omega_p^2 \sin\phi = \omega_p^2 I(t)/I_c
    \label{eq:rcsj2}
\end{align}
where \( 2\Gamma = 1/RC \) is the damping rate, and \( \omega_p = 1/\sqrt{L_J^{\text{lin}} C} \) is the linear resonance frequency, often called the Josephson plasma frequency.
\cref{eq:rcsj2} can be interpreted as the equation of motion for a fictitious "phase particle" moving in an anharmonic potential, subject to damping and external drive. This analogy provides deep insight into the nonlinear dynamics of the system. For small-amplitude oscillations, one often linearizes the sine term using a Taylor expansion: \( \sin\phi \approx \phi - {\phi^3}/{3!}, \) which allows for approximate analytic treatment of the phase particle's trajectory~\cite{vijay_invited_2009}. The table in Fig. c of the second figure summarizes the key features and the control parameters of JJs that control the strength of the critical current.
\end{multicols}
\end{naturebox}

\cleardoublepage
\bibliography{References}

@article{golubov_current-phase_2004,
	title = {The current-phase relation in {Josephson} junctions},
	volume = {76},
	issn = {0034-6861, 1539-0756},
	urlp = {https://link.aps.org/doip/10.1103/RevModPhys.76.411},
	doip = {10.1103/RevModPhys.76.411},
	language = {en},
	number = {2},
	urlpdate = {2022-01-27},
	journal = {Reviews of Modern Physics},
	author = {Golubov, A. A. and Kupriyanov, M. Yu. and Il’ichev, E.},
	month = apr,
	year = {2004},
	pages = {411--469},
}

@article{schmidt_ballistic_2018,
	title = {A ballistic graphene superconducting microwave circuit},
	volume = {9},
	issn = {2041-1723},
	urlp = {http://www.nature.com/articles/s41467-018-06595-2},
	doip = {10.1038/s41467-018-06595-2},
	language = {en},
	number = {1},
	urlpdate = {2022-01-27},
	journal = {Nature Communications},
	author = {Schmidt, Felix E. and Jenkins, Mark D. and Watanabe, Kenji and Taniguchi, Takashi and Steele, Gary A.},
	month = dec,
	year = {2018},
	pages = {4069},
}

@article{kroll_magnetic_2018,
	title = {Magnetic field compatible circuit quantum electrodynamics with graphene {Josephson} junctions},
	volume = {9},
	issn = {2041-1723},
	urlp = {http://www.nature.com/articles/s41467-018-07124-x},
	doip = {10.1038/s41467-018-07124-x},
	language = {en},
	number = {1},
	urlpdate = {2022-01-27},
	journal = {Nature Communications},
	author = {Kroll, J. G. and Uilhoorn, W. and van der Enden, K. L. and de Jong, D. and Watanabe, K. and Taniguchi, T. and Goswami, S. and Cassidy, M. C. and Kouwenhoven, L. P.},
	month = dec,
	year = {2018},
	pages = {4615},
}

@article{wang_coherent_2019,
	title = {Coherent control of a hybrid superconducting circuit made with graphene-based van der {Waals} heterostructures},
	volume = {14},
	issn = {1748-3387, 1748-3395},
	urlp = {http://www.nature.com/articles/s41565-018-0329-2},
	doip = {10.1038/s41565-018-0329-2},
	language = {en},
	number = {2},
	urlpdate = {2022-01-27},
	journal = {Nature Nanotechnology},
	author = {Wang, Joel I-Jan and Rodan-Legrain, Daniel and Bretheau, Landry and Campbell, Daniel L. and Kannan, Bharath and Kim, David and Kjaergaard, Morten and Krantz, Philip and Samach, Gabriel O. and Yan, Fei and Yoder, Jonilyn L. and Watanabe, Kenji and Taniguchi, Takashi and Orlando, Terry P. and Gustavsson, Simon and Jarillo-Herrero, Pablo and Oliver, William D.},
	month = feb,
	year = {2019},
	pages = {120--125},
}

@article{lee_ultimately_2015,
	title = {Ultimately short ballistic vertical graphene {Josephson} junctions},
	volume = {6},
	issn = {2041-1723},
	urlp = {http://www.nature.com/articles/ncomms7181},
	doip = {10.1038/ncomms7181},
	language = {en},
	number = {1},
	urlpdate = {2022-01-27},
	journal = {Nature Communications},
	author = {Lee, Gil-Ho and Kim, Sol and Jhi, Seung-Hoon and Lee, Hu-Jong},
	month = may,
	year = {2015},
	pages = {6181},
}

@article{bergeal_phase-preserving_2010,
	title = {Phase-preserving amplification near the quantum limit with a {Josephson} ring modulator},
	volume = {465},
	issn = {0028-0836, 1476-4687},
	urlp = {http://www.nature.com/articles/nature09035},
	doip = {10.1038/nature09035},
	language = {en},
	number = {7294},
	urlpdate = {2022-01-27},
	journal = {Nature},
	author = {Bergeal, N. and Schackert, F. and Metcalfe, M. and Vijay, R. and Manucharyan, V. E. and Frunzio, L. and Prober, D. E. and Schoelkopf, R. J. and Girvin, S. M. and Devoret, M. H.},
	month = may,
	year = {2010},
	pages = {64--68},
}

@article{hatridge_dispersive_2011,
	title = {Dispersive magnetometry with a quantum limited {SQUID} parametric amplifier},
	volume = {83},
	issn = {1098-0121, 1550-235X},
	urlp = {https://link.aps.org/doip/10.1103/PhysRevB.83.134501},
	doip = {10.1103/PhysRevB.83.134501},
	language = {en},
	number = {13},
	urlpdate = {2022-01-27},
	journal = {Physical Review B},
	author = {Hatridge, M. and Vijay, R. and Slichter, D. H. and Clarke, John and Siddiqi, I.},
	month = apr,
	year = {2011},
	pages = {134501},
}

@article{braine_extended_2020,
	title = {Extended {Search} for the {Invisible} {Axion} with the {Axion} {Dark} {Matter} {Experiment}},
	volume = {124},
	issn = {0031-9007, 1079-7114},
	urlp = {https://link.aps.org/doip/10.1103/PhysRevLett.124.101303},
	doip = {10.1103/PhysRevLett.124.101303},
	language = {en},
	number = {10},
	urlpdate = {2022-01-27},
	journal = {Physical Review Letters},
	author = {Braine, T. and Cervantes, R. and Crisosto, N. and Du, N. and Kimes, S. and Rosenberg, L. J. and Rybka, G. and Yang, J. and Bowring, D. and Chou, A. S. and Khatiwada, R. and Sonnenschein, A. and Wester, W. and Carosi, G. and Woollett, N. and Duffy, L. D. and Bradley, R. and Boutan, C. and Jones, M. and LaRoque, B. H. and Oblath, N. S. and Taubman, M. S. and Clarke, J. and Dove, A. and Eddins, A. and O’Kelley, S. R. and Nawaz, S. and Siddiqi, I. and Stevenson, N. and Agrawal, A. and Dixit, A. V. and Gleason, J. R. and Jois, S. and Sikivie, P. and Solomon, J. A. and Sullivan, N. S. and Tanner, D. B. and Lentz, E. and Daw, E. J. and Buckley, J. H. and Harrington, P. M. and Henriksen, E. A. and Murch, K. W. and {ADMX Collaboration}},
	month = mar,
	year = {2020},
	pages = {101303},
}

@article{lee_graphene-based_2020,
	title = {Graphene-based {Josephson} junction microwave bolometer},
	volume = {586},
	issn = {0028-0836, 1476-4687},
	urlp = {https://www.nature.com/articles/s41586-020-2752-4},
	doip = {10.1038/s41586-020-2752-4},
	language = {en},
	number = {7827},
	urlpdate = {2022-01-27},
	journal = {Nature},
	author = {Lee, Gil-Ho and Efetov, Dmitri K. and Jung, Woochan and Ranzani, Leonardo and Walsh, Evan D. and Ohki, Thomas A. and Taniguchi, Takashi and Watanabe, Kenji and Kim, Philip and Englund, Dirk and Fong, Kin Chung},
	month = oct,
	year = {2020},
	pages = {42--46},
}

@article{walsh_graphene-based_2017,
	title = {Graphene-{Based} {Josephson}-{Junction} {Single}-{Photon} {Detector}},
	volume = {8},
	issn = {2331-7019},
	urlp = {https://link.aps.org/doip/10.1103/PhysRevApplied.8.024022},
	doip = {10.1103/PhysRevApplied.8.024022},
	language = {en},
	number = {2},
	urlpdate = {2022-01-27},
	journal = {Physical Review Applied},
	author = {Walsh, Evan D. and Efetov, Dmitri K. and Lee, Gil-Ho and Heuck, Mikkel and Crossno, Jesse and Ohki, Thomas A. and Kim, Philip and Englund, Dirk and Fong, Kin Chung},
	month = aug,
	year = {2017},
	pages = {024022},
}

@article{walsh_josephson_2021,
	title = {Josephson junction infrared single-photon detector},
	volume = {372},
	issn = {0036-8075, 1095-9203},
	urlp = {https://www.science.org/doip/10.1126/science.abf5539},
	doip = {10.1126/science.abf5539},
	language = {en},
	number = {6540},
	urlpdate = {2022-01-27},
	journal = {Science},
	author = {Walsh, Evan D. and Jung, Woochan and Lee, Gil-Ho and Efetov, Dmitri K. and Wu, Bae-Ian and Huang, K.-F. and Ohki, Thomas A. and Taniguchi, Takashi and Watanabe, Kenji and Kim, Philip and Englund, Dirk and Fong, Kin Chung},
	month = apr,
	year = {2021},
	pages = {409--412},
}

@article{wang_hexagonal_2022,
	title = {Hexagonal boron nitride as a low-loss dielectric for superconducting quantum circuits and qubits},
	issn = {1476-1122, 1476-4660},
	urlp = {https://www.nature.com/articles/s41563-021-01187-w},
	doip = {10.1038/s41563-021-01187-w},
	language = {en},
	urlpdate = {2022-01-31},
	journal = {Nature Materials},
	author = {Wang, Joel I-J. and Yamoah, Megan A. and Li, Qing and Karamlou, Amir H. and Dinh, Thao and Kannan, Bharath and Braumüller, Jochen and Kim, David and Melville, Alexander J. and Muschinske, Sarah E. and Niedzielski, Bethany M. and Serniak, Kyle and Sung, Youngkyu and Winik, Roni and Yoder, Jonilyn L. and Schwartz, Mollie E. and Watanabe, Kenji and Taniguchi, Takashi and Orlando, Terry P. and Gustavsson, Simon and Jarillo-Herrero, Pablo and Oliver, William D.},
	month = jan,
	year = {2022},
}

@article{vijay_invited_2009,
	title = {Invited {Review} {Article}: {The} {Josephson} bifurcation amplifier},
	volume = {80},
	issn = {0034-6748, 1089-7623},
	shorttitle = {Invited {Review} {Article}},
	urlp = {http://aip.scitation.org/doip/10.1063/1.3224703},
	doip = {10.1063/1.3224703},
	language = {en},
	number = {11},
	urldate = {2022-02-08},
	journal = {Review of Scientific Instruments},
	author = {Vijay, R. and Devoret, M. H. and Siddiqi, I.},
	month = nov,
	year = {2009},
	pages = {111101},
}

@article{nanda_current-phase_2017,
	title = {Current-{Phase} {Relation} of {Ballistic} {Graphene} {Josephson} {Junctions}},
	volume = {17},
	issn = {1530-6984, 1530-6992},
	urlp = {https://pubs.acs.org/doip/10.1021/acs.nanolett.7b00097},
	doip = {10.1021/acs.nanolett.7b00097},
	language = {en},
	number = {6},
	urldate = {2022-02-13},
	journal = {Nano Letters},
	author = {Nanda, G. and Aguilera-Servin, J. L. and Rakyta, P. and Kormányos, A. and Kleiner, R. and Koelle, D. and Watanabe, K. and Taniguchi, T. and Vandersypen, L. M. K. and Goswami, S.},
	month = jun,
	year = {2017},
	pages = {3396--3401},
}

@article{kokkoniemi_bolometer_2020,
	title = {Bolometer operating at the threshold for circuit quantum electrodynamics},
	volume = {586},
	issn = {0028-0836, 1476-4687},
	urlP = {https://www.nature.com/articles/s41586-020-2753-3},
	doip = {10.1038/s41586-020-2753-3},
	language = {en},
	number = {7827},
	urldate = {2022-02-16},
	journal = {Nature},
	author = {Kokkoniemi, R. and Girard, J.-P. and Hazra, D. and Laitinen, A. and Govenius, J. and Lake, R. E. and Sallinen, I. and Vesterinen, V. and Partanen, M. and Tan, J. Y. and Chan, K. W. and Tan, K. Y. and Hakonen, P. and Möttönen, M.},
	month = oct,
	year = {2020},
	pages = {47--51},
}

@article{devoret_superconducting_2013,
	title = {Superconducting {Circuits} for {Quantum} {Information}: {An} {Outlook}},
	volume = {339},
	issn = {0036-8075, 1095-9203},
	shorttitle = {Superconducting {Circuits} for {Quantum} {Information}},
	urlp = {https://www.science.org/doip/10.1126/science.1231930},
	doip = {10.1126/science.1231930},
	language = {en},
	number = {6124},
	urldate = {2022-02-17},
	journal = {Science},
	author = {Devoret, M. H. and Schoelkopf, R. J.},
	month = mar,
	year = {2013},
	pages = {1169--1174},
}

@article{aumentado_superconducting_2020,
	title = {Superconducting {Parametric} {Amplifiers}: {The} {State} of the {Art} in {Josephson} {Parametric} {Amplifiers}},
	volume = {21},
	issn = {1527-3342, 1557-9581},
	shorttitle = {Superconducting {Parametric} {Amplifiers}},
	urlp = {https://ieeexplore.ieee.org/document/9134828/},
	doip = {10.1109/MMM.2020.2993476},
	language = {en},
	number = {8},
	urldate = {2022-03-01},
	journal = {IEEE Microwave Magazine},
	author = {Aumentado, Jose},
	month = aug,
	year = {2020},
	pages = {45--59},
}

@article{larsen_semiconductor-nanowire-based_2015,
	title = {Semiconductor-{Nanowire}-{Based} {Superconducting} {Qubit}},
	volume = {115},
	issn = {0031-9007, 1079-7114},
	urlP = {https://link.aps.org/doip/10.1103/PhysRevLett.115.127001},
	doip = {10.1103/PhysRevLett.115.127001},
	language = {en},
	number = {12},
	urldate = {2022-03-07},
	journal = {Physical Review Letters},
	author = {Larsen, T. W. and Petersson, K. D. and Kuemmeth, F. and Jespersen, T. S. and Krogstrup, P. and Nygård, J. and Marcus, C. M.},
	month = sep,
	year = {2015},
	pages = {127001},
}

@article{larsen_parity-protected_2020,
	title = {Parity-{Protected} {Superconductor}-{Semiconductor} {Qubit}},
	volume = {125},
	issn = {0031-9007, 1079-7114},
	urlP = {https://link.aps.org/doip/10.1103/PhysRevLett.125.056801},
	doip = {10.1103/PhysRevLett.125.056801},
	language = {en},
	number = {5},
	urldate = {2022-03-07},
	journal = {Physical Review Letters},
	author = {Larsen, T. W. and Gershenson, M. E. and Casparis, L. and Kringhøj, A. and Pearson, N. J. and McNeil, R. P. G. and Kuemmeth, F. and Krogstrup, P. and Petersson, K. D. and Marcus, C. M.},
	month = jul,
	year = {2020},
	pages = {056801},
}

@article{de_lange_realization_2015,
	title = {Realization of {Microwave} {Quantum} {Circuits} {Using} {Hybrid} {Superconducting}-{Semiconducting} {Nanowire} {Josephson} {Elements}},
	volume = {115},
	issn = {0031-9007, 1079-7114},
	urlP = {https://link.aps.org/doip/10.1103/PhysRevLett.115.127002},
	doip = {10.1103/PhysRevLett.115.127002},
	language = {en},
	number = {12},
	urldate = {2022-03-07},
	journal = {Physical Review Letters},
	author = {de Lange, G. and van Heck, B. and Bruno, A. and van Woerkom, D. J. and Geresdi, A. and Plissard, S. R. and Bakkers, E. P. A. M. and Akhmerov, A. R. and DiCarlo, L.},
	month = sep,
	year = {2015},
	pages = {127002},
}

@article{dou_microwave_2021,
	title = {Microwave photoassisted dissipation and supercurrent of a phase-biased graphene-superconductor ring},
	volume = {3},
	issn = {2643-1564},
	urlp = {https://link.aps.org/doip/10.1103/PhysRevResearch.3.L032009},
	doip = {10.1103/PhysRevResearch.3.L032009},
	language = {en},
	number = {3},
	urldate = {2022-03-11},
	journal = {Physical Review Research},
	author = {Dou, Ziwei and Wakamura, Taro and Virtanen, Pauli and Wu, Nian-Jheng and Deblock, Richard and Autier-Laurent, Sandrine and Watanabe, Kenji and Taniguchi, Takashi and Guéron, Sophie and Bouchiat, Hélène and Ferrier, Meydi},
	month = jul,
	year = {2021},
	pages = {L032009},
}

@article{haller_phase-dependent_2022,
	title = {Phase-dependent microwave response of a graphene {Josephson} junction},
	volume = {4},
	issn = {2643-1564},
	urlp = {https://link.aps.org/doip/10.1103/PhysRevResearch.4.013198},
	doip = {10.1103/PhysRevResearch.4.013198},
	language = {en},
	number = {1},
	urldate = {2022-03-20},
	journal = {Physical Review Research},
	author = {Haller, R. and Fülöp, G. and Indolese, D. and Ridderbos, J. and Kraft, R. and Cheung, L. Y. and Ungerer, J. H. and Watanabe, K. and Taniguchi, T. and Beckmann, D. and Danneau, R. and Virtanen, P. and Schönenberger, C.},
	month = mar,
	year = {2022},
	pages = {013198},
}

@article{heersche_bipolar_2007,
	title = {Bipolar supercurrent in graphene},
	volume = {446},
	issn = {0028-0836, 1476-4687},
	urlp = {http://www.nature.com/articles/nature05555},
	doip = {10.1038/nature05555},
	language = {en},
	number = {7131},
	urldate = {2022-06-20},
	journal = {Nature},
	author = {Heersche, Hubert B. and Jarillo-Herrero, Pablo and Oostinga, Jeroen B. and Vandersypen, Lieven M. K. and Morpurgo, Alberto F.},
	month = mar,
	year = {2007},
	pages = {56--59},
}

@article{borzenets_ballistic_2016,
	title = {Ballistic {Graphene} {Josephson} {Junctions} from the {Short} to the {Long} {Junction} {Regimes}},
	volume = {117},
	issn = {0031-9007, 1079-7114},
	urlp = {https://link.aps.org/doip/10.1103/PhysRevLett.117.237002},
	doip = {10.1103/PhysRevLett.117.237002},
	language = {en},
	number = {23},
	urldate = {2022-05-24},
	journal = {Physical Review Letters},
	author = {Borzenets, I. V. and Amet, F. and Ke, C. T. and Draelos, A. W. and Wei, M. T. and Seredinski, A. and Watanabe, K. and Taniguchi, T. and Bomze, Y. and Yamamoto, M. and Tarucha, S. and Finkelstein, G.},
	month = dec,
	year = {2016},
	pages = {237002},
}

@article{el_fatimy_epitaxial_2016,
	title = {Epitaxial graphene quantum dots for high-performance terahertz bolometers},
	volume = {11},
	issn = {1748-3387, 1748-3395},
	urlp = {http://www.nature.com/articles/nnano.2015.303},
	doip = {10.1038/nnano.2015.303},
	language = {en},
	number = {4},
	urlpdate = {2022-08-07},
	journal = {Nature Nanotechnology},
	author = {El Fatimy, Abdel and Myers-Ward, Rachael L. and Boyd, Anthony K. and Daniels, Kevin M. and Gaskill, D. Kurt and Barbara, Paola},
	month = apr,
	year = {2016},
	pages = {335--338},
}

@article{katti_hot_2023,
	title = {Hot {Carrier} {Thermalization} and {Josephson} {Inductance} {Thermometry} in a {Graphene}-{Based} {Microwave} {Circuit}},
	volume = {23},
	copyright = {https://doi.org/10.15223/policy-029},
	issn = {1530-6984, 1530-6992},
	urlp = {https://pubs.acs.org/doi/10.1021/acs.nanolett.2c04791},
	doip = {10.1021/acs.nanolett.2c04791},
	language = {en},
	number = {10},
	urldate = {2025-03-21},
	journal = {Nano Letters},
	author = {Katti, Raj and Arora, Harpreet Singh and Saira, Olli-Pentti and Watanabe, Kenji and Taniguchi, Takashi and Schwab, Keith C. and Roukes, Michael Lee and Nadj-Perge, Stevan},
	month = may,
	year = {2023},
	pages = {4136--4141},
}

@article{sarkar_quantum-noise-limited_2022,
	title = {Quantum-noise-limited microwave amplification using a graphene {Josephson} junction},
	volume = {17},
	issn = {1748-3387, 1748-3395},
	urlp = {https://www.nature.com/articles/s41565-022-01223-z},
	doip = {10.1038/s41565-022-01223-z},
	language = {en},
	number = {11},
	urlpdate = {2023-09-22},
	journal = {Nature Nanotechnology},
	author = {Sarkar, Joydip and Salunkhe, Kishor V. and Mandal, Supriya and Ghatak, Subhamoy and Marchawala, Alisha H. and Das, Ipsita and Watanabe, Kenji and Taniguchi, Takashi and Vijay, R. and Deshmukh, Mandar M.},
	month = nov,
	year = {2022},
	pages = {1147--1152},
}

@article{opremcak_measurement_2018,
	title = {Measurement of a superconducting qubit with a microwave photon counter},
	volume = {361},
	issn = {0036-8075, 1095-9203},
	urlp = {https://www.science.org/doip/10.1126/science.aat4625},
	doip = {10.1126/science.aat4625},
	language = {en},
	number = {6408},
	urlpdate = {2023-09-25},
	journal = {Science},
	author = {Opremcak, A. and Pechenezhskiy, I. V. and Howington, C. and Christensen, B. G. and Beck, M. A. and Leonard, E. and Suttle, J. and Wilen, C. and Nesterov, K. N. and Ribeill, G. J. and Thorbeck, T. and Schlenker, F. and Vavilov, M. G. and Plourde, B. L. T. and McDermott, R.},
	month = sep,
	year = {2018},
	pages = {1239--1242},
}

@article{benz_application_2004,
	title = {Application of the {Josephson} effect to voltage metrology},
	volume = {92},
	issn = {0018-9219},
	urlp = {http://ieeexplore.ieee.org/document/1335552/},
	doip = {10.1109/JPROC.2004.833671},
	language = {en},
	number = {10},
	urldate = {2023-12-20},
	journal = {Proceedings of the IEEE},
	author = {Benz, S.P. and Hamilton, C.A.},
	month = oct,
	year = {2004},
	pages = {1617--1629},
}

@article{yan_dual-gated_2012,
	title = {Dual-gated bilayer graphene hot-electron bolometer},
	volume = {7},
	copyright = {http://www.springer.com/tdm},
	issn = {1748-3387, 1748-3395},
	urlp = {https://www.nature.com/articles/nnano.2012.88},
	doip = {10.1038/nnano.2012.88},
	language = {en},
	number = {7},
	urldate = {2024-06-17},
	journal = {Nature Nanotechnology},
	author = {Yan, Jun and Kim, M-H. and Elle, J. A. and Sushkov, A. B. and Jenkins, G. S. and Milchberg, H. M. and Fuhrer, M. S. and Drew, H. D.},
	month = jul,
	year = {2012},
	pages = {472--478},
}

@article{cai_sensitive_2014,
	title = {Sensitive room-temperature terahertz detection via the photothermoelectric effect in graphene},
	volume = {9},
	issn = {1748-3387, 1748-3395},
	urlp = {https://www.nature.com/articles/nnano.2014.182},
	doip = {10.1038/nnano.2014.182},
	language = {en},
	number = {10},
	urldate = {2024-06-17},
	journal = {Nature Nanotechnology},
	author = {Cai, Xinghan and Sushkov, Andrei B. and Suess, Ryan J. and Jadidi, Mohammad M. and Jenkins, Gregory S. and Nyakiti, Luke O. and Myers-Ward, Rachael L. and Li, Shanshan and Yan, Jun and Gaskill, D. Kurt and Murphy, Thomas E. and Drew, H. Dennis and Fuhrer, Michael S.},
	month = oct,
	year = {2014},
	pages = {814--819},
}

@article{paolucci_fully_2021,
	title = {Fully {Superconducting} {Josephson} {Bolometers} for {Gigahertz} {Astronomy}},
	volume = {11},
	copyright = {https://creativecommons.org/licenses/by/4.0/},
	issn = {2076-3417},
	urlp = {https://www.mdpi.com/2076-3417/11/2/746},
	doip = {10.3390/app11020746},
	language = {en},
	number = {2},
	urldate = {2024-06-17},
	journal = {Applied Sciences},
	author = {Paolucci, Federico and Ligato, Nadia and Germanese, Gaia and Buccheri, Vittorio and Giazotto, Francesco},
	month = jan,
	year = {2021},
	pages = {746},
}

@article{dean_boron_2010,
	title = {Boron nitride substrates for high-quality graphene electronics},
	volume = {5},
	copyright = {http://www.springer.com/tdm},
	issn = {1748-3387, 1748-3395},
	urlp = {https://www.nature.com/articles/nnano.2010.172},
	doip = {10.1038/nnano.2010.172},
	language = {en},
	number = {10},
	urldate = {2024-06-27},
	journal = {Nature Nanotechnology},
	author = {Dean, C. R. and Young, A. F. and Meric, I. and Lee, C. and Wang, L. and Sorgenfrei, S. and Watanabe, K. and Taniguchi, T. and Kim, P. and Shepard, K. L. and Hone, J.},
	month = oct,
	year = {2010},
	pages = {722--726},
}

@article{tombros_electronic_2007,
	title = {Electronic spin transport and spin precession in single graphene layers at room temperature},
	volume = {448},
	copyright = {http://www.springer.com/tdm},
	issn = {0028-0836, 1476-4687},
	urlp = {https://www.nature.com/articles/nature06037},
	doip = {10.1038/nature06037},
	language = {en},
	number = {7153},
	urldate = {2024-07-12},
	journal = {Nature},
	author = {Tombros, Nikolaos and Jozsa, Csaba and Popinciuc, Mihaita and Jonkman, Harry T. and Van Wees, Bart J.},
	month = aug,
	year = {2007},
	pages = {571--574},
}

@article{geim_van_2013,
	title = {Van der {Waals} heterostructures},
	volume = {499},
	copyright = {http://www.springer.com/tdm},
	issn = {0028-0836, 1476-4687},
	urlp = {https://www.nature.com/articles/nature12385},
	doip = {10.1038/nature12385},
	language = {en},
	number = {7459},
	urldate = {2024-07-17},
	journal = {Nature},
	author = {Geim, A. K. and Grigorieva, I. V.},
	month = jul,
	year = {2013},
	pages = {419--425},
}

@article{liu_two-dimensional_2020,
	title = {Two-dimensional materials for next-generation computing technologies},
	volume = {15},
	issn = {1748-3387, 1748-3395},
	urlp = {https://www.nature.com/articles/s41565-020-0724-3},
	doip = {10.1038/s41565-020-0724-3},
	language = {en},
	number = {7},
	urldate = {2024-07-17},
	journal = {Nature Nanotechnology},
	author = {Liu, Chunsen and Chen, Huawei and Wang, Shuiyuan and Liu, Qi and Jiang, Yu-Gang and Zhang, David Wei and Liu, Ming and Zhou, Peng},
	month = jul,
	year = {2020},
	pages = {545--557},
}

@article{montblanch_layered_2023,
	title = {Layered materials as a platform for quantum technologies},
	volume = {18},
	issn = {1748-3387, 1748-3395},
	urlp = {https://www.nature.com/articles/s41565-023-01354-x},
	doip = {10.1038/s41565-023-01354-x},
	language = {en},
	number = {6},
	urldate = {2024-07-17},
	journal = {Nature Nanotechnology},
	author = {Montblanch, Alejandro R.-P. and Barbone, Matteo and Aharonovich, Igor and Atatüre, Mete and Ferrari, Andrea C.},
	month = jun,
	year = {2023},
	pages = {555--571},
}

@article{browaeys_many-body_2020,
	title = {Many-body physics with individually controlled {Rydberg} atoms},
	volume = {16},
	issn = {1745-2473, 1745-2481},
	urlp = {https://www.nature.com/articles/s41567-019-0733-z},
	doip = {10.1038/s41567-019-0733-z},
	language = {en},
	number = {2},
	urlpdate = {2024-07-17},
	journal = {Nature Physics},
	author = {Browaeys, Antoine and Lahaye, Thierry},
	month = feb,
	year = {2020},
	pages = {132--142},
}

@article{casparis_superconducting_2018,
	title = {Superconducting gatemon qubit based on a proximitized two-dimensional electron gas},
	volume = {13},
	issn = {1748-3387, 1748-3395},
	urlp = {https://www.nature.com/articles/s41565-018-0207-y},
	doip = {10.1038/s41565-018-0207-y},
	language = {en},
	number = {10},
	urldate = {2024-07-26},
	journal = {Nature Nanotechnology},
	author = {Casparis, Lucas and Connolly, Malcolm R. and Kjaergaard, Morten and Pearson, Natalie J. and Kringhøj, Anders and Larsen, Thorvald W. and Kuemmeth, Ferdinand and Wang, Tiantian and Thomas, Candice and Gronin, Sergei and Gardner, Geoffrey C. and Manfra, Michael J. and Marcus, Charles M. and Petersson, Karl D.},
	month = oct,
	year = {2018},
	pages = {915--919},
}

@article{arute_quantum_2019,
	title = {Quantum supremacy using a programmable superconducting processor},
	volume = {574},
	issn = {0028-0836, 1476-4687},
	urlp = {https://www.nature.com/articles/s41586-019-1666-5},
	doip = {10.1038/s41586-019-1666-5},
	language = {en},
	number = {7779},
	urldate = {2024-07-26},
	journal = {Nature},
	author = {Arute, Frank and Arya, Kunal and Babbush, Ryan and Bacon, Dave and Bardin, Joseph C. and Barends, Rami and Biswas, Rupak and Boixo, Sergio and Brandao, Fernando G. S. L. and Buell, David A. and Burkett, Brian and Chen, Yu and Chen, Zijun and Chiaro, Ben and Collins, Roberto and Courtney, William and Dunsworth, Andrew and Farhi, Edward and Foxen, Brooks and Fowler, Austin and Gidney, Craig and Giustina, Marissa and Graff, Rob and Guerin, Keith and Habegger, Steve and Harrigan, Matthew P. and Hartmann, Michael J. and Ho, Alan and Hoffmann, Markus and Huang, Trent and Humble, Travis S. and Isakov, Sergei V. and Jeffrey, Evan and Jiang, Zhang and Kafri, Dvir and Kechedzhi, Kostyantyn and Kelly, Julian and Klimov, Paul V. and Knysh, Sergey and Korotkov, Alexander and Kostritsa, Fedor and Landhuis, David and Lindmark, Mike and Lucero, Erik and Lyakh, Dmitry and Mandrà, Salvatore and McClean, Jarrod R. and McEwen, Matthew and Megrant, Anthony and Mi, Xiao and Michielsen, Kristel and Mohseni, Masoud and Mutus, Josh and Naaman, Ofer and Neeley, Matthew and Neill, Charles and Niu, Murphy Yuezhen and Ostby, Eric and Petukhov, Andre and Platt, John C. and Quintana, Chris and Rieffel, Eleanor G. and Roushan, Pedram and Rubin, Nicholas C. and Sank, Daniel and Satzinger, Kevin J. and Smelyanskiy, Vadim and Sung, Kevin J. and Trevithick, Matthew D. and Vainsencher, Amit and Villalonga, Benjamin and White, Theodore and Yao, Z. Jamie and Yeh, Ping and Zalcman, Adam and Neven, Hartmut and Martinis, John M.},
	month = oct,
	year = {2019},
	pages = {505--510},
}

@article{deshmukh_simple_nodate,
	title = {Simple solids mimic complex electronic states},
        volume = {609},
	pages = {470--471},
	language = {en},
	year = {2022},
	journal = {Nature},
	author = {Deshmukh, Mandar M},
}

@article{andrei_marvels_2021,
	title = {The marvels of moiré materials},
	volume = {6},
	issn = {2058-8437},
	urlp = {https://www.nature.com/articles/s41578-021-00284-1},
	doip = {10.1038/s41578-021-00284-1},
	language = {en},
	number = {3},
	urldate = {2024-07-26},
	journal = {Nature Reviews Materials},
	author = {Andrei, Eva Y. and Efetov, Dmitri K. and Jarillo-Herrero, Pablo and MacDonald, Allan H. and Mak, Kin Fai and Senthil, T. and Tutuc, Emanuel and Yazdani, Ali and Young, Andrea F.},
	month = mar,
	year = {2021},
	pages = {201--206},
}

@article{geim_rise_2007,
	title = {The rise of graphene},
	volume = {6},
	copyright = {2007 Springer Nature Limited},
	issn = {1476-4660},
	urlp = {https://www.nature.com/articles/nmat1849},
	doip = {10.1038/nmat1849},
	language = {en},
	number = {3},
	urldate = {2024-07-26},
	journal = {Nature Materials},
	author = {Geim, A. K. and Novoselov, K. S.},
	month = mar,
	year = {2007},
	pages = {183--191},
}

@article{place_new_2021,
	title = {New material platform for superconducting transmon qubits with coherence times exceeding 0.3 milliseconds},
	volume = {12},
	issn = {2041-1723},
	urlp = {https://www.nature.com/articles/s41467-021-22030-5},
	doip = {10.1038/s41467-021-22030-5},
	language = {en},
	number = {1},
	urldate = {2024-07-26},
	journal = {Nature Communications},
	author = {Place, Alexander P. M. and Rodgers, Lila V. H. and Mundada, Pranav and Smitham, Basil M. and Fitzpatrick, Mattias and Leng, Zhaoqi and Premkumar, Anjali and Bryon, Jacob and Vrajitoarea, Andrei and Sussman, Sara and Cheng, Guangming and Madhavan, Trisha and Babla, Harshvardhan K. and Le, Xuan Hoang and Gang, Youqi and Jäck, Berthold and Gyenis, András and Yao, Nan and Cava, Robert J. and De Leon, Nathalie P. and Houck, Andrew A.},
	month = mar,
	year = {2021},
	pages = {1779},
}

@article{lee_two-dimensional_2019,
	title = {Two-{Dimensional} {Material} {Tunnel} {Barrier} for {Josephson} {Junctions} and {Superconducting} {Qubits}},
	volume = {19},
	issn = {1530-6984},
	urlp = {https://doip.org/10.1021/acs.nanolett.9b03886},
	doip = {10.1021/acs.nanolett.9b03886},
	number = {11},
	urldate = {2024-07-26},
	journal = {Nano Letters},
	author = {Lee, Kan-Heng and Chakram, Srivatsan and Kim, Shi En and Mujid, Fauzia and Ray, Ariana and Gao, Hui and Park, Chibeom and Zhong, Yu and Muller, David A. and Schuster, David I. and Park, Jiwoong},
	month = nov,
	year = {2019},
	pages = {8287--8293},
}

@article{google_ai_quantum_and_collaborators_hartree-fock_2020,
	title = {Hartree-{Fock} on a superconducting qubit quantum computer},
	volume = {369},
	issn = {0036-8075, 1095-9203},
	urlp = {https://www.science.org/doip/10.1126/science.abb9811},
	number = {6507},
	urldate = {2024-07-27},
	journal = {Science},
	author = {{Google AI Quantum and Collaborators} and Arute, Frank and Arya, Kunal and Babbush, Ryan and Bacon, Dave and Bardin, Joseph C. and Barends, Rami and Boixo, Sergio and Broughton, Michael and Buckley, Bob B. and Buell, David A. and Burkett, Brian and Bushnell, Nicholas and Chen, Yu and Chen, Zijun and Chiaro, Benjamin and Collins, Roberto and Courtney, William and Demura, Sean and Dunsworth, Andrew and Farhi, Edward and Fowler, Austin and Foxen, Brooks and Gidney, Craig and Giustina, Marissa and Graff, Rob and Habegger, Steve and Harrigan, Matthew P. and Ho, Alan and Hong, Sabrina and Huang, Trent and Huggins, William J. and Ioffe, Lev and Isakov, Sergei V. and Jeffrey, Evan and Jiang, Zhang and Jones, Cody and Kafri, Dvir and Kechedzhi, Kostyantyn and Kelly, Julian and Kim, Seon and Klimov, Paul V. and Korotkov, Alexander and Kostritsa, Fedor and Landhuis, David and Laptev, Pavel and Lindmark, Mike and Lucero, Erik and Martin, Orion and Martinis, John M. and McClean, Jarrod R. and McEwen, Matt and Megrant, Anthony and Mi, Xiao and Mohseni, Masoud and Mruczkiewicz, Wojciech and Mutus, Josh and Naaman, Ofer and Neeley, Matthew and Neill, Charles and Neven, Hartmut and Niu, Murphy Yuezhen and O’Brien, Thomas E. and Ostby, Eric and Petukhov, Andre and Putterman, Harald and Quintana, Chris and Roushan, Pedram and Rubin, Nicholas C. and Sank, Daniel and Satzinger, Kevin J. and Smelyanskiy, Vadim and Strain, Doug and Sung, Kevin J. and Szalay, Marco and Takeshita, Tyler Y. and Vainsencher, Amit and White, Theodore and Wiebe, Nathan and Yao, Z. Jamie and Yeh, Ping and Zalcman, Adam},
	month = aug,
	year = {2020},
	pages = {1084--1089},
}

@article{antony_miniaturizing_2021,
	title = {Miniaturizing {Transmon} {Qubits} {Using} van der {Waals} {Materials}},
	volume = {21},
	issn = {1530-6984},
	urlp = {https://doi.org/10.1021/acs.nanolett.1c04160},
	doip = {10.1021/acs.nanolett.1c04160},
	number = {23},
	journal = {Nano Letters},
	author = {Antony, Abhinandan and Gustafsson, Martin V. and Ribeill, Guilhem J. and Ware, Matthew and Rajendran, Anjaly and Govia, Luke C. G. and Ohki, Thomas A. and Taniguchi, Takashi and Watanabe, Kenji and Hone, James and Fong, Kin Chung},
	month = dec,
	year = {2021},
	pages = {10122--10126},
}

@article{ben_shalom_quantum_2016,
	title = {Quantum oscillations of the critical current and high-field superconducting proximity in ballistic graphene},
	volume = {12},
	issn = {1745-2473, 1745-2481},
	urlp = {http://www.nature.com/articles/nphys3592},
	doip = {10.1038/nphys3592},
	language = {en},
	number = {4},
	urldate = {2022-05-24},
	journal = {Nature Physics},
	author = {Ben Shalom, M. and Zhu, M. J. and Fal’ko, V. I. and Mishchenko, A. and Kretinin, A. V. and Novoselov, K. S. and Woods, C. R. and Watanabe, K. and Taniguchi, T. and Geim, A. K. and Prance, J. R.},
	month = apr,
	year = {2016},
	pages = {318--322},
}

@article{messelot_direct_2024,
	title = {Direct {Measurement} of a sin ( 2 $\phi$ ) {Current} {Phase} {Relation} in a {Graphene} {Superconducting} {Quantum} {Interference} {Device}},
	volume = {133},
	issn = {0031-9007, 1079-7114},
	urlp = {https://link.aps.org/doi/10.1103/PhysRevLett.133.106001},
	doip = {10.1103/PhysRevLett.133.106001},
	language = {en},
	number = {10},
	urldate = {2025-04-24},
	journal = {Physical Review Letters},
	author = {Messelot, Simon and Aparicio, Nicolas and De Seze, Elie and Eyraud, Eric and Coraux, Johann and Watanabe, Kenji and Taniguchi, Takashi and Renard, Julien},
	month = sep,
	year = {2024},
	pages = {106001},
}

@article{allen_spatially_2016,
	title = {Spatially resolved edge currents and guided-wave electronic states in graphene},
	volume = {12},
	issn = {1745-2473, 1745-2481},
	urlp = {https://www.nature.com/articles/nphys3534},
	doip = {10.1038/nphys3534},
	language = {en},
	number = {2},
	urlpdate = {2025-04-24},
	journal = {Nature Physics},
	author = {Allen, M. T. and Shtanko, O. and Fulga, I. C. and Akhmerov, A. R. and Watanabe, K. and Taniguchi, T. and Jarillo-Herrero, P. and Levitov, L. S. and Yacoby, A.},
	month = feb,
	year = {2016},
	pages = {128--133},
}

@article{choi_complete_2013,
	title = {Complete gate control of supercurrent in graphene p–n junctions},
	volume = {4},
	issn = {2041-1723},
	urlp = {https://www.nature.com/articles/ncomms3525},
	doip = {10.1038/ncomms3525},
	language = {en},
	number = {1},
	urlpdate = {2025-04-24},
	journal = {Nature Communications},
	author = {Choi, Jae-Hyun and Lee, Gil-Ho and Park, Sunghun and Jeong, Dongchan and Lee, Jeong-O and Sim, H.-S. and Doh, Yong-Joo and Lee, Hu-Jong},
	month = sep,
	year = {2013},
	pages = {2525},
}

@article{mizuno_ballistic-like_2013,
	title = {Ballistic-like supercurrent in suspended graphene {Josephson} weak links},
	volume = {4},
	issn = {2041-1723},
	urlp = {https://www.nature.com/articles/ncomms3716},
	doip = {10.1038/ncomms3716},
	language = {en},
	number = {1},
	urlpdate = {2025-04-24},
	journal = {Nature Communications},
	author = {Mizuno, Naomi and Nielsen, Bent and Du, Xu},
	month = nov,
	year = {2013},
	pages = {2716},
}

@article{komatsu_superconducting_2012,
	title = {Superconducting proximity effect in long superconductor/graphene/superconductor junctions: {From} specular {Andreev} reflection at zero field to the quantum {Hall} regime},
	volume = {86},
	copyright = {http://link.aps.org/licenses/aps-default-license},
	issn = {1098-0121, 1550-235X},
	shorttitle = {Superconducting proximity effect in long superconductor/graphene/superconductor junctions},
	urlp = {https://link.aps.org/doip/10.1103/PhysRevB.86.115412},
	doip = {10.1103/PhysRevB.86.115412},
	language = {en},
	number = {11},
	urlpdate = {2025-04-24},
	journal = {Physical Review B},
	author = {Komatsu, Katsuyoshi and Li, Chuan and Autier-Laurent, S. and Bouchiat, H. and Guéron, S.},
	month = sep,
	year = {2012},
	pages = {115412},
}

@article{borzenets_phase_2011,
	title = {Phase {Diffusion} in {Graphene}-{Based} {Josephson} {Junctions}},
	volume = {107},
	copyright = {http://link.aps.org/licenses/aps-default-license},
	issn = {0031-9007, 1079-7114},
	urlp = {https://link.aps.org/doip/10.1103/PhysRevLett.107.137005},
	doip = {10.1103/PhysRevLett.107.137005},
	language = {en},
	number = {13},
	urlpdate = {2025-04-24},
	journal = {Physical Review Letters},
	author = {Borzenets, I. V. and Coskun, U. C. and Jones, S. J. and Finkelstein, G.},
	month = sep,
	year = {2011},
	pages = {137005},
}

@article{girit_tunable_2009,
	title = {Tunable {Graphene} dc {Superconducting} {Quantum} {Interference} {Device}},
	volume = {9},
	issn = {1530-6984},
	urlp = {https://doip.org/10.1021/nl802765x},
	doip = {10.1021/nl802765x},
	number = {1},
	urlpdate = {2025-04-24},
	journal = {Nano Letters},
	author = {Girit, Caglar and Bouchiat, V. and Naaman, O. and Zhang, Y. and Crommie, M. F. and Zettl, A. and Siddiqi, I.},
	month = jan,
	year = {2009},
	note = {Publisher: American Chemical Society},
	pages = {198--199},
}

@article{ojeda-aristizabal_tuning_2009,
	title = {Tuning the proximity effect in a superconductor-graphene-superconductor junction},
	volume = {79},
	copyright = {http://link.aps.org/licenses/aps-default-license},
	issn = {1098-0121, 1550-235X},
	urlp = {https://link.aps.org/doip/10.1103/PhysRevB.79.165436},
	doip = {10.1103/PhysRevB.79.165436},
	language = {en},
	number = {16},
	urlpdate = {2025-04-24},
	journal = {Physical Review B},
	author = {Ojeda-Aristizabal, C. and Ferrier, M. and Guéron, S. and Bouchiat, H.},
	month = apr,
	year = {2009},
	pages = {165436},
}

@article{du_josephson_2008,
	title = {Josephson current and multiple {Andreev} reflections in graphene {SNS} junctions},
	volume = {77},
	copyright = {http://link.aps.org/licenses/aps-default-license},
	issn = {1098-0121, 1550-235X},
	urlp = {https://link.aps.org/doip/10.1103/PhysRevB.77.184507},
	doip = {10.1103/PhysRevB.77.184507},
	language = {en},
	number = {18},
	urlpdate = {2025-04-24},
	journal = {Physical Review B},
	author = {Du, Xu and Skachko, Ivan and Andrei, Eva Y.},
	month = may,
	year = {2008},
	pages = {184507},
}

@article{borzenets_phonon_2013,
	title = {Phonon {Bottleneck} in {Graphene}-{Based} {Josephson} {Junctions} at {Millikelvin} {Temperatures}},
	volume = {111},
	copyright = {http://link.aps.org/licenses/aps-default-license},
	issn = {0031-9007, 1079-7114},
	urlp = {https://link.aps.org/doip/10.1103/PhysRevLett.111.027001},
	doip = {10.1103/PhysRevLett.111.027001},
	language = {en},
	number = {2},
	urlpdate = {2025-04-24},
	journal = {Physical Review Letters},
	author = {Borzenets, I. V. and Coskun, U. C. and Mebrahtu, H. T. and Bomze, Yu. V. and Smirnov, A. I. and Finkelstein, G.},
	month = jul,
	year = {2013},
	pages = {027001},
}

@misc{huang_graphene_2024,
	title = {Graphene calorimetric single-photon detector},
	urlp = {http://arxiv.org/abs/2410.22433},
	doip = {10.48550/arXiv.2410.22433},
	urldate = {2024-12-31},
	publisher = {arXiv},
	author = {Huang, Bevin and Arnault, Ethan G. and Jung, Woochan and Fried, Caleb and Russell, B. Jordan and Watanabe, Kenji and Taniguchi, Takashi and Henriksen, Erik A. and Englund, Dirk and Lee, Gil-Ho and Fong, Kin Chun},
	month = oct,
	year = {2024},
	note = {arXiv:2410.22433 [cond-mat]},
}

@article{patel_d_2024,
	title = {d -{Mon}: {A} {Transmon} with {Strong} {Anharmonicity} {Based} on {Planar} c -{Axis} {Tunneling} {Junction} between d -{Wave} and s -{Wave} {Superconductors}},
	volume = {132},
	issn = {0031-9007, 1079-7114},
	shorttitle = {d -{Mon}},
	urlp = {https://link.aps.org/doi/10.1103/PhysRevLett.132.017002},
	doip = {10.1103/PhysRevLett.132.017002},
	language = {en},
	number = {1},
	urldate = {2025-04-24},
	journal = {Physical Review Letters},
	author = {Patel, Hrishikesh and Pathak, Vedangi and Can, Oguzhan and Potter, Andrew C. and Franz, Marcel},
	month = jan,
	year = {2024},
	pages = {017002},
}

@article{jang_engineering_2024,
	title = {Engineering {Superconducting} {Contacts} {Transparent} to a {Bipolar} {Graphene}},
	volume = {24},
	issn = {1530-6984},
	urlp = {https://doi.org/10.1021/acs.nanolett.4c03767},
	doip = {10.1021/acs.nanolett.4c03767},
	number = {49},
	urldate = {2025-04-24},
	journal = {Nano Letters},
	author = {Jang, Seong and Park, Geon-Hyoung and Park, Sein and Jeong, Hyeon-Woo and Watanabe, Kenji and Taniguchi, Takashi and Lee, Gil-Ho},
	month = dec,
	year = {2024},
	note = {Publisher: American Chemical Society},
	pages = {15582--15587},
}

@article{patil_pick-up_2024,
	title = {Pick-up and assembling of chemically sensitive van der {Waals} heterostructures using dry cryogenic exfoliation},
	volume = {14},
	issn = {2045-2322},
	urlp = {https://www.nature.com/articles/s41598-024-58935-6},
	doip = {10.1038/s41598-024-58935-6},
	language = {en},
	number = {1},
	urldate = {2025-04-24},
	journal = {Scientific Reports},
	author = {Patil, Vilas and Ghosh, Sanat and Basu, Amit and {Kuldeep} and Dutta, Achintya and Agrawal, Khushabu and Bhatia, Neha and Shah, Amit and Jangade, Digambar A. and Kulkarni, Ruta and Thamizhavel, A. and Deshmukh, Mandar M.},
	month = may,
	year = {2024},
	pages = {11097},
}

@article{Likharev_1979, title={Superconducting weak links}, volume={51}, ISSN={0034-6861}, url={http://dx.doi.org/10.1103/RevModPhys.51.101}, DOI={10.1103/revmodphys.51.101}, number={1}, journal={Reviews of Modern Physics}, publisher={American Physical Society (APS)}, author={Likharev, K. K.}, year={1979}, month=jan, pages={101–159} }

@article{nature2024abs,
  author    = {A. Fornieri and F. Giazotto},
  title     = {Andreev molecules and the switching current of Josephson junctions},
  journal   = {Nature Physics},
  year      = {2024},
  note      = {doi:10.1038/s41567-024-02400-8},
  doi       = {10.1038/s41567-024-02400-8}
}

@article{royalsociety2015, title={Theory of Andreev bound states in S-F-S junctions and S-F proximity devices}, volume={376}, ISSN={1471-2962}, url={http://dx.doi.org/10.1098/rsta.2015.0149}, DOI={10.1098/rsta.2015.0149}, number={2125}, journal={Philosophical Transactions of the Royal Society A: Mathematical, Physical and Engineering Sciences}, publisher={The Royal Society}, author={Eschrig, M.}, year={2018}, month=jun, pages={20150149} }

@article{royalsociety2018, title={Andreev bound states and their signatures}, volume={376}, ISSN={1471-2962}, url={http://dx.doi.org/10.1098/rsta.2018.0140}, DOI={10.1098/rsta.2018.0140}, number={2125}, journal={Philosophical Transactions of the Royal Society A: Mathematical, Physical and Engineering Sciences}, publisher={The Royal Society}, author={Sauls, J. A.}, year={2018}, month=jun, pages={20180140} }

@book{pearson2011superconductivity,
  author    = {Charles P. Poole and Horacio A. Farach and Richard J. Creswick and Ruslan Prozorov},
  title     = {Superconductivity},
  edition   = {3rd},
  publisher = {Academic Press (Elsevier)},
  year      = {2014},
  isbn      = {978-0-12-409509-0},
  url       = {https://shop.elsevier.com/books/superconductivity/poole/978-0-12-409509-0}
}

@article{Golubov_2004, title={The current-phase relation in Josephson junctions}, volume={76}, ISSN={1539-0756}, url={http://dx.doi.org/10.1103/RevModPhys.76.411}, DOI={10.1103/revmodphys.76.411}, number={2}, journal={Reviews of Modern Physics}, publisher={American Physical Society (APS)}, author={Golubov, A. A. and Kupriyanov, M. Yu. and Il’ichev, E.}, year={2004}, month=apr, pages={411–469} }

@article{Vignaud_2023, title={Evidence for chiral supercurrent in quantum Hall Josephson junctions}, volume={624}, ISSN={1476-4687}, url={http://dx.doi.org/10.1038/s41586-023-06764-4}, DOI={10.1038/s41586-023-06764-4}, number={7992}, journal={Nature}, publisher={Springer Science and Business Media LLC}, author={Vignaud, Hadrien and Perconte, David and Yang, Wenmin and Kousar, Bilal and Wagner, Edouard and Gay, Frédéric and Watanabe, Kenji and Taniguchi, Takashi and Courtois, Hervé and Han, Zheng and Sellier, Hermann and Sacépé, Benjamin}, year={2023}, month=nov, pages={545–550} }

@article{Barrier_2024, title={One-dimensional proximity superconductivity in the quantum Hall regime}, volume={628}, ISSN={1476-4687}, url={http://dx.doi.org/10.1038/s41586-024-07271-w}, DOI={10.1038/s41586-024-07271-w}, number={8009}, journal={Nature}, publisher={Springer Science and Business Media LLC}, author={Barrier, Julien and Kim, Minsoo and Kumar, Roshan Krishna and Xin, Na and Kumaravadivel, P. and Hague, Lee and Nguyen, E. and Berdyugin, A. I. and Moulsdale, Christian and Enaldiev, V. V. and Prance, J. R. and Koppens, F. H. L. and Gorbachev, R. V. and Watanabe, K. and Taniguchi, T. and Glazman, L. I. and Grigorieva, I. V. and Fal’ko, V. I. and Geim, A. K.}, year={2024}, month=apr, pages={741–745} }

@article{nam2017lattice,
  title={Lattice relaxation and energy band modulation in twisted bilayer graphene},
  author={Nam, Nguyen NT and Koshino, Mikito},
  journal={Physical Review B},
  volume={96},
  number={7},
  pages={075311},
  year={2017},
  publisher={APS}
}

@article{turkel2022orderly,
  title={Orderly disorder in magic-angle twisted trilayer graphene},
  author={Turkel, Simon and Swann, Joshua and Zhu, Ziyan and Christos, Maine and Watanabe, K and Taniguchi, T and Sachdev, Subir and Scheurer, Mathias S and Kaxiras, Efthimios and Dean, Cory R and others},
  journal={Science},
  volume={376},
  number={6589},
  pages={193--199},
  year={2022},
  publisher={American Association for the Advancement of Science}
}

@article{cao2018unconventional,
  title={Unconventional superconductivity in magic-angle graphene superlattices},
  author={Cao, Yuan and Fatemi, Valla and Fang, Shiang and Watanabe, Kenji and Taniguchi, Takashi and Kaxiras, Efthimios and Jarillo-Herrero, Pablo},
  journal={Nature},
  volume={556},
  number={7699},
  pages={43--50},
  year={2018},
  publisher={Nature Publishing Group UK London}
}

@article{yankowitz2019tuning,
  title={Tuning superconductivity in twisted bilayer graphene},
  author={Yankowitz, Matthew and Chen, Shaowen and Polshyn, Hryhoriy and Zhang, Yuxuan and Watanabe, K and Taniguchi, T and Graf, David and Young, Andrea F and Dean, Cory R},
  journal={Science},
  volume={363},
  number={6431},
  pages={1059--1064},
  year={2019},
  publisher={American Association for the Advancement of Science}
}

@article{park2021tunable,
  title={Tunable strongly coupled superconductivity in magic-angle twisted trilayer graphene},
  author={Park, Jeong Min and Cao, Yuan and Watanabe, Kenji and Taniguchi, Takashi and Jarillo-Herrero, Pablo},
  journal={Nature},
  volume={590},
  number={7845},
  pages={249--255},
  year={2021},
  publisher={Nature Publishing Group UK London}
}

@article{hao2021electric,
  title={Electric field--tunable superconductivity in alternating-twist magic-angle trilayer graphene},
  author={Hao, Zeyu and Zimmerman, AM and Ledwith, Patrick and Khalaf, Eslam and Najafabadi, Danial Haie and Watanabe, Kenji and Taniguchi, Takashi and Vishwanath, Ashvin and Kim, Philip},
  journal={Science},
  volume={371},
  number={6534},
  pages={1133--1138},
  year={2021},
  publisher={American Association for the Advancement of Science}
}

@article{mukherjee2025superconducting,
  title={Superconducting magic-angle twisted trilayer graphene with competing magnetic order and moir{\'e} inhomogeneities},
  author={Mukherjee, Ayshi and Layek, Surat and Sinha, Subhajit and Kundu, Ritajit and Marchawala, Alisha H and Hingankar, Mahesh and Sarkar, Joydip and Sangani, LD and Agarwal, Heena and Ghosh, Sanat and others},
  journal={Nature Materials},
  pages={1--7},
  year={2025},
  publisher={Nature Publishing Group}
}

@article{de2021gate,
  title={Gate-defined Josephson junctions in magic-angle twisted bilayer graphene},
  author={de Vries, Folkert K and Portol{\'e}s, El{\'\i}as and Zheng, Giulia and Taniguchi, Takashi and Watanabe, Kenji and Ihn, Thomas and Ensslin, Klaus and Rickhaus, Peter},
  journal={Nature Nanotechnology},
  volume={16},
  number={7},
  pages={760--763},
  year={2021},
  publisher={Nature Publishing Group UK London}
}

@article{diez2023symmetry,
  title={Symmetry-broken Josephson junctions and superconducting diodes in magic-angle twisted bilayer graphene},
  author={Diez-Merida, Jaime and D{\'\i}ez-Carl{\'o}n, Andr{\'e}s and Yang, SY and Xie, Y-M and Gao, X-J and Senior, Jorden and Watanabe, K and Taniguchi, T and Lu, X and Higginbotham, Andrew P and others},
  journal={Nature Communications},
  volume={14},
  number={1},
  pages={2396},
  year={2023},
  publisher={Nature Publishing Group UK London}
}

@article{diez2025probing,
  title={Probing the flat-band limit of the superconducting proximity effect in Twisted Bilayer Graphene Josephson junctions},
  author={Diez-Carlon, A and Diez-Merida, J and Rout, P and Sedov, D and Virtanen, P and Banerjee, S and Penttila, RPS and Altpeter, P and Watanabe, K and Taniguchi, T and others},
  journal={arXiv preprint arXiv:2502.04785},
  year={2025}
}

@article{portoles2022tunable,
  title={A tunable monolithic SQUID in twisted bilayer graphene},
  author={Portol{\'e}s, El{\'\i}as and Iwakiri, Shuichi and Zheng, Giulia and Rickhaus, Peter and Taniguchi, Takashi and Watanabe, Kenji and Ihn, Thomas and Ensslin, Klaus and de Vries, Folkert K},
  journal={Nature Nanotechnology},
  volume={17},
  number={11},
  pages={1159--1164},
  year={2022},
  publisher={Nature Publishing Group UK London}
}

@article{jha2025large,
  title={Large tunable kinetic inductance in a twisted graphene superconductor},
  author={Jha, Rounak and Endres, Martin and Watanabe, Kenji and Taniguchi, Takashi and Banerjee, Mitali and Sch{\"o}nenberger, Christian and Karnatak, Paritosh},
  journal={Physical Review Letters},
  volume={134},
  number={21},
  pages={216001},
  year={2025},
  publisher={APS}
}

@article{banerjee2025superfluid,
  title={Superfluid stiffness of twisted trilayer graphene superconductors},
  author={Banerjee, Abhishek and Hao, Zeyu and Kreidel, Mary and Ledwith, Patrick and Phinney, Isabelle and Park, Jeong Min and Zimmerman, Andrew and Wesson, Marie E and Watanabe, Kenji and Taniguchi, Takashi and others},
  journal={Nature},
  volume={638},
  number={8049},
  pages={93--98},
  year={2025},
  publisher={Nature Publishing Group UK London}
}

@article{tanaka2025superfluid,
  title={Superfluid stiffness of magic-angle twisted bilayer graphene},
  author={Tanaka, Miuko and Wang, Joel {\^I}-j and Dinh, Thao H and Rodan-Legrain, Daniel and Zaman, Sameia and Hays, Max and Almanakly, Aziza and Kannan, Bharath and Kim, David K and Niedzielski, Bethany M and others},
  journal={Nature},
  volume={638},
  number={8049},
  pages={99--105},
  year={2025},
  publisher={Nature Publishing Group UK London}
}

@article{adak2024tunable,
  title={Tunable moir{\'e} materials for probing Berry physics and topology},
  author={Adak, Pratap Chandra and Sinha, Subhajit and Agarwal, Amit and Deshmukh, Mandar M},
  journal={Nature Reviews Materials},
  volume={9},
  number={7},
  pages={481--498},
  year={2024},
  publisher={Nature Publishing Group UK London}
}

@article{wang2024two,
  title={Two-dimensional van der Waals superconductor heterostructures: Josephson junctions and beyond},
  author={Wang, Chao and Zhou, Zhenjia and Gao, Libo},
  journal={Precision Chemistry},
  volume={2},
  number={7},
  pages={273--281},
  year={2024},
  publisher={ACS Publications}
}

@article{choi2020evidence,
  title={Evidence of higher-order topology in multilayer WTe2 from Josephson coupling through anisotropic hinge states},
  author={Choi, Yong-Bin and Xie, Yingming and Chen, Chui-Zhen and Park, Jinho and Song, Su-Beom and Yoon, Jiho and Kim, Bum Joon and Taniguchi, Takashi and Watanabe, Kenji and Kim, Jonghwan and others},
  journal={Nature Materials},
  volume={19},
  number={9},
  pages={974--979},
  year={2020},
  publisher={Nature Publishing Group UK London}
}

@article{kononov2020one,
  title={One-dimensional edge transport in few-layer WTe2},
  author={Kononov, Artem and Abulizi, Gulibusitan and Qu, Kejian and Yan, Jiaqiang and Mandrus, David and Watanabe, Kenji and Taniguchi, Takashi and Schönenberger, Christian},
  journal={Nano letters},
  volume={20},
  number={6},
  pages={4228--4233},
  year={2020},
  publisher={ACS Publications}
}

@article{huang2020edge,
  title={Edge superconductivity in multilayer WTe2 Josephson junction},
  author={Huang, Ce and Narayan, Awadhesh and Zhang, Enze and Xie, Xiaoyi and Ai, Linfeng and Liu, Shanshan and Yi, Changjiang and Shi, Youguo and Sanvito, Stefano and Xiu, Faxian},
  journal={National Science Review},
  volume={7},
  number={9},
  pages={1468--1475},
  year={2020},
  publisher={Oxford University Press}
}

@article{endres2022transparent,
  title={Transparent Josephson junctions in higher-order topological insulator WTe 2 via Pd diffusion},
  author={Endres, Martin and Kononov, Artem and Stiefel, Michael and Wyss, Marcus and Arachchige, Hasitha Suriya and Yan, Jiaqiang and Mandrus, David and Watanabe, Kenji and Taniguchi, Takashi and Sch{\"o}nenberger, Christian},
  journal={Physical Review Materials},
  volume={6},
  number={8},
  pages={L081201},
  year={2022},
  publisher={APS}
}

@article{chen2024asymmetric,
  title={Asymmetric edge supercurrents in MoTe 2 Josephson junctions},
  author={Chen, Pingbo and Wang, Jinhua and Wang, Gongqi and Ye, Bicong and Zhou, Liang and Wang, Le and Wang, Jiannong and Zhang, Wenqing and Chen, Weiqiang and Mei, Jiawei and others},
  journal={Nanoscale Advances},
  volume={6},
  number={2},
  pages={690--696},
  year={2024},
  publisher={Royal Society of Chemistry}
}

@article{rodan-legrain_highly_2021,
	title = {Highly tunable junctions and non-local {Josephson} effect in magic-angle graphene tunnelling devices},
	volume = {16},
	issn = {1748-3387, 1748-3395},
	url = {https://www.nature.com/articles/s41565-021-00894-4},
	doi = {10.1038/s41565-021-00894-4},
	language = {en},
	number = {7},
	urldate = {2025-04-28},
	journal = {Nature Nanotechnology},
	author = {Rodan-Legrain, Daniel and Cao, Yuan and Park, Jeong Min and De La Barrera, Sergio C. and Randeria, Mallika T. and Watanabe, Kenji and Taniguchi, Takashi and Jarillo-Herrero, Pablo},
	month = jul,
	year = {2021},
	pages = {769--775},
}

@article{lee_twisted_2021,
	title = {Twisted van der {Waals} {Josephson} {Junction} {Based} on a {High}-{Tc} {Superconductor}},
	volume = {21},
	issn = {1530-6984},
	url = {https://doi.org/10.1021/acs.nanolett.1c03906},
	doi = {10.1021/acs.nanolett.1c03906},
	number = {24},
	urldate = {2025-04-28},
	journal = {Nano Letters},
	author = {Lee, Jongyun and Lee, Wonjun and Kim, Gi-Yeop and Choi, Yong-Bin and Park, Jinho and Jang, Seong and Gu, Genda and Choi, Si-Young and Cho, Gil Young and Lee, Gil-Ho and Lee, Hu-Jong},
	month = dec,
	year = {2021},
	note = {Publisher: American Chemical Society},
	pages = {10469--10477},
}

@article{amet_supercurrent_2016,
	title = {Supercurrent in the quantum {Hall} regime},
	volume = {352},
	copyright = {http://www.sciencemag.org/about/science-licenses-journal-article-reuse},
	issn = {0036-8075, 1095-9203},
	url = {https://www.science.org/doi/10.1126/science.aad6203},
	doi = {10.1126/science.aad6203},
	language = {en},
	number = {6288},
	urldate = {2025-06-17},
	journal = {Science},
	author = {Amet, F. and Ke, C. T. and Borzenets, I. V. and Wang, J. and Watanabe, K. and Taniguchi, T. and Deacon, R. S. and Yamamoto, M. and Bomze, Y. and Tarucha, S. and Finkelstein, G.},
	month = may,
	year = {2016},
	pages = {966--969},
}

@article{kang_van_2022,
	title = {van der {Waals} $\Pi$ {Josephson} {Junctions}},
	volume = {22},
	issn = {1530-6984},
	url = {https://doi.org/10.1021/acs.nanolett.2c01640},
	doi = {10.1021/acs.nanolett.2c01640},
	number = {13},
	urldate = {2025-06-17},
	journal = {Nano Letters},
	author = {Kang, Kaifei and Berger, Helmuth and Watanabe, Kenji and Taniguchi, Takashi and Forró, László and Shan, Jie and Mak, Kin Fai},
	month = jul,
	year = {2022},
	note = {Publisher: American Chemical Society},
	pages = {5510--5515},
}

@article{wu_field-free_2022,
	title = {The field-free {Josephson} diode in a van der {Waals} heterostructure},
	volume = {604},
	issn = {0028-0836, 1476-4687},
	url = {https://www.nature.com/articles/s41586-022-04504-8},
	doi = {10.1038/s41586-022-04504-8},
	language = {en},
	number = {7907},
	urldate = {2025-06-17},
	journal = {Nature},
	author = {Wu, Heng and Wang, Yaojia and Xu, Yuanfeng and Sivakumar, Pranava K. and Pasco, Chris and Filippozzi, Ulderico and Parkin, Stuart S. P. and Zeng, Yu-Jia and McQueen, Tyrel and Ali, Mazhar N.},
	month = apr,
	year = {2022},
	pages = {653--656},
}

@article{can_high-temperature_2021,
	title = {High-temperature topological superconductivity in twisted double-layer copper oxides},
	volume = {17},
	issn = {1745-2473, 1745-2481},
	url = {http://www.nature.com/articles/s41567-020-01142-7},
	doi = {10.1038/s41567-020-01142-7},
	language = {en},
	number = {4},
	urldate = {2022-06-08},
	journal = {Nature Physics},
	author = {Can, Oguzhan and Tummuru, Tarun and Day, Ryan P. and Elfimov, Ilya and Damascelli, Andrea and Franz, Marcel},
	month = apr,
	year = {2021},
	pages = {519--524},
}

@article{ghosh_high-temperature_2024,
	title = {High-temperature {Josephson} diode},
	issn = {1476-1122, 1476-4660},
	url = {https://www.nature.com/articles/s41563-024-01804-4},
	doi = {10.1038/s41563-024-01804-4},
	language = {en},
	urldate = {2024-02-19},
	journal = {Nature Materials},
	author = {Ghosh, Sanat and Patil, Vilas and Basu, Amit and {Kuldeep} and Dutta, Achintya and Jangade, Digambar A. and Kulkarni, Ruta and Thamizhavel, A. and Steiner, Jacob F. and Von Oppen, Felix and Deshmukh, Mandar M.},
	month = feb,
	year = {2024},
}

@article{zhao_time-reversal_2023,
	title = {Time-reversal symmetry breaking superconductivity between twisted cuprate superconductors},
	volume = {382},
	issn = {0036-8075, 1095-9203},
	url = {https://www.science.org/doi/10.1126/science.abl8371},
	doi = {10.1126/science.abl8371},
	language = {en},
	number = {6677},
	urldate = {2024-07-20},
	journal = {Science},
	author = {Zhao, S. Y. Frank and Cui, Xiaomeng and Volkov, Pavel A. and Yoo, Hyobin and Lee, Sangmin and Gardener, Jules A. and Akey, Austin J. and Engelke, Rebecca and Ronen, Yuval and Zhong, Ruidan and Gu, Genda and Plugge, Stephan and Tummuru, Tarun and Kim, Miyoung and Franz, Marcel and Pixley, Jedediah H. and Poccia, Nicola and Kim, Philip},
	month = dec,
	year = {2023},
	pages = {1422--1427},
}

@article{farrar_superconducting_2021,
	title = {Superconducting {Quantum} {Interference} in {Twisted} van der {Waals} {Heterostructures}},
	volume = {21},
	copyright = {https://creativecommons.org/licenses/by/4.0/},
	issn = {1530-6984, 1530-6992},
	url = {https://pubs.acs.org/doi/10.1021/acs.nanolett.1c00152},
	doi = {10.1021/acs.nanolett.1c00152},
	language = {en},
	number = {16},
	urldate = {2024-07-20},
	journal = {Nano Letters},
	author = {Farrar, Liam S. and Nevill, Aimee and Lim, Zhen Jieh and Balakrishnan, Geetha and Dale, Sara and Bending, Simon J.},
	month = aug,
	year = {2021},
	pages = {6725--6731},
}

@article{lee_encapsulating_2023,
	title = {Encapsulating {High}‐{Temperature} {Superconducting} {Twisted} van der {Waals} {Heterostructures} {Blocks} {Detrimental} {Effects} of {Disorder}},
	issn = {0935-9648, 1521-4095},
	url = {https://onlinelibrary.wiley.com/doi/10.1002/adma.202209135},
	doi = {10.1002/adma.202209135},
	language = {en},
	urldate = {2024-07-20},
	journal = {Advanced Materials},
	author = {Lee, Yejin and Martini, Mickey and Confalone, Tommaso and Shokri, Sanaz and Saggau, Christian N. and Wolf, Daniel and Gu, Genda and Watanabe, Kenji and Taniguchi, Takashi and Montemurro, Domenico and Vinokur, Valerii M. and Nielsch, Kornelius and Poccia, Nicola},
	month = jan,
	year = {2023},
	pages = {2209135},
}

@article{endres_currentphase_2023,
	title = {Current–{Phase} {Relation} of a {WTe} $_{\textrm{2}}$ {Josephson} {Junction}},
	volume = {23},
	copyright = {https://creativecommons.org/licenses/by/4.0/},
	issn = {1530-6984, 1530-6992},
	url = {https://pubs.acs.org/doi/10.1021/acs.nanolett.3c01416},
	doi = {10.1021/acs.nanolett.3c01416},
	language = {en},
	number = {10},
	urldate = {2024-10-25},
	journal = {Nano Letters},
	author = {Endres, Martin and Kononov, Artem and Arachchige, Hasitha Suriya and Yan, Jiaqiang and Mandrus, David and Watanabe, Kenji and Taniguchi, Takashi and Schönenberger, Christian},
	month = may,
	year = {2023},
	pages = {4654--4659},
}

@article{fong_graphene_2022,
	title = {Graphene amplifier reaches the quantum limit},
	volume = {17},
	issn = {1748-3387, 1748-3395},
	url = {https://www.nature.com/articles/s41565-022-01239-5},
	doi = {10.1038/s41565-022-01239-5},
	language = {en},
	number = {11},
	urldate = {2025-06-17},
	journal = {Nature Nanotechnology},
	author = {Fong, Kin Chung},
	month = nov,
	year = {2022},
	pages = {1128--1129},
}

@article{zhu_presence_2021,
	title = {Presence of s -{Wave} {Pairing} in {Josephson} {Junctions} {Made} of {Twisted} {Ultrathin} {Bi} 2 {Sr} 2 {CaCu} 2 {O} 8 + x {Flakes}},
	volume = {11},
	issn = {2160-3308},
	url = {https://link.aps.org/doi/10.1103/PhysRevX.11.031011},
	doi = {10.1103/PhysRevX.11.031011},
	language = {en},
	number = {3},
	urldate = {2024-07-20},
	journal = {Physical Review X},
	author = {Zhu, Yuying and Liao, Menghan and Zhang, Qinghua and Xie, Hong-Yi and Meng, Fanqi and Liu, Yaowu and Bai, Zhonghua and Ji, Shuaihua and Zhang, Jin and Jiang, Kaili and Zhong, Ruidan and Schneeloch, John and Gu, Genda and Gu, Lin and Ma, Xucun and Zhang, Ding and Xue, Qi-Kun},
	month = jul,
	year = {2021},
	pages = {031011},
}

@article{butseraen_gate-tunable_2022,
	title = {A gate-tunable graphene {Josephson} parametric amplifier},
	volume = {17},
	issn = {1748-3387, 1748-3395},
	urlp = {https://www.nature.com/articles/s41565-022-01235-9},
	doip = {10.1038/s41565-022-01235-9},
	language = {en},
	number = {11},
	urldate = {2025-06-17},
	journal = {Nature Nanotechnology},
	author = {Butseraen, Guilliam and Ranadive, Arpit and Aparicio, Nicolas and Rafsanjani Amin, Kazi and Juyal, Abhishek and Esposito, Martina and Watanabe, Kenji and Taniguchi, Takashi and Roch, Nicolas and Lefloch, François and Renard, Julien},
	month = nov,
	year = {2022},
	pages = {1153--1158},
}

@article{huang_manipulation_2024,
	title = {Manipulation of {Cooper}‐{Pair} {Tunneling} via {Domain} {Structure} in a van der {Waals} {Ferromagnetic} {Josephson} {Junction}},
	volume = {36},
	issn = {0935-9648, 1521-4095},
	url = {https://onlinelibrary.wiley.com/doi/10.1002/adma.202314190},
	doi = {10.1002/adma.202314190},
	language = {en},
	number = {33},
	urldate = {2025-06-24},
	journal = {Advanced Materials},
	author = {Huang, Junwei and Li, Zeya and Bi, Xiangyu and Tang, Ming and Qiu, Caiyu and Qin, Feng and Yuan, Hongtao},
	month = aug,
	year = {2024},
	pages = {2314190},
}

@article{yabuki_supercurrent_2016,
	title = {Supercurrent in van der {Waals} {Josephson} junction},
	volume = {7},
	issn = {2041-1723},
	url = {https://www.nature.com/articles/ncomms10616},
	doi = {10.1038/ncomms10616},
	language = {en},
	number = {1},
	urldate = {2025-06-24},
	journal = {Nature Communications},
	author = {Yabuki, Naoto and Moriya, Rai and Arai, Miho and Sata, Yohta and Morikawa, Sei and Masubuchi, Satoru and Machida, Tomoki},
	month = feb,
	year = {2016},
	pages = {10616},
}

@article{ai_van_2021,
	title = {Van der {Waals} ferromagnetic {Josephson} junctions},
	volume = {12},
	issn = {2041-1723},
	url = {https://www.nature.com/articles/s41467-021-26946-w},
	doi = {10.1038/s41467-021-26946-w},
	language = {en},
	number = {1},
	urldate = {2025-06-24},
	journal = {Nature Communications},
	author = {Ai, Linfeng and Zhang, Enze and Yang, Jinshan and Xie, Xiaoyi and Yang, Yunkun and Jia, Zehao and Zhang, Yuda and Liu, Shanshan and Li, Zihan and Leng, Pengliang and Cao, Xiangyu and Sun, Xingdan and Zhang, Tongyao and Kou, Xufeng and Han, Zheng and Xiu, Faxian and Dong, Shaoming},
	month = nov,
	year = {2021},
	pages = {6580},
}

@article{idzuchi_unconventional_2021,
	title = {Unconventional supercurrent phase in {Ising} superconductor {Josephson} junction with atomically thin magnetic insulator},
	volume = {12},
	issn = {2041-1723},
	url = {https://www.nature.com/articles/s41467-021-25608-1},
	doi = {10.1038/s41467-021-25608-1},
	language = {en},
	number = {1},
	urldate = {2025-06-24},
	journal = {Nature Communications},
	author = {Idzuchi, H. and Pientka, F. and Huang, K.-F. and Harada, K. and Gül, Ö. and Shin, Y. J. and Nguyen, L. T. and Jo, N. H. and Shindo, D. and Cava, R. J. and Canfield, P. C. and Kim, P.},
	month = sep,
	year = {2021},
	pages = {5332},
}

@misc{kudriashov_non-reciprocal_2025,
	title = {Non-{Reciprocal} {Current}-{Phase} {Relation} and {Superconducting} {Diode} {Effect} in {Topological}-{Insulator}-{Based} {Josephson} {Junctions}},
	url = {http://arxiv.org/abs/2502.08527},
	doi = {10.48550/arXiv.2502.08527},
	language = {en},
	urldate = {2025-06-24},
	publisher = {arXiv},
	author = {Kudriashov, A. and Zhou, X. and Hovhannisyan, R. A. and Frolov, A. and Elesin, L. and Wang, Y. and Zharkova, E. V. and Taniguchi, T. and Watanabe, K. and Yashina, L. A. and Liu, Z. and Zhou, Xin and Novoselov, K. S. and Bandurin, D. A.},
	month = feb,
	year = {2025},
	note = {arXiv:2502.08527 [cond-mat]},
}

@article{Bretheau_2017, title={Tunnelling spectroscopy of Andreev states in graphene}, volume={13}, ISSN={1745-2481}, url={http://dx.doi.org/10.1038/nphys4110}, DOI={10.1038/nphys4110}, number={8}, journal={Nature Physics}, publisher={Springer Science and Business Media LLC}, author={Bretheau, Landry and Wang, Joel I-Jan and Pisoni, Riccardo and Watanabe, Kenji and Taniguchi, Takashi and Jarillo-Herrero, Pablo}, year={2017}, month=may, pages={756–760} }

@article{Heide_2021, title={Electronic Coherence and Coherent Dephasing in the Optical Control of Electrons in Graphene}, volume={21}, ISSN={1530-6992}, url={http://dx.doi.org/10.1021/acs.nanolett.1c02538}, DOI={10.1021/acs.nanolett.1c02538}, number={22}, journal={Nano Letters}, publisher={American Chemical Society (ACS)}, author={Heide, Christian and Eckstein, Timo and Boolakee, Tobias and Gerner, Constanze and Weber, Heiko B. and Franco, Ignacio and Hommelhoff, Peter}, year={2021}, month=nov, pages={9403–9409} }

@article{Merboldt_2025, title={Observation of Floquet states in graphene}, ISSN={1745-2481}, url={http://dx.doi.org/10.1038/s41567-025-02889-7}, DOI={10.1038/s41567-025-02889-7}, journal={Nature Physics}, publisher={Springer Science and Business Media LLC}, author={Merboldt, Marco and Schüler, Michael and Schmitt, David and Bange, Jan Philipp and Bennecke, Wiebke and Gadge, Karun and Pierz, Klaus and Schumacher, Hans Werner and Momeni, Davood and Steil, Daniel and Manmana, Salvatore R. and Sentef, Michael A. and Reutzel, Marcel and Mathias, Stefan}, year={2025}, month=may }

@article{Park_2022, title={Steady Floquet–Andreev states in graphene Josephson junctions}, volume={603}, ISSN={1476-4687}, url={http://dx.doi.org/10.1038/s41586-021-04364-8}, DOI={10.1038/s41586-021-04364-8}, number={7901}, journal={Nature}, publisher={Springer Science and Business Media LLC}, author={Park, Sein and Lee, Wonjun and Jang, Seong and Choi, Yong-Bin and Park, Jinho and Jung, Woochan and Watanabe, Kenji and Taniguchi, Takashi and Cho, Gil Young and Lee, Gil-Ho}, year={2022}, month=mar, pages={421–426} }

@article{Haxell_2023, title={Microwave-induced conductance replicas in hybrid Josephson junctions without Floquet—Andreev states}, volume={14}, ISSN={2041-1723}, url={http://dx.doi.org/10.1038/s41467-023-42357-5}, DOI={10.1038/s41467-023-42357-5}, number={1}, journal={Nature Communications}, publisher={Springer Science and Business Media LLC}, author={Haxell, Daniel Z. and Coraiola, Marco and Sabonis, Deividas and Hinderling, Manuel and ten Kate, Sofieke C. and Cheah, Erik and Krizek, Filip and Schott, Rüdiger and Wegscheider, Werner and Belzig, Wolfgang and Cuevas, Juan Carlos and Nichele, Fabrizio}, year={2023}, month=oct }

@article{wallraff_strong_2004,
	title = {Strong coupling of a single photon to a superconducting qubit using circuit quantum electrodynamics},
	volume = {431},
	copyright = {http://www.springer.com/tdm},
	issn = {0028-0836, 1476-4687},
	urlp = {https://www.nature.com/articles/nature02851},
	doip = {10.1038/nature02851},
	language = {en},
	number = {7005},
	urldate = {2025-06-30},
	journal = {Nature},
	author = {Wallraff, A. and Schuster, D. I. and Blais, A. and Frunzio, L. and Huang, R.- S. and Majer, J. and Kumar, S. and Girvin, S. M. and Schoelkopf, R. J.},
	month = sep,
	year = {2004},
	pages = {162--167},
}

@article{zhu_edge_2017,
	title = {Edge currents shunt the insulating bulk in gapped graphene},
	volume = {8},
	issn = {2041-1723},
	urlp = {https://www.nature.com/articles/ncomms14552},
	doip = {10.1038/ncomms14552},
	language = {en},
	number = {1},
	urldate = {2025-07-02},
	journal = {Nature Communications},
	author = {Zhu, M. J. and Kretinin, A. V. and Thompson, M. D. and Bandurin, D. A. and Hu, S. and Yu, G. L. and Birkbeck, J. and Mishchenko, A. and Vera-Marun, I. J. and Watanabe, K. and Taniguchi, T. and Polini, M. and Prance, J. R. and Novoselov, K. S. and Geim, A. K. and Ben Shalom, M.},
	month = feb,
	year = {2017},
	pages = {14552},
}

@article{simon_theory_2018,
	title = {Theory of the {Josephson} {Junction} {Laser}},
	volume = {121},
	issn = {0031-9007, 1079-7114},
	url = {https://link.aps.org/doi/10.1103/PhysRevLett.121.027004},
	doi = {10.1103/PhysRevLett.121.027004},
	language = {en},
	number = {2},
	urldate = {2025-07-03},
	journal = {Physical Review Letters},
	author = {Simon, Steven H. and Cooper, Nigel R.},
	month = jul,
	year = {2018},
	pages = {027004},
}

@article{cassidy_demonstration_2017,
	title = {Demonstration of an ac {Josephson} junction laser},
	volume = {355},
	copyright = {http://www.sciencemag.org/about/science-licenses-journal-article-reuse},
	issn = {0036-8075, 1095-9203},
	url = {https://www.science.org/doi/10.1126/science.aah6640},
	doi = {10.1126/science.aah6640},
	language = {en},
	number = {6328},
	urldate = {2025-07-03},
	journal = {Science},
	author = {Cassidy, M. C. and Bruno, A. and Rubbert, S. and Irfan, M. and Kammhuber, J. and Schouten, R. N. and Akhmerov, A. R. and Kouwenhoven, L. P.},
	month = mar,
	year = {2017},
	pages = {939--942},
}

@article{diez-merida_symmetry-broken_2023,
	title = {Symmetry-broken {Josephson} junctions and superconducting diodes in magic-angle twisted bilayer graphene},
	volume = {14},
	issn = {2041-1723},
	url = {https://www.nature.com/articles/s41467-023-38005-7},
	doi = {10.1038/s41467-023-38005-7},
	language = {en},
	number = {1},
	urldate = {2025-07-03},
	journal = {Nature Communications},
	author = {Díez-Mérida, J. and Díez-Carlón, A. and Yang, S. Y. and Xie, Y.-M. and Gao, X.-J. and Senior, J. and Watanabe, K. and Taniguchi, T. and Lu, X. and Higginbotham, A. P. and Law, K. T. and Efetov, Dmitri K.},
	month = apr,
	year = {2023},
	pages = {2396},
}

@article{brosco_superconducting_2024,
	title = {Superconducting {Qubit} {Based} on {Twisted} {Cuprate} {Van} der {Waals} {Heterostructures}},
	volume = {132},
	issn = {0031-9007, 1079-7114},
	url = {https://link.aps.org/doi/10.1103/PhysRevLett.132.017003},
	doi = {10.1103/PhysRevLett.132.017003},
	language = {en},
	number = {1},
	urldate = {2025-07-03},
	journal = {Physical Review Letters},
	author = {Brosco, Valentina and Serpico, Giuseppe and Vinokur, Valerii and Poccia, Nicola and Vool, Uri},
	month = jan,
	year = {2024},
	pages = {017003},
}

@article{zhao_merged-element_2020,
	title = {Merged-{Element} {Transmon}},
	volume = {14},
	issn = {2331-7019},
	url = {https://link.aps.org/doi/10.1103/PhysRevApplied.14.064006},
	doi = {10.1103/PhysRevApplied.14.064006},
	language = {en},
	number = {6},
	urldate = {2025-07-03},
	journal = {Physical Review Applied},
	author = {Zhao, R. and Park, S. and Zhao, T. and Bal, M. and McRae, C.R.H. and Long, J. and Pappas, D.P.},
	month = dec,
	year = {2020},
	pages = {064006},
}

@article{balgley_crystalline_2025,
	title = {Crystalline superconductor-semiconductor {Josephson} junctions for compact superconducting qubits},
	volume = {24},
	issn = {2331-7019},
	url = {https://link.aps.org/doi/10.1103/3ssz-jjt6},
	doi = {10.1103/3ssz-jjt6},
	language = {en},
	number = {3},
	urldate = {2025-10-31},
	journal = {Physical Review Applied},
	author = {Balgley, Jesse and Park, Jinho and Chu, Xuanjing and Arnault, Ethan G. and Gustafsson, Martin V. and Ranzani, Leonardo and Holbrook, Madisen and He, Yangchen and Watanabe, Kenji and Taniguchi, Takashi and Rhodes, Daniel and Perebeinos, Vasili and Hone, James and Fong, Kin Chung},
	month = sep,
	year = {2025},
	pages = {034016},
}

@article{gunyho_single-shot_2024,
	title = {Single-shot readout of a superconducting qubit using a thermal detector},
	volume = {7},
	issn = {2520-1131},
	url = {https://www.nature.com/articles/s41928-024-01147-7},
	doi = {10.1038/s41928-024-01147-7},
	language = {en},
	number = {4},
	urldate = {2025-07-03},
	journal = {Nature Electronics},
	author = {Gunyhó, András M. and Kundu, Suman and Ma, Jian and Liu, Wei and Niemelä, Sakari and Catto, Giacomo and Vadimov, Vasilii and Vesterinen, Visa and Singh, Priyank and Chen, Qiming and Möttönen, Mikko},
	month = apr,
	year = {2024},
	pages = {288--298},
}

@article{koch_charge-insensitive_2007,
	title = {Charge-insensitive qubit design derived from the {Cooper} pair box},
	volume = {76},
	copyright = {http://link.aps.org/licenses/aps-default-license},
	issn = {1050-2947, 1094-1622},
	url = {https://link.aps.org/doi/10.1103/PhysRevA.76.042319},
	doi = {10.1103/PhysRevA.76.042319},
	language = {en},
	number = {4},
	urldate = {2025-07-03},
	journal = {Physical Review A},
	author = {Koch, Jens and Yu, Terri M. and Gambetta, Jay and Houck, A. A. and Schuster, D. I. and Majer, J. and Blais, Alexandre and Devoret, M. H. and Girvin, S. M. and Schoelkopf, R. J.},
	month = oct,
	year = {2007},
	pages = {042319},
}

@article{walter_mkid_2020,
	title = {The {MKID} {Exoplanet} {Camera} for {Subaru} {SCExAO}},
	volume = {132},
	copyright = {http://iopscience.iop.org/info/page/text-and-data-mining},
	issn = {1538-3873},
	urlp = {https://iopscience.iop.org/article/10.1088/1538-3873/abc60f},
	doip = {10.1088/1538-3873/abc60f},
	language = {en},
	number = {1018},
	urldate = {2025-07-07},
	journal = {Publications of the Astronomical Society of the Pacific},
	author = {Walter, Alexander B. and Fruitwala, Neelay and Steiger, Sarah and Bailey, John I. and Zobrist, Nicholas and Swimmer, Noah and Lipartito, Isabel and Smith, Jennifer Pearl and Meeker, Seth R. and Bockstiegel, Clint and Coiffard, Gregoire and Dodkins, Rupert and Szypryt, Paul and Davis, Kristina K. and Daal, Miguel and Bumble, Bruce and Collura, Giulia and Guyon, Olivier and Lozi, Julien and Vievard, Sebastien and Jovanovic, Nemanja and Martinache, Frantz and Currie, Thayne and Mazin, Benjamin A.},
	month = nov,
	year = {2020},
	notep = {Publisher: IOP Publishing},
	pages = {125005},
}

@article{oripov_superconducting_2023,
	title = {A superconducting nanowire single-photon camera with 400,000 pixels},
	volume = {622},
	copyright = {https://www.springernature.com/gp/researchers/text-and-data-mining},
	issn = {0028-0836, 1476-4687},
	urlp = {https://www.nature.com/articles/s41586-023-06550-2},
	doip = {10.1038/s41586-023-06550-2},
	language = {en},
	number = {7984},
	urldate = {2025-07-07},
	journal = {Nature},
	author = {Oripov, B. G. and Rampini, D. S. and Allmaras, J. and Shaw, M. D. and Nam, S. W. and Korzh, B. and McCaughan, A. N.},
	month = oct,
	year = {2023},
	notep = {Publisher: Springer Science and Business Media LLC},
	pages = {730--734},

}

@article{mounet_two-dimensional_2018,
	title = {Two-dimensional materials from high-throughput computational exfoliation of experimentally known compounds},
	volume = {13},
	copyright = {2018 © The Author (s) 2017, under exclusive licence to Macmillan Publishers Limited, part of Springer Nature},
	issn = {1748-3395},
	url = {https://www.nature.com/articles/s41565-017-0035-5},
	doi = {10.1038/s41565-017-0035-5},
	language = {en},
	number = {3},
	urldate = {2025-07-14},
	journal = {Nature Nanotechnology},
	author = {Mounet, Nicolas and Gibertini, Marco and Schwaller, Philippe and Campi, Davide and Merkys, Andrius and Marrazzo, Antimo and Sohier, Thibault and Castelli, Ivano Eligio and Cepellotti, Andrea and Pizzi, Giovanni and Marzari, Nicola},
	month = mar,
	year = {2018},
	note = {Publisher: Nature Publishing Group},
	pages = {246--252},
}

@article{PhysRevResearch.7.023273,
  title = {Flat band Josephson junctions with quantum metric},
  author = {Li, Zhong C. F. and Deng, Yuxuan and Chen, Shuai A. and Efetov, Dmitri K. and Law, K. T.},
  journal = {Phys. Rev. Res.},
  volume = {7},
  issue = {2},
  pages = {023273},
  numpages = {7},
  year = {2025},
  month = {Jun},
  publisher = {American Physical Society},
  doi = {10.1103/PhysRevResearch.7.023273},
  url = {https://link.aps.org/doi/10.1103/PhysRevResearch.7.023273}
}

@article{virtanen2024superconducting,
  title = {Superconducting junctions with flat bands},
  author = {Virtanen, P. and Penttil\"a, R. P. S. and T\"orm\"a, P. and D\'{\i}ez-Carl\'on, A. and Efetov, D. K. and Heikkil\"a, T. T.},
  journal = {Phys. Rev. B},
  volume = {112},
  issue = {10},
  pages = {L100502},
  numpages = {6},
  year = {2025},
  month = {Sep},
  publisher = {American Physical Society},
  doi = {10.1103/7f1y-kjyl},
  url = {https://link.aps.org/doi/10.1103/7f1y-kjyl}
}

@article{nadeem_superconducting_2023,
	title = {The superconducting diode effect},
	volume = {5},
	copyright = {2023 Springer Nature Limited},
	issn = {2522-5820},
	url = {https://www.nature.com/articles/s42254-023-00632-w},
	doi = {10.1038/s42254-023-00632-w},
	language = {en},
	number = {10},
	urldate = {2025-07-19},
	journal = {Nature Reviews Physics},
	author = {Nadeem, Muhammad and Fuhrer, Michael S. and Wang, Xiaolin},
	month = oct,
	year = {2023},
	note = {Publisher: Nature Publishing Group},
	pages = {558--577},
}

@article{ingla-aynes_efficient_2025,
	title = {Efficient superconducting diodes and rectifiers for quantum circuitry},
	volume = {8},
	copyright = {2025 The Author(s), under exclusive licence to Springer Nature Limited},
	issn = {2520-1131},
	url = {https://www.nature.com/articles/s41928-025-01375-5},
	doi = {10.1038/s41928-025-01375-5},
	language = {en},
	number = {5},
	urldate = {2025-07-19},
	journal = {Nature Electronics},
	author = {Ingla-Aynés, Josep and Hou, Yasen and Wang, Sarah and Chu, En-De and Mukhanov, Oleg A. and Wei, Peng and Moodera, Jagadeesh S.},
	month = may,
	year = {2025},
	note = {Publisher: Nature Publishing Group},
	pages = {411--416},
}

@article{castellani_superconducting_2025,
	title = {A superconducting full-wave bridge rectifier},
	volume = {8},
	copyright = {2025 The Author(s), under exclusive licence to Springer Nature Limited},
	issn = {2520-1131},
	url = {https://www.nature.com/articles/s41928-025-01376-4},
	doi = {10.1038/s41928-025-01376-4},
	language = {en},
	number = {5},
	urldate = {2025-07-19},
	journal = {Nature Electronics},
	author = {Castellani, Matteo and Medeiros, Owen and Buzzi, Alessandro and Foster, Reed A. and Colangelo, Marco and Berggren, Karl K.},
	month = may,
	year = {2025},
	note = {Publisher: Nature Publishing Group},
	pages = {417--425},
}

@article{mamin_merged-element_2021,
	title = {Merged-{Element} {Transmons}: {Design} and {Qubit} {Performance}},
	volume = {16},
	copyright = {https://link.aps.org/licenses/aps-default-license},
	issn = {2331-7019},
	shorttitle = {Merged-{Element} {Transmons}},
	url = {https://link.aps.org/doi/10.1103/PhysRevApplied.16.024023},
	doi = {10.1103/physrevapplied.16.024023},
	language = {en},
	number = {2},
	urldate = {2025-07-22},
	journal = {Physical Review Applied},
	author = {Mamin, H.J. and Huang, E. and Carnevale, S. and Rettner, C.T. and Arellano, N. and Sherwood, M.H. and Kurter, C. and Trimm, B. and Sandberg, M. and Shelby, R.M. and Mueed, M.A. and Madon, B.A. and Pushp, A. and Steffen, M. and Rugar, D.},
	month = aug,
	year = {2021},
	note = {Publisher: American Physical Society (APS)},
}

@article{Ambegaokar_1963, title={Tunneling Between Superconductors}, volume={10}, ISSN={0031-9007}, url={http://dx.doi.org/10.1103/PhysRevLett.10.486}, DOI={10.1103/physrevlett.10.486}, number={11}, journal={Physical Review Letters}, publisher={American Physical Society (APS)}, author={Ambegaokar, Vinay and Baratoff, Alexis}, year={1963}, month=jun, pages={486–489} }

@article{pandey_ballistic_2021,
	title = {Ballistic {Graphene} {Cooper} {Pair} {Splitter}},
	volume = {126},
	copyright = {https://link.aps.org/licenses/aps-default-license},
	issn = {0031-9007, 1079-7114},
	url = {https://link.aps.org/doi/10.1103/PhysRevLett.126.147701},
	doi = {10.1103/physrevlett.126.147701},
	language = {en},
	number = {14},
	urldate = {2025-07-23},
	journal = {Physical Review Letters},
	author = {Pandey, P. and Danneau, R. and Beckmann, D.},
	month = apr,
	year = {2021},
	note = {Publisher: American Physical Society (APS)},
}

@article{Stern_2013, title={Topological Quantum Computation—From Basic Concepts to First Experiments}, volume={339}, ISSN={1095-9203}, url={http://dx.doi.org/10.1126/science.1231473}, DOI={10.1126/science.1231473}, number={6124}, journal={Science}, publisher={American Association for the Advancement of Science (AAAS)}, author={Stern, Ady and Lindner, Netanel H.}, year={2013}, month=mar, pages={1179–1184} }

@article{Alicea_2015, title={Designer non-Abelian anyon platforms: from Majorana to Fibonacci}, volume={T164}, ISSN={1402-4896}, url={http://dx.doi.org/10.1088/0031-8949/2015/T164/014006}, DOI={10.1088/0031-8949/2015/t164/014006}, journal={Physica Scripta}, publisher={IOP Publishing}, author={Alicea, Jason and Stern, Ady}, year={2015}, month=aug, pages={014006} }

@article{Lindner_2012, title={Fractionalizing Majorana Fermions: Non-Abelian Statistics on the Edges of Abelian Quantum Hall States}, volume={2}, ISSN={2160-3308}, url={http://dx.doi.org/10.1103/PhysRevX.2.041002}, DOI={10.1103/physrevx.2.041002}, number={4}, journal={Physical Review X}, publisher={American Physical Society (APS)}, author={Lindner, Netanel H. and Berg, Erez and Refael, Gil and Stern, Ady}, year={2012}, month=oct }

@article{Clarke_2013, title={Exotic non-Abelian anyons from conventional fractional quantum Hall states}, volume={4}, ISSN={2041-1723}, url={http://dx.doi.org/10.1038/ncomms2340}, DOI={10.1038/ncomms2340}, number={1}, journal={Nature Communications}, publisher={Springer Science and Business Media LLC}, author={Clarke, David J. and Alicea, Jason and Shtengel, Kirill}, year={2013}, month=jan }

@article{Mong_2014, title={Universal Topological Quantum Computation from a Superconductor-Abelian Quantum Hall Heterostructure}, volume={4}, ISSN={2160-3308}, url={http://dx.doi.org/10.1103/PhysRevX.4.011036}, DOI={10.1103/physrevx.4.011036}, number={1}, journal={Physical Review X}, publisher={American Physical Society (APS)}, author={Mong, Roger S. K. and Clarke, David J. and Alicea, Jason and Lindner, Netanel H. and Fendley, Paul and Nayak, Chetan and Oreg, Yuval and Stern, Ady and Berg, Erez and Shtengel, Kirill and Fisher, Matthew P. A.}, year={2014}, month=mar }

@article{San_Jose_2015, title={Majorana Zero Modes in Graphene}, volume={5}, ISSN={2160-3308}, url={http://dx.doi.org/10.1103/PhysRevX.5.041042}, DOI={10.1103/physrevx.5.041042}, number={4}, journal={Physical Review X}, publisher={American Physical Society (APS)}, author={San-Jose, P. and Lado, J. L. and Aguado, R. and Guinea, F. and Fernández-Rossier, J.}, year={2015}, month=dec }

@article{sarkar_kerr_2025,
	title = {Kerr non-linearity enhances the response of a graphene {Josephson} bolometer},
	volume = {16},
	issn = {2041-1723},
	url = {https://www.nature.com/articles/s41467-025-62480-9},
	doi = {10.1038/s41467-025-62480-9},
	language = {en},
	number = {1},
	urldate = {2025-08-04},
	journal = {Nature Communications},
	author = {Sarkar, Joydip and Maji, Krishnendu and Sunamudi, Abhishek and Agarwal, Heena and Samanta, Priyanka and Bhattacharjee, Anirban and Rajkhowa, Rishiraj and Patankar, Meghan P. and Watanabe, Kenji and Taniguchi, Takashi and Deshmukh, Mandar M.},
	month = jul,
	year = {2025},
	pages = {7043},
}

@article{park_controllable_2024,
	title = {Controllable {Andreev} {Bound} {States} in {Bilayer} {Graphene} {Josephson} {Junctions} from {Short} to {Long} {Junction} {Limits}},
	volume = {132},
	issn = {0031-9007, 1079-7114},
	url = {https://link.aps.org/doi/10.1103/PhysRevLett.132.226301},
	doi = {10.1103/PhysRevLett.132.226301},
	language = {en},
	number = {22},
	urldate = {2025-08-05},
	journal = {Physical Review Letters},
	author = {Park, Geon-Hyoung and Lee, Wonjun and Park, Sein and Watanabe, Kenji and Taniguchi, Takashi and Cho, Gil Young and Lee, Gil-Ho},
	month = may,
	year = {2024},
	pages = {226301},
}

@article{jung_tunneling_2025,
	title = {Tunneling spectroscopy of {Andreev} bands in multiterminal graphene-based {Josephson} junctions},
	language = {en},
        volume = {11},
	pages = {eads0342},
	journal = {Science Advances},
	author = {Jung, Woochan and Jin, Seyoung and Park, Sein and Taniguchi, Takashi and Cho, Gil Young and Lee, Ho},
	year = {2025},
}

@article{PhysRevLett.122.247001,
  title = {Sign-Reversing Hall Effect in Atomically Thin High-Temperature ${\mathrm{Bi}}_{2.1}{\mathrm{Sr}}_{1.9}{\mathrm{CaCu}}_{2.0}{\mathrm{O}}_{8+\ensuremath{\delta}}$ Superconductors},
  author = {Zhao, S. Y. Frank and Poccia, Nicola and Panetta, Margaret G. and Yu, Cyndia and Johnson, Jedediah W. and Yoo, Hyobin and Zhong, Ruidan and Gu, G. D. and Watanabe, Kenji and Taniguchi, Takashi and Postolova, Svetlana V. and Vinokur, Valerii M. and Kim, Philip},
  journal = {Phys. Rev. Lett.},
  volume = {122},
  issue = {24},
  pages = {247001},
  numpages = {6},
  year = {2019},
  month = {Jun},
  publisher = {American Physical Society},
  doi = {10.1103/PhysRevLett.122.247001},
  url = {https://link.aps.org/doi/10.1103/PhysRevLett.122.247001}
}

@misc{bland_2d_2025,
	title = {{2D} transmons with lifetimes and coherence times exceeding 1 millisecond},
	url = {http://arxiv.org/abs/2503.14798},
	doi = {10.48550/arXiv.2503.14798},
	urldate = {2025-08-05},
	publisher = {arXiv},
	author = {Bland, Matthew P. and Bahrami, Faranak and Martinez, Jeronimo G. C. and Prestegaard, Paal H. and Smitham, Basil M. and Joshi, Atharv and Hedrick, Elizabeth and Pakpour-Tabrizi, Alex and Kumar, Shashwat and Jindal, Apoorv and Chang, Ray D. and Yang, Ambrose and Cheng, Guangming and Yao, Nan and Cava, Robert J. and Leon, Nathalie P. de and Houck, Andrew A.},
	month = mar,
	year = {2025},
	note = {arXiv:2503.14798 [quant-ph]},
}

@article{nayak_non-abelian_2008,
	title = {Non-{Abelian} anyons and topological quantum computation},
	volume = {80},
	copyright = {http://link.aps.org/licenses/aps-default-license},
	issn = {0034-6861, 1539-0756},
	url = {https://link.aps.org/doi/10.1103/RevModPhys.80.1083},
	doi = {10.1103/RevModPhys.80.1083},
	language = {en},
	number = {3},
	urldate = {2025-08-11},
	journal = {Reviews of Modern Physics},
	author = {Nayak, Chetan and Simon, Steven H. and Stern, Ady and Freedman, Michael and Das Sarma, Sankar},
	month = sep,
	year = {2008},
	pages = {1083--1159},
}

@article{google_quantum_ai_and_collaborators_non-abelian_2023,
	title = {Non-{Abelian} braiding of graph vertices in a superconducting processor},
	volume = {618},
	issn = {0028-0836, 1476-4687},
	url = {https://www.nature.com/articles/s41586-023-05954-4},
	doi = {10.1038/s41586-023-05954-4},
	language = {en},
	number = {7964},
	urldate = {2025-08-11},
	journal = {Nature},
	author = {{Google Quantum AI and Collaborators} and Andersen, T. I. and Lensky, Y. D. and Kechedzhi, K. and Drozdov, I. K. and Bengtsson, A. and Hong, S. and Morvan, A. and Mi, X. and Opremcak, A. and Acharya, R. and Allen, R. and Ansmann, M. and Arute, F. and Arya, K. and Asfaw, A. and Atalaya, J. and Babbush, R. and Bacon, D. and Bardin, J. C. and Bortoli, G. and Bourassa, A. and Bovaird, J. and Brill, L. and Broughton, M. and Buckley, B. B. and Buell, D. A. and Burger, T. and Burkett, B. and Bushnell, N. and Chen, Z. and Chiaro, B. and Chik, D. and Chou, C. and Cogan, J. and Collins, R. and Conner, P. and Courtney, W. and Crook, A. L. and Curtin, B. and Debroy, D. M. and Del Toro Barba, A. and Demura, S. and Dunsworth, A. and Eppens, D. and Erickson, C. and Faoro, L. and Farhi, E. and Fatemi, R. and Ferreira, V. S. and Burgos, L. F. and Forati, E. and Fowler, A. G. and Foxen, B. and Giang, W. and Gidney, C. and Gilboa, D. and Giustina, M. and Gosula, R. and Dau, A. G. and Gross, J. A. and Habegger, S. and Hamilton, M. C. and Hansen, M. and Harrigan, M. P. and Harrington, S. D. and Heu, P. and Hilton, J. and Hoffmann, M. R. and Huang, T. and Huff, A. and Huggins, W. J. and Ioffe, L. B. and Isakov, S. V. and Iveland, J. and Jeffrey, E. and Jiang, Z. and Jones, C. and Juhas, P. and Kafri, D. and Khattar, T. and Khezri, M. and KieferovÃ¡, M. and Kim, S. and Kitaev, A. and Klimov, P. V. and Klots, A. R. and Korotkov, A. N. and Kostritsa, F. and Kreikebaum, J. M. and Landhuis, D. and Laptev, P. and Lau, K.-M. and Laws, L. and Lee, J. and Lee, K. W. and Lester, B. J. and Lill, A. T. and Liu, W. and Locharla, A. and Lucero, E. and Malone, F. D. and Martin, O. and McClean, J. R. and McCourt, T. and McEwen, M. and Miao, K. C. and Mieszala, A. and Mohseni, M. and Montazeri, S. and Mount, E. and Movassagh, R. and Mruczkiewicz, W. and Naaman, O. and Neeley, M. and Neill, C. and Nersisyan, A. and Newman, M. and Ng, J. H. and Nguyen, A. and Nguyen, M. and Niu, M. Y. and Oâ€™Brien, T. E. and Omonije, S. and Petukhov, A. and Potter, R. and Pryadko, L. P. and Quintana, C. and Rocque, C. and Rubin, N. C. and Saei, N. and Sank, D. and Sankaragomathi, K. and Satzinger, K. J. and Schurkus, H. F. and Schuster, C. and Shearn, M. J. and Shorter, A. and Shutty, N. and Shvarts, V. and Skruzny, J. and Smith, W. C. and Somma, R. and Sterling, G. and Strain, D. and Szalay, M. and Torres, A. and Vidal, G. and Villalonga, B. and Heidweiller, C. V. and White, T. and Woo, B. W. K. and Xing, C. and Yao, Z. J. and Yeh, P. and Yoo, J. and Young, G. and Zalcman, A. and Zhang, Y. and Zhu, N. and Zobrist, N. and Neven, H. and Boixo, S. and Megrant, A. and Kelly, J. and Chen, Y. and Smelyanskiy, V. and Kim, E.-A. and Aleiner, I. and Roushan, P.},
	month = jun,
	year = {2023},
	pages = {264--269},
}

@article{breunig_opportunities_2021,
	title = {Opportunities in topological insulator devices},
	volume = {4},
	issn = {2522-5820},
	url = {https://www.nature.com/articles/s42254-021-00402-6},
	doi = {10.1038/s42254-021-00402-6},
	language = {en},
	number = {3},
	urldate = {2025-08-11},
	journal = {Nature Reviews Physics},
	author = {Breunig, Oliver and Ando, Yoichi},
	month = dec,
	year = {2021},
	pages = {184--193},
}

@article{hasan_colloquium_2010,
	title = {\textit{{Colloquium}} : {Topological} insulators},
	volume = {82},
	copyright = {http://link.aps.org/licenses/aps-default-license},
	issn = {0034-6861, 1539-0756},
	shorttitle = {\textit{{Colloquium}}},
	url = {https://link.aps.org/doi/10.1103/RevModPhys.82.3045},
	doi = {10.1103/RevModPhys.82.3045},
	language = {en},
	number = {4},
	urldate = {2025-08-11},
	journal = {Reviews of Modern Physics},
	author = {Hasan, M. Z. and Kane, C. L.},
	month = nov,
	year = {2010},
	pages = {3045--3067},
}

@article{qi_topological_2011,
	title = {Topological insulators and superconductors},
	volume = {83},
	copyright = {http://link.aps.org/licenses/aps-default-license},
	issn = {0034-6861, 1539-0756},
	url = {https://link.aps.org/doi/10.1103/RevModPhys.83.1057},
	doi = {10.1103/RevModPhys.83.1057},
	language = {en},
	number = {4},
	urldate = {2025-08-11},
	journal = {Reviews of Modern Physics},
	author = {Qi, Xiao-Liang and Zhang, Shou-Cheng},
	month = oct,
	year = {2011},
	pages = {1057--1110},
}

@article{das_sarma_topological_2006,
	title = {Topological quantum computation},
	volume = {59},
	issn = {0031-9228, 1945-0699},
	url = {https://pubs.aip.org/physicstoday/article/59/7/32/1040851/Topological-quantum-computationThe-search-for-a},
	doi = {10.1063/1.2337825},
	language = {en},
	number = {7},
	urldate = {2025-08-11},
	journal = {Physics Today},
	author = {Das Sarma, Sankar and Freedman, Michael and Nayak, Chetan},
	month = jul,
	year = {2006},
	pages = {32--38},
}

@article{dvir_planar_2021,
	title = {Planar graphene- {NbSe} 2 {Josephson} junctions in a parallel magnetic field},
	volume = {103},
	issn = {2469-9950, 2469-9969},
	url = {https://link.aps.org/doi/10.1103/PhysRevB.103.115401},
	doi = {10.1103/PhysRevB.103.115401},
	language = {en},
	number = {11},
	urldate = {2025-08-11},
	journal = {Physical Review B},
	author = {Dvir, Tom and Zalic, Ayelet and Fyhn, Eirik Holm and Amundsen, Morten and Taniguchi, Takashi and Watanabe, Kenji and Linder, Jacob and Steinberg, Hadar},
	month = mar,
	year = {2021},
	pages = {115401},
}

@article{kim_strong_2017,
	title = {Strong {Proximity} {Josephson} {Coupling} in {Vertically} {Stacked} {NbSe}$_{\textrm{2}}$ –{Graphene}–{NbSe}$_{\textrm{2}}$ van der {Waals} {Junctions}},
	volume = {17},
	issn = {1530-6984, 1530-6992},
	url = {https://pubs.acs.org/doi/10.1021/acs.nanolett.7b02707},
	doi = {10.1021/acs.nanolett.7b02707},
	language = {en},
	number = {10},
	urldate = {2025-08-11},
	journal = {Nano Letters},
	author = {Kim, Minsoo and Park, Geon-Hyoung and Lee, Jongyun and Lee, Jae Hyeong and Park, Jinho and Lee, Hyunwoo and Lee, Gil-Ho and Lee, Hu-Jong},
	month = oct,
	year = {2017},
	pages = {6125--6130},
}

@article{halbertal_imaging_2017,
	title = {Imaging resonant dissipation from individual atomic defects in graphene},
	volume = {358},
	issn = {0036-8075, 1095-9203},
	urlp = {https://www.science.org/doi/10.1126/science.aan0877},
	doi = {10.1126/science.aan0877},
	language = {en},
	number = {6368},
	urldate = {2024-08-29},
	journal = {Science},
	author = {Halbertal, Dorri and Ben Shalom, Moshe and Uri, Aviram and Bagani, Kousik and Meltzer, Alexander Y. and Marcus, Ido and Myasoedov, Yuri and Birkbeck, John and Levitov, Leonid S. and Geim, Andre K. and Zeldov, Eli},
	month = dec,
	year = {2017},
	pages = {1303--1306},
}

@article{huang_study_2022,
	title = {The study of contact properties in edge-contacted graphene–aluminum {Josephson} junctions},
	volume = {121},
	issn = {0003-6951, 1077-3118},
	url = {https://pubs.aip.org/apl/article/121/24/243503/2834860/The-study-of-contact-properties-in-edge-contacted},
	doi = {10.1063/5.0135034},
	language = {en},
	number = {24},
	urldate = {2025-08-11},
	journal = {Applied Physics Letters},
	author = {Huang, Zhujun and Lotfizadeh, Neda and Elfeky, Bassel H. and Kisslinger, Kim and Cuniberto, Edoardo and Yu, Peng and Hatefipour, Mehdi and Taniguchi, Takashi and Watanabe, Kenji and Shabani, Javad and Shahrjerdi, Davood},
	month = dec,
	year = {2022},
	pages = {243503},
}

@article{Riwar_2016, title={Multi-terminal Josephson junctions as topological matter}, volume={7}, ISSN={2041-1723}, url={http://dx.doi.org/10.1038/ncomms11167}, DOI={10.1038/ncomms11167}, number={1}, journal={Nature Communications}, publisher={Springer Science and Business Media LLC}, author={Riwar, Roman-Pascal and Houzet, Manuel and Meyer, Julia S. and Nazarov, Yuli V.}, year={2016}, month=apr }

@article{Draelos_2019, title={Supercurrent Flow in Multiterminal Graphene Josephson Junctions}, volume={19}, ISSN={1530-6992}, url={http://dx.doi.org/10.1021/acs.nanolett.8b04330}, DOI={10.1021/acs.nanolett.8b04330}, number={2}, journal={Nano Letters}, publisher={American Chemical Society (ACS)}, author={Draelos, Anne W. and Wei, Ming-Tso and Seredinski, Andrew and Li, Hengming and Mehta, Yash and Watanabe, Kenji and Taniguchi, Takashi and Borzenets, Ivan V. and Amet, François and Finkelstein, Gleb}, year={2019}, month=jan, pages={1039–1043} }

@article{Arnault_2021, title={Multiterminal Inverse AC Josephson Effect}, volume={21}, ISSN={1530-6992}, url={http://dx.doi.org/10.1021/acs.nanolett.1c03474}, DOI={10.1021/acs.nanolett.1c03474}, number={22}, journal={Nano Letters}, publisher={American Chemical Society (ACS)}, author={Arnault, Ethan G. and Larson, Trevyn F. Q. and Seredinski, Andrew and Zhao, Lingfei and Idris, Sara and McConnell, Aeron and Watanabe, Kenji and Taniguchi, Takashi and Borzenets, Ivan and Amet, François and Finkelstein, Gleb}, year={2021}, month=nov, pages={9668–9674} }

@article{Calado_2015, title={Ballistic Josephson junctions in edge-contacted graphene}, volume={10}, ISSN={1748-3395}, url={http://dx.doi.org/10.1038/nnano.2015.156}, DOI={10.1038/nnano.2015.156}, number={9}, journal={Nature Nanotechnology}, publisher={Springer Science and Business Media LLC}, author={Calado, V. E. and Goswami, S. and Nanda, G. and Diez, M. and Akhmerov, A. R. and Watanabe, K. and Taniguchi, T. and Klapwijk, T. M. and Vandersypen, L. M. K.}, year={2015}, month=jul, pages={761–764} }

@article{Coraiola_2023, title={Phase-engineering the Andreev band structure of a three-terminal Josephson junction}, volume={14}, ISSN={2041-1723}, url={http://dx.doi.org/10.1038/s41467-023-42356-6}, DOI={10.1038/s41467-023-42356-6}, number={1}, journal={Nature Communications}, publisher={Springer Science and Business Media LLC}, author={Coraiola, Marco and Haxell, Daniel Z. and Sabonis, Deividas and Weisbrich, Hannes and Svetogorov, Aleksandr E. and Hinderling, Manuel and ten Kate, Sofieke C. and Cheah, Erik and Krizek, Filip and Schott, Rüdiger and Wegscheider, Werner and Cuevas, Juan Carlos and Belzig, Wolfgang and Nichele, Fabrizio}, year={2023}, month=oct }

@article{Jung_2025, title={Tunneling spectroscopy of Andreev bands in multiterminal graphene-based Josephson junctions}, volume={11}, ISSN={2375-2548}, url={http://dx.doi.org/10.1126/sciadv.ads0342}, DOI={10.1126/sciadv.ads0342}, number={21}, journal={Science Advances}, publisher={American Association for the Advancement of Science (AAAS)}, author={Jung, Woochan and Jin, Seyoung and Park, Sein and Shin, Seung-Hyun and Watanabe, Kenji and Taniguchi, Takashi and Cho, Gil Young and Lee, Gil-Ho}, year={2025}, month=may }

@article{Prosko_2024, title={Flux-tunable Josephson effect in a four-terminal junction}, volume={110}, ISSN={2469-9969}, url={http://dx.doi.org/10.1103/PhysRevB.110.064518}, DOI={10.1103/physrevb.110.064518}, number={6}, journal={Physical Review B}, publisher={American Physical Society (APS)}, author={Prosko, Christian G. and Huisman, Wietze D. and Kulesh, Ivan and Xiao, Di and Thomas, Candice and Manfra, Michael J. and Goswami, Srijit}, year={2024}, month=aug }

@article{clarke_quantum_1988,
	title = {Quantum {Mechanics} of a {Macroscopic} {Variable}: {The} {Phase} {Difference} of a {Josephson} {Junction}},
	volume = {239},
	issn = {0036-8075, 1095-9203},
	shorttitle = {Quantum {Mechanics} of a {Macroscopic} {Variable}},
	url = {https://www.science.org/doi/10.1126/science.239.4843.992},
	doi = {10.1126/science.239.4843.992},
	language = {en},
	number = {4843},
	urldate = {2025-08-11},
	journal = {Science},
	author = {Clarke, John and Cleland, Andrew N. and Devoret, Michel H. and Esteve, Daniel and Martinis, John M.},
	month = feb,
	year = {1988},
	pages = {992--997},
}

@article{krantz_quantum_2019,
	title = {A quantum engineer's guide to superconducting qubits},
	volume = {6},
	issn = {1931-9401},
	url = {https://pubs.aip.org/apr/article/6/2/021318/570326/A-quantum-engineer-s-guide-to-superconducting},
	doi = {10.1063/1.5089550},
	language = {en},
	number = {2},
	urldate = {2025-08-11},
	journal = {Applied Physics Reviews},
	author = {Krantz, P. and Kjaergaard, M. and Yan, F. and Orlando, T. P. and Gustavsson, S. and Oliver, W. D.},
	month = jun,
	year = {2019},
	pages = {021318},
}

@article{blais_circuit_2021,
	title = {Circuit quantum electrodynamics},
	volume = {93},
	issn = {0034-6861, 1539-0756},
	url = {https://link.aps.org/doi/10.1103/RevModPhys.93.025005},
	doi = {10.1103/RevModPhys.93.025005},
	language = {en},
	number = {2},
	urldate = {2025-08-11},
	journal = {Reviews of Modern Physics},
	author = {Blais, Alexandre and Grimsmo, Arne L. and Girvin, S. M. and Wallraff, Andreas},
	month = may,
	year = {2021},
	pages = {025005},
}

@article{van_damme_advanced_2024,
	title = {Advanced {CMOS} manufacturing of superconducting qubits on 300 mm wafers},
	volume = {634},
	issn = {0028-0836, 1476-4687},
	url = {https://www.nature.com/articles/s41586-024-07941-9},
	doi = {10.1038/s41586-024-07941-9},
	language = {en},
	number = {8032},
	urldate = {2025-08-11},
	journal = {Nature},
	author = {Van Damme, J. and Massar, S. and Acharya, R. and Ivanov, Ts. and Perez Lozano, D. and Canvel, Y. and Demarets, M. and Vangoidsenhoven, D. and Hermans, Y. and Lai, J. G. and Vadiraj, A. M. and Mongillo, M. and Wan, D. and De Boeck, J. and Potočnik, A. and De Greve, K.},
	month = oct,
	year = {2024},
	pages = {74--79},
}

@article{rudner_band_2020,
	title = {Band structure engineering and non-equilibrium dynamics in {Floquet} topological insulators},
	volume = {2},
	issn = {2522-5820},
	url = {https://www.nature.com/articles/s42254-020-0170-z},
	doi = {10.1038/s42254-020-0170-z},
	language = {en},
	number = {5},
	urldate = {2025-08-11},
	journal = {Nature Reviews Physics},
	author = {Rudner, Mark S. and Lindner, Netanel H.},
	month = may,
	year = {2020},
	pages = {229--244},
}

@article{pillet_andreev_2010,
	title = {Andreev bound states in supercurrent-carrying carbon nanotubes revealed},
	volume = {6},
	issn = {1745-2473, 1745-2481},
	url = {https://www.nature.com/articles/nphys1811},
	doi = {10.1038/nphys1811},
	language = {en},
	number = {12},
	urldate = {2025-08-11},
	journal = {Nature Physics},
	author = {Pillet, J-D. and Quay, C. H. L. and Morfin, P. and Bena, C. and Yeyati, A. Levy and Joyez, P.},
	month = dec,
	year = {2010},
	pages = {965--969},
}

@article{xia_superconductivity_2025,
	title = {Superconductivity in twisted bilayer {WSe2}},
	volume = {637},
	issn = {0028-0836, 1476-4687},
	url = {https://www.nature.com/articles/s41586-024-08116-2},
	doi = {10.1038/s41586-024-08116-2},
	language = {en},
	number = {8047},
	urldate = {2025-08-11},
	journal = {Nature},
	author = {Xia, Yiyu and Han, Zhongdong and Watanabe, Kenji and Taniguchi, Takashi and Shan, Jie and Mak, Kin Fai},
	month = jan,
	year = {2025},
	pages = {833--838},
}

@article{guo_superconductivity_2025,
	title = {Superconductivity in 5.0° twisted bilayer {WSe2}},
	volume = {637},
	issn = {0028-0836, 1476-4687},
	url = {https://www.nature.com/articles/s41586-024-08381-1},
	doi = {10.1038/s41586-024-08381-1},
	language = {en},
	number = {8047},
	urldate = {2025-08-11},
	journal = {Nature},
	author = {Guo, Yinjie and Pack, Jordan and Swann, Joshua and Holtzman, Luke and Cothrine, Matthew and Watanabe, Kenji and Taniguchi, Takashi and Mandrus, David G. and Barmak, Katayun and Hone, James and Millis, Andrew J. and Pasupathy, Abhay and Dean, Cory R.},
	month = jan,
	year = {2025},
	pages = {839--845},
}

@article{buzdin_proximity_2005,
	title = {Proximity effects in superconductor-ferromagnet heterostructures},
	volume = {77},
	copyright = {http://link.aps.org/licenses/aps-default-license},
	issn = {0034-6861, 1539-0756},
	url = {https://link.aps.org/doi/10.1103/RevModPhys.77.935},
	doi = {10.1103/RevModPhys.77.935},
	language = {en},
	number = {3},
	urldate = {2025-08-11},
	journal = {Reviews of Modern Physics},
	author = {Buzdin, A. I.},
	month = sep,
	year = {2005},
	pages = {935--976},
}

@article{gao_reaction_1987,
	title = {Reaction and intermixing at metal-superconductor interfaces: {Fe}/{YBa2Cu3O6}.9},
	volume = {51},
	issn = {0003-6951, 1077-3118},
	shorttitle = {Reaction and intermixing at metal-superconductor interfaces},
	url = {https://pubs.aip.org/apl/article/51/13/1032/52781/Reaction-and-intermixing-at-metal-superconductor},
	doi = {10.1063/1.98769},
	language = {en},
	number = {13},
	urldate = {2025-08-11},
	journal = {Applied Physics Letters},
	author = {Gao, Y. and Wagener, T. J. and Weaver, J. H. and Flandermeyer, B. and Capone, D. W.},
	month = sep,
	year = {1987},
	pages = {1032--1034},
}

@article{degen_quantum_2017,
	title = {Quantum sensing},
	volume = {89},
	copyright = {http://link.aps.org/licenses/aps-default-license},
	issn = {0034-6861, 1539-0756},
	url = {http://link.aps.org/doi/10.1103/RevModPhys.89.035002},
	doi = {10.1103/RevModPhys.89.035002},
	language = {en},
	number = {3},
	urldate = {2025-08-11},
	journal = {Reviews of Modern Physics},
	author = {Degen, C. L. and Reinhard, F. and Cappellaro, P.},
	month = jul,
	year = {2017},
	pages = {035002},
}

@article{noauthor_qubits_2021,
	title = {Qubits meet materials science},
	volume = {6},
	issn = {2058-8437},
	url = {https://www.nature.com/articles/s41578-021-00378-w},
	doi = {10.1038/s41578-021-00378-w},
	language = {en},
	number = {10},
	urldate = {2025-08-11},
	journal = {Nature Reviews Materials},
	month = oct,
	year = {2021},
	pages = {869--869},
}

@article{siddiqi_engineering_2021,
	title = {Engineering high-coherence superconducting qubits},
	volume = {6},
	issn = {2058-8437},
	url = {https://www.nature.com/articles/s41578-021-00370-4},
	doi = {10.1038/s41578-021-00370-4},
	language = {en},
	number = {10},
	urldate = {2025-08-11},
	journal = {Nature Reviews Materials},
	author = {Siddiqi, Irfan},
	month = sep,
	year = {2021},
	pages = {875--891},
}

@article{drung_integrated_1996,
	title = {Integrated {YBa$_2$Cu$_3$O$_{7-x}$} magnetometer for biomagnetic measurements},
	volume = {68},
	issn = {0003-6951, 1077-3118},
	url = {https://pubs.aip.org/apl/article/68/10/1421/777387/Integrated-YBa2Cu3O7-x-magnetometer-for},
	doi = {10.1063/1.116100},
	language = {en},
	number = {10},
	urldate = {2025-08-11},
	journal = {Applied Physics Letters},
	author = {Drung, D. and Ludwig, F. and Müller, W. and Steinhoff, U. and Trahms, L. and Koch, H. and Shen, Y. Q. and Jensen, M. B. and Vase, P. and Holst, T. and Freltoft, T. and Curio, G.},
	month = mar,
	year = {1996},
	pages = {1421--1423},
}

@article{vasyukov_scanning_2013,
	title = {A scanning superconducting quantum interference device with single electron spin sensitivity},
	volume = {8},
	copyright = {http://www.springer.com/tdm},
	issn = {1748-3387, 1748-3395},
	url = {https://www.nature.com/articles/nnano.2013.169},
	doi = {10.1038/nnano.2013.169},
	language = {en},
	number = {9},
	urldate = {2025-08-11},
	journal = {Nature Nanotechnology},
	author = {Vasyukov, Denis and Anahory, Yonathan and Embon, Lior and Halbertal, Dorri and Cuppens, Jo and Neeman, Lior and Finkler, Amit and Segev, Yehonathan and Myasoedov, Yuri and Rappaport, Michael L. and Huber, Martin E. and Zeldov, Eli},
	month = sep,
	year = {2013},
	pages = {639--644},
}

@misc{kim_josephson_2025,
	title = {Josephson {Junctions} in the {Age} of {Quantum} {Discovery}},
	url = {http://arxiv.org/abs/2505.12724},
	doi = {10.48550/arXiv.2505.12724},
	urldate = {2025-08-11},
	publisher = {arXiv},
	author = {Kim, Hyunseong and Jang, Gyunghyun and Jin, Seungwon and Shin, Dongbin and Shin, Hyeon-Jin and Luo, Jie and Siddiqi, Irfan and Kim, Yosep and Yoon, Hoon Hahn and Nguyen, Long B.},
	month = may,
	year = {2025},
	note = {arXiv:2505.12724 [quant-ph]},
}

@article{wang_two-dimensional_2024,
	title = {Two-{Dimensional} van der {Waals} {Superconductor} {Heterostructures}: {Josephson} {Junctions} and {Beyond}},
	volume = {2},
	copyright = {https://creativecommons.org/licenses/by-nc-nd/4.0/},
	issn = {2771-9316, 2771-9316},
	shorttitle = {Two-{Dimensional} van der {Waals} {Superconductor} {Heterostructures}},
	url = {https://pubs.acs.org/doi/10.1021/prechem.3c00126},
	doi = {10.1021/prechem.3c00126},
	language = {en},
	number = {7},
	urldate = {2025-08-11},
	journal = {Precision Chemistry},
	author = {Wang, Chao and Zhou, Zhenjia and Gao, Libo},
	month = jul,
	year = {2024},
	pages = {273--281},
}

@article{tian_josephson_2021,
	title = {A {Josephson} junction with h-{BN} tunnel barrier: observation of low critical current noise},
	volume = {33},
	issn = {0953-8984, 1361-648X},
	shorttitle = {A {Josephson} junction with h-{BN} tunnel barrier},
	url = {https://iopscience.iop.org/article/10.1088/1361-648X/ac268f},
	doi = {10.1088/1361-648X/ac268f},
	language = {en},
	number = {49},
	urldate = {2025-08-11},
	journal = {Journal of Physics: Condensed Matter},
	author = {Tian, Jifa and Jauregui, Luis A and Wilen, C D and Rigosi, Albert F and Newell, David B and McDermott, R and Chen, Yong P},
	month = dec,
	year = {2021},
	pages = {495301},
}

@article{schmidt_anisotropic_2025,
	title = {Anisotropic supercurrent suppression and revivals in a graphene-based {Josephson} junction under in-plane magnetic fields},
	volume = {111},
	issn = {2469-9950, 2469-9969},
	url = {https://link.aps.org/doi/10.1103/PhysRevB.111.245301},
	doi = {10.1103/PhysRevB.111.245301},
	language = {en},
	number = {24},
	urldate = {2025-08-11},
	journal = {Physical Review B},
	author = {Schmidt, Philipp and Stanojević, Katarina and Watanabe, Kenji and Taniguchi, Takashi and Beschoten, Bernd and Mourik, Vincent and Stampfer, Christoph},
	month = jun,
	year = {2025},
	pages = {245301},
}

@article{kalantre_anomalous_2020,
	title = {Anomalous phase dynamics of driven graphene {Josephson} junctions},
	volume = {2},
	issn = {2643-1564},
	url = {https://link.aps.org/doi/10.1103/PhysRevResearch.2.023093},
	doi = {10.1103/PhysRevResearch.2.023093},
	language = {en},
	number = {2},
	urldate = {2025-08-11},
	journal = {Physical Review Research},
	author = {Kalantre, S. S. and Yu, F. and Wei, M. T. and Watanabe, K. and Taniguchi, T. and Hernandez-Rivera, M. and Amet, F. and Williams, J. R.},
	month = apr,
	year = {2020},
	pages = {023093},
}

@article{larson_zero_2020,
	title = {Zero {Crossing} {Steps} and {Anomalous} {Shapiro} {Maps} in {Graphene} {Josephson} {Junctions}},
	volume = {20},
	copyright = {https://doi.org/10.15223/policy-029},
	issn = {1530-6984, 1530-6992},
	url = {https://pubs.acs.org/doi/10.1021/acs.nanolett.0c01598},
	doi = {10.1021/acs.nanolett.0c01598},
	language = {en},
	number = {10},
	urldate = {2025-08-11},
	journal = {Nano Letters},
	author = {Larson, Trevyn F. Q. and Zhao, Lingfei and Arnault, Ethan G. and Wei, Ming-Tso and Seredinski, Andrew and Li, Henming and Watanabe, Kenji and Taniguchi, Takashi and Amet, François and Finkelstein, Gleb},
	month = oct,
	year = {2020},
	pages = {6998--7003},
}

@article{huang_observation_2023,
	title = {Observation of half-integer {Shapiro} steps in graphene {Josephson} junctions},
	volume = {122},
	issn = {0003-6951, 1077-3118},
	url = {https://pubs.aip.org/apl/article/122/26/262601/2900233/Observation-of-half-integer-Shapiro-steps-in},
	doi = {10.1063/5.0153646},
	language = {en},
	number = {26},
	urldate = {2025-08-11},
	journal = {Applied Physics Letters},
	author = {Huang, Zhujun and Elfeky, Bassel Heiba and Taniguchi, Takashi and Watanabe, Kenji and Shabani, Javad and Shahrjerdi, Davood},
	month = jun,
	year = {2023},
	pages = {262601},
}

@article{
doi:10.1073/pnas.1108174108,
author = {Rafi Bistritzer  and Allan H. MacDonald },
title = {Moiré bands in twisted double-layer graphene},
journal = {Proceedings of the National Academy of Sciences},
volume = {108},
number = {30},
pages = {12233-12237},
year = {2011},
doi = {10.1073/pnas.1108174108},
URL = {https://www.pnas.org/doi/abs/10.1073/pnas.1108174108},
eprint = {https://www.pnas.org/doi/pdf/10.1073/pnas.1108174108}}

@article{confalone_cuprate_2025,
	title = {Cuprate {Twistronics} for {Quantum} {Hardware}},
	issn = {2511-9044, 2511-9044},
	url = {https://advanced.onlinelibrary.wiley.com/doi/10.1002/qute.202500203},
	doi = {10.1002/qute.202500203},
	language = {en},
	urldate = {2025-08-12},
	journal = {Advanced Quantum Technologies},
	author = {Confalone, Tommaso and Lo Sardo, Flavia and Lee, Yejin and Shokri, Sanaz and Serpico, Giuseppe and Coppo, Alessandro and Chirolli, Luca and Vinokur, Valerii M. and Brosco, Valentina and Vool, Uri and Montemurro, Domenico and Tafuri, Francesco and Nielsch, Kornelius and Haider, Golam and Poccia, Nicola},
	month = may,
	year = {2025},
	pages = {2500203},
}

@book{weinstock_squid_1996,
	address = {Dordrecht},
	title = {{SQUID} {Sensors}: {Fundamentals}, {Fabrication} and {Applications}},
	copyright = {http://www.springer.com/tdm},
	isbn = {978-94-010-6393-7 978-94-011-5674-5},
	shorttitle = {{SQUID} {Sensors}},
	url = {http://link.springer.com/10.1007/978-94-011-5674-5},
	language = {en},
	urldate = {2025-08-12},
	publisher = {Springer Netherlands},
	editor = {Weinstock, Harold},
	year = {1996},
	doi = {10.1007/978-94-011-5674-5},
}

@article{sun_broadband_2025,
	title = {Broadband merged-element {Josephson} parametric amplifier},
	volume = {126},
	issn = {0003-6951, 1077-3118},
	url = {https://pubs.aip.org/apl/article/126/24/244005/3350084/Broadband-merged-element-Josephson-parametric},
	doi = {10.1063/5.0259358},
	language = {en},
	number = {24},
	urldate = {2025-08-12},
	journal = {Applied Physics Letters},
	author = {Sun, Yuting and Li, Xianke and Wang, Qingyu and Bai, Tairong and Liao, Xudong and Lan, Dong and Zhao, Jie and Yu, Yang},
	month = jun,
	year = {2025},
	pages = {244005},
}

@article{sliwa_reconfigurable_2015,
	title = {Reconfigurable {Josephson} {Circulator}/{Directional} {Amplifier}},
	volume = {5},
	copyright = {http://creativecommons.org/licenses/by/3.0/},
	issn = {2160-3308},
	url = {https://link.aps.org/doi/10.1103/PhysRevX.5.041020},
	doi = {10.1103/PhysRevX.5.041020},
	language = {en},
	number = {4},
	urldate = {2025-08-12},
	journal = {Physical Review X},
	author = {Sliwa, K. M. and Hatridge, M. and Narla, A. and Shankar, S. and Frunzio, L. and Schoelkopf, R. J. and Devoret, M. H.},
	month = nov,
	year = {2015},
	pages = {041020},
}

@article{navarathna_passive_2023,
	title = {Passive {Superconducting} {Circulator} on a {Chip}},
	volume = {130},
	issn = {0031-9007, 1079-7114},
	url = {https://link.aps.org/doi/10.1103/PhysRevLett.130.037001},
	doi = {10.1103/PhysRevLett.130.037001},
	language = {en},
	number = {3},
	urldate = {2025-08-12},
	journal = {Physical Review Letters},
	author = {Navarathna, Rohit and Le, Dat Thanh and Hamann, Andrés Rosario and Nguyen, Hien Duy and Stace, Thomas M. and Fedorov, Arkady},
	month = jan,
	year = {2023},
	pages = {037001},
}

@article{deutsch_harnessing_2020,
	title = {Harnessing the {Power} of the {Second} {Quantum} {Revolution}},
	volume = {1},
	issn = {2691-3399},
	url = {https://link.aps.org/doi/10.1103/PRXQuantum.1.020101},
	doi = {10.1103/PRXQuantum.1.020101},
	language = {en},
	number = {2},
	urldate = {2025-08-15},
	journal = {PRX Quantum},
	author = {Deutsch, Ivan H.},
	month = nov,
	year = {2020},
	pages = {020101},
}

\clearpage
\noindent\textbf{Acknowledgements}\\
The authors thank S. Gueron, E. Henriksen, P. Karnatak, V. Ranjan, C. Usha, S. Sinha, and P. Adak for providing helpful feedback on the article. 
M.M.D. acknowledges support from the Air Force Office of Scientific Research under award number FA2386-23-1-4031 and FA2386-25-1-4027, J.C. Bose Fellowship JCB/2022/000045 from the Department of Science and Technology of India, and Department of Atomic Energy of the Government of India under award number 12-R\&D-TFR-5.10-0100.\\

\noindent\textbf{Author contributions}\\
M.M.D. ideated and supervised the writing of this review. All authors discussed and contributed to the writing.\\

\noindent\textbf{Competing interests}\\
The authors declare no competing interests.

\end{document}